\documentclass[preprint,12pt,sort&compress]{elsarticle}
\usepackage[hidelinks]{hyperref}	
\usepackage{amsmath}
\usepackage{mathtools}
\usepackage{amssymb}
\usepackage{mathrsfs}
\usepackage{color}
\usepackage{verbatim}
\usepackage{sistyle}
\usepackage{subcaption}
\usepackage[nameinlink,capitalise]{cleveref} 

\usepackage{bbm}
\usepackage{overpic} 
\usepackage[mathscr]{eucal}
\usepackage{mathrsfs}
\usepackage{amscd}

\usepackage{lineno}
\usepackage[top=1in,bottom=1in,left=1in,right=1in]{geometry}

\usepackage{rotating}

\DeclareMathOperator{\tr}{{{tr}}}

\newcommand{\f}[1]{{\boldsymbol{#1}}}
\newcommand{\baf}[1]{{\bar{\boldsymbol{#1}}}}
\newcommand{\C}[1]{{\mathcal{#1}}}

\newcommand{\R}[1]{{{\rm{#1}}}}
\newcommand{\bafR}[1]{{\bar{\boldsymbol{\rm{#1}}}}}
\newcommand{\baR}[1]{{\bar{\rm{#1}}}}
\newcommand{\sub}{\subset}
\newcommand{\bEq}{\begin{equation}}
\newcommand{\eEq}{\end{equation}}
\newcommand{\beq}{\begin{equation*}}
\newcommand{\eeq}{\end{equation*}}
\newcommand{\vphi}{\varphi}
\newcommand{\car}{\times}

\newcommand{\emp}{\emph}

\newcommand{\der}{\partial}
\newcommand{\fR}[1]{{\mathbf{#1}}}

\newcommand{\hfR}[1]{\hat{\mathbf{#1}}}

\newcommand{\lf}{\left}
\newcommand{\rg}{\right}

\newcommand{\sS}[1]{{\scriptscriptstyle {#1}}}
\newcommand{\Tra}{^{\mathsf{\sS\!T}}}
\newcommand{\Rn}{\text{I\!R}}

\newcommand{\bAl}{\begin{align}}
\newcommand{\eAl}{\end{align}}

\newcommand{\dt}[1]{{\dot{#1}}}

\newcommand{\ome}{\omega}
\newcommand{\lam}{\lambda}

\newcommand{\veps}{\varepsilon}
\newcommand{\sig}{\sigma}

\newcommand{\del}{\delta}

\newcommand{\tht}{\theta}

\newcommand{\fr}[2]{\frac{#1}{#2}\,}

\newcommand{\hf}[1]{\hat{\boldsymbol{#1}}} 

\newcommand{\haf}[1]{{\hat{\boldsymbol{#1}}}}

\newcommand{\haR}[1]{{\hat{\rm{#1}}}}

\newcommand{\alp}{\alpha}
\newcommand{\bet}{\beta}
\newcommand{\bdg}{\beq\begin{diagram}}
\newcommand{\edg}{\end{diagram}\eeq}

\renewcommand{\r}[1]{{\color{black}{#1}}} 

\begin{document}

\begin{frontmatter}

\title{Automated identification of linear viscoelastic constitutive laws with EUCLID} 

\author[a1]{Enzo Marino\corref{cor1}}
\cortext[cor1]{Corresponding author}
\ead{enzo.marino@unifi.it}

\author[a2]{Moritz Flaschel}
\author[a3]{Siddhant Kumar}
\author[a2]{Laura De Lorenzis}

\address[a1]{Department of Civil and Environmental Engineering, University of Florence, Firenze, Italy}
\address[a2]{Department of Mechanical and Process Engineering, ETH Z{\"u}rich, 8092 Z{\"u}rich, Switzerland}
\address[a3]{Department of Materials Science and Engineering, Delft University of Technology, 2628 CD Delft, The Netherlands}

\begin{abstract}
We extend EUCLID, a computational strategy for automated material model discovery and identification, to linear viscoelasticity. For this case, we perform \textit{a priori} model selection  by adopting a generalized Maxwell model expressed by a Prony series, and deploy EUCLID for identification. The methodology is based on four ingredients: i. full-field displacement and net force data; ii. a very wide material model library - in our case, a very large number of terms in the Prony series; iii. the linear momentum balance constraint; iv. the sparsity constraint. 
The devised strategy comprises two stages. Stage 1 relies on sparse regression; it enforces momentum balance on the data and exploits sparsity-promoting regularization to drastically reduce the number of terms in the Prony series and identify the material parameters. Stage 2 relies on k-means clustering; starting from the reduced set of terms from stage 1, it further reduces their number by grouping together Maxwell elements with very close relaxation times and summing the corresponding moduli. 
Automated procedures are proposed for the choice of the regularization parameter in stage 1 and of the number of clusters in stage 2. 
The overall strategy is demonstrated on artificial numerical data, both without and with the addition of noise, and shown to efficiently and accurately identify a linear viscoelastic model with five relaxation times across four orders of magnitude, out of a library with several hundreds of terms spanning relaxation times across seven orders of magnitude.
\end{abstract}


\end{frontmatter}


\section{Introduction}\label{sec.intro}
The mechanical behavior of linear viscoelastic materials can be described by convolutional constitutive equations in which the stress tensor is a function of the strain history. The relaxation functions of the constitutive integrals are generally well represented by the generalized Maxwell model expressed through a Prony series \cite{Tschoegl1989,Christensen2013}, where the unknown parameters are the shear and bulk moduli and their corresponding relaxation times, and the number of terms in the series is itself unknown. The identification of all these parameters requires the solution of a non-linear regression problem with non-negativity constraints \cite{Gerlach&Matzenmiller2005}. 
If the relaxation times are known {\em a priori}, the identification task is drastically simplified since the associated regression problem becomes linear. 
There is a vast literature proposing methods for both identification scenarios, see also the review in \cite{Tschoegl1989}. An important challenge is that the identification problem is known to be ill-posed \cite{Honerkamp1989}, meaning that the solution may not be unique and that small perturbations in the measured data can produce high variations on the identified parameters. Among the approaches in which the relaxation times are chosen upfront and the corresponding bulk and shear moduli are identified, we mention the collocation method by Schapery \cite{Schapery1962}, with its more recent developments and applications in \cite{Kraus&Niederwald2017,Kraus_etal2017}, the windowing technique \cite{Emri&Tschoegl1993,Tschoegl&Emri1993}, and the multidata method \cite{Cost&Becker1970, Bradshaw&Brinson1997}. A performance comparison of some of these methods is presented in \cite{Gerlach&Matzenmiller2005}.
Ill-posedness is addressed e.g. using Tikhonov regularization (also known as ridge regression) \cite{Honerkamp&Weese1990,Elster_etal1992,Weese1993,Diani_etal2012,Diebels_etal2018}, or the maximum entropy method  \cite{Elster&Honerkamp1991}. Among the approaches which solve for both material parameters and relaxation times, we mention \cite{Baumgaertel&Winter1989}, \cite{Jalocha_etal2015}, \cite{Babaei_etal2016}, and more recently   \cite{Yue_etal2021} and \cite{Monaco_etal2022}, which respectively use Bayesian inference and multi-objective optimization. Linear and non-linear regression methods are compared in \cite{Orbey&Dealy1991}.

The vast majority of the available approaches make use of experimental data obtained through Dynamic Mechanical Analysis (DMA) and quasi-static creep or relaxation curves. These tests do not exploit the wealth of local information contained in full-field displacement/strain data, nowadays readily accessible through measurement technologies such as Digital Image Correlation (DIC) and Digital Volume Correlation (DVC). In \cite{Pagnacco_etal2007}, full-field displacement data are deployed for viscoelastic material identification by minimizing the difference in the forces obtained from the measured displacements and from finite element analysis. The Virtual Field Method, a method specifically designed to solve inverse problems of material identification based on full-field data \cite{Grediac_etal2008,Pierron&Grediac2012,Avril_etal2004}, is applied to viscoelastic materials in \cite{Connesson_etal2015,Hoshino_etal2020}. 

Assuming a viscoelastic material model \textit{a priori} and calibrating its parameters by leveraging available experimental information may fail to result in an accurate description of the material response \r{if the model is not chosen well to interpret the data.} This observation has prompted the emergence of data-driven approaches as more versatile alternatives to classical material models.
E.g. neural networks are powerful for describing complex mathematical relations due to their flexible architecture and large number of tunable parameters.
First applied in the context of material modeling by \cite{ghaboussi_knowledgebased_1991},
they were more recently developed for viscoelastic material behavior; e.g. they are used
in \cite{jung_neural_2006} to learn the viscoelastic stress update,
in \cite{al-haik_prediction_2006,kopal_modeling_2017,jordan_neural_2020} to learn temperature-dependent viscoelastic material behavior, and in
\cite{linka_unraveling_2021} to predict the Prony parameters of a viscoelastic material at finite strains. In \cite{huang_variational_2022}, input convex neural networks are employed to learn the thermodynamic potentials of viscoelastic materials which govern the material response.
Importantly, \cite{huang_variational_2022} and \cite{xu_learning_2021} train the neural networks for viscoelasticity by leveraging indirect data, which are easier to acquire through experimental testing than labeled stress-strain data tuples.
Other authors depart from pure machine-learning models in favor of physics-augmented approaches; e.g. in \cite{gonzalez_learning_2019} data are used for learning viscoelastic corrections to conventional hyperelastic material models.
Another stream of research bypasses material modeling altogether by running finite element simulations that are directly informed by the data \cite{kirchdoerfer_data-driven_2016,chinesta_data-driven_2017}, an idea which was recently extended to viscoelastic material behavior in the frequency domain \cite{salahshoor_model-free_2023}.
\r{For both machine-learning-based and model-free approaches, the material behavior is not amenable to physical interpretation nor to mathematical analysis, as it is encoded in a black-box tool (the trained neural network) or in the raw data set.}

We recently proposed a new method for automated discovery of material models based on full-field displacement and global force data, which we denote as EUCLID (Efficient Unsupervised Constitutive Law Identification and Discovery). The idea behind EUCLID is to start from a very large modeling space (a “library" or “catalogue" of material models), and to simultaneously perform model selection and  parameter identification by enforcing balance of linear momentum along with sparsity-promoting regularization. The outcome is a parsimonious and interpretable expression for the material model.
Thus far, EUCLID was applied to hyperelastic \cite{Flaschel_etal2021} and elastoplastic materials \cite{Flaschel_etal2022}, and more recently generalized to the wide class of standard dissipative materials \cite{flaschel_automated_2022}. This contribution included viscoelasticity; however, the focus was on evaluating the ability of EUCLID to automatically discriminate between different categories of constitutive behavior (e.g. elasticity, plasticity with different types of hardening, viscoelasticity, viscoplasticity), and for each category catalogues of relatively limited extent were adopted (including a simple linear viscoelastic model with only one Maxwell element). 
For hyperelasticity, we also developed versions of EUCLID relying on  Bayesian regression \cite{joshi_bayesian-euclid_2022} and on input-convex neural networks \cite{thakolkaran_nn-euclid_2022}.

In this paper, we extend EUCLID to viscoelasticity. For this case, we perform model selection a priori and target linear viscoelasticity with a generalized Maxwell model. This implies no significant limitation, as a Prony series with a sufficient number of terms is known to be able to approximate a very general linear viscoelastic behavior. Thus, we deploy EUCLID for the identification procedure and aim at exploiting its favorable features to solve the aforementioned issues with identification of linear viscoelastic models.
The determination of the relaxation times is addressed by starting with an extremely large catalogue of possible values, which has no significant impact on the overall efficiency of the method (and is facilitated by the frequency domain formulation). To automatically select only a few relevant features, we use Lasso (or $l_1$) regularization \cite{Tibshirani1996}, which preserves the stability of the ridge ($l_2$) regression while promoting sparsity in the set of the Prony series terms.
The non-uniqueness of the solution, which may manifest itself with two or more Maxwell elements being associated with very similar relaxation times, is addressed through an automatic clustering stage.
\r{Thus, compared to previous versions of EUCLID, the main novelty aspects lie in i. the formulation of the problem in the frequency domain, leading to a different expression of the physics-driven loss function; ii. the use of Lasso regularization, which leads to a convex minimization problem and thus significantly enhances the efficiency; the introduction of the clustering phase, which is a powerful tool if model features in the library are highly correlated.} 

The remainder of this paper is organized as follows. After a brief review of the linear viscoelastic problem in Section 2, Section 3 formulates the inverse problem of material identification in the frequency domain. In Section 4 we present our two-stage identification strategy, which is tested and discussed in Section 5. Finally, Section 6 draws the main conclusions.

\section{Brief review of the linear viscoelastic problem}\label{sec:formulation}
As follows, we introduce some simple relationships valid for linear viscoelasticity, in the continuum and discretized frameworks, that are useful for the subsequent developments.
\subsection{Linear viscoelastic constitutive laws in the time and frequency domains}
Let $\C B \sub \Rn^3$ be our physical domain and $T \sub \Rn$ the time interval of interest. For any $\mathbf{x}\in \C B$ and $t \in T$, 
we preliminarily write the volumetric-deviatoric decomposition of  the Cauchy stress tensor as $\f \sig(\mathbf{x},t) = \fR s(\mathbf{x},t) + {\rm p}(\mathbf{x},t)\mathbf{I}$, where $\fR s(\mathbf{x},t)$ is the deviatoric stress tensor, $\R p = \fr 1 3 \tr(\f\sig)$ is the pressure, and $\mathbf{I}$ denotes the identity tensor.
Similarly, for the infinitesimal strain tensor we have $\f \veps(\mathbf{x},t) = \fR e(\mathbf{x},t)+ \tht(\mathbf{x},t)\mathbf{I}$,  where $\tht  = \fr 1 3 \tr(\f\veps)$ is the volumetric strain, and $\fR e(\mathbf{x},t)$ is the deviatoric strain tensor.  
The stress-strain relations for a linear isotropic viscoelastic material can be expressed as 
\begin{align}
\fR s(\mathbf{x},t)  & = \int_{-\infty}^t G(t-\tau)  \dt{\fR e}(\mathbf{x},\tau) d \tau \,, \label{eq:relax_td_G}\\
\R p(\mathbf{x},t) & = \int_{-\infty}^t K(t-\tau) \dt{\tht}(\mathbf{x},\tau) d \tau\,, \label{eq:relax_td_K}
\end{align} 
where $G(t)$ and $K(t)$ are independent functions referred to as relaxation functions \cite{Christensen2013}, and we denote the time derivative with a superposed dot.

Let us now consider as (steady-state) strain history a harmonic function of time with circular frequency $\ome$ and phase angle $\phi$,  i.e.
\begin{align}
\fR e(\mathbf{x},t) &= \bafR e(\mathbf{x},\ome)\exp (i\phi) \exp (i\ome t) = \hfR e(\mathbf{x},\ome)\exp (i\ome t)   \,,\label{eq:e_harmonic} \\
\tht(\mathbf{x},t)  &=  \baR \tht(\mathbf{x},\ome)\exp (i \phi) \exp (i\ome t) = 
 \haR\tht(\mathbf{x},\ome)\exp (i\ome t) \,,\label{eq:tht_harmonic}
\end{align} 
where  
$\bafR e(\mathbf{x},\ome)$ and $\baR\tht(\mathbf{x},\ome)$ are the (complex) deviatoric and volumetric strain moduli. 
Note that we have set 
$\hfR e(\mathbf{x},\ome) = \bafR e(\mathbf{x},\ome)\exp (i\phi)$ and 
$\haR \tht(\mathbf{x},\ome) = \baR \tht(\mathbf{x},\ome)\exp (i\phi)$. 
Correspondingly, the deviatoric and volumetric stresses at steady state must be of the form
\begin{align} 
\fR s(\mathbf{x},t) & = \bafR s(\mathbf{x},\ome)\exp (i\vphi)\exp (i\ome t) = \hfR s(\mathbf{x},\ome)\exp (i\ome t)\,,\label{eq:s_harmonic}\\
\R p(\mathbf{x},t)  & = \baR p(\mathbf{x},\ome)\exp (i\vphi)\exp (i\ome t) = \haR p(\mathbf{x},\ome)\exp (i\ome t)\,,\label{eq:p_harmonic}
\end{align} 
where $ \bafR s(\mathbf{x},\ome)$ and $\baR p(\mathbf{x},\ome)$ are the (complex) deviatoric and volumetric stress moduli and $\vphi$ is the stress phase angle. As for the strains, we have set $\hfR s(\mathbf{x},\ome) = \bafR s(\mathbf{x},\ome)\exp (i\vphi)$ and $\haR p(\mathbf{x},\ome) = \baR p(\mathbf{x},\ome)\exp (i\vphi)$. 
The linear viscoelastic constitutive model in the frequency domain can then be expressed by introducing the two complex transfer functions $G^\ast(i\ome)$ and $K^\ast(i\ome)$ \cite{Christensen2013}, such that  
\begin{align}
\hfR  s(\mathbf{x},\ome)     & = G^\ast(i\ome)\, \hfR e(\mathbf{x},\ome) \,,\label{eq:s_harmonic2}\\
\haR p(\mathbf{x},\ome)     & = K^\ast(i\ome)\, \hfR \tht(\mathbf{x},\ome) \,.\label{eq:p_harmonic2}
\end{align} 
The functions $G^\ast(i\ome)$ and $K^\ast(i\ome)$ are the Fourier transforms of $G(t)$ and $K(t)$ \cite{Christensen2013} and can be decomposed into real and imaginary parts as follows
\begin{align*}
G^\ast(i\ome) &= G^s(\ome) + i G^l(\ome)   \,,\\
K^\ast(i\ome) &= K^s(\ome) + i K^l(\ome) \,,
\end{align*} 
where $G^s(\ome)$ and $K^s(\ome)$ are often denoted as shear and bulk \emp{storage} moduli, respectively, whereas $G^l(\ome)$ and $K^l(\ome)$ are the shear and bulk \emp{loss} moduli.

The relaxation functions $G(t)$ and $K(t)$, or equivalently their transforms $G^\ast(i\ome)$ and $K^\ast(i\ome)$, entirely characterize the viscoelastic material response. 

\subsection{Discrete weak form of linear momentum balance in the frequency domain}
\label{weak_form}
Neglecting body forces and inertial effects, the weak form of linear momentum balance in the frequency domain can be written as
\bEq
\int_{\C B} \haf \sig : \del \haf\veps\, dV  = \int_{\der \C B_t} \hf{t} \cdot \del \haf u\, dS
\eEq
where
$\hat{\f \sig}(\mathbf{x},\omega) = \hat{\fR s}(\mathbf{x},\omega) + {\hat{\rm p}}(\mathbf{x},\omega)\mathbf{I}$, $\der \C B_t$ is the Neumann portion of the domain boundary $\der \C B$ with imposed traction $\hf{t}(\mathbf{x},\omega)$ (zero in our case as we assume displacement-controlled loading), and the equality has to hold for all admissible test functions $\del \haf u$, i.e. for all those that are sufficiently regular and vanish on the Dirichlet boundary $\der \C B_u$. 
By introducing a spatial discretization in $n_e$ linear three-node finite elements $\C B = \bigcup_{e=1}^{n_e} \C B_e$ for a plane strain problem, the (complex) internal force vector associated with element $\C B_e$ is obtained as
\begin{align}
\f f^{int}_e = &  \int_{\C B_e} \fR B_e\Tra \lf[ \hfR s^h(\mathbf{x},\ome)  + \haR p^h(\mathbf{x},\ome)\fR m\rg] dV  = \nonumber \\
                    = &\,  \lf[ G^\ast(i\ome) \int_{\C B_e} \fR B_e\Tra \fR D\fR B_D\, dV + K^\ast(i\ome) \int_{\C B_e} \fR b\Tra \fR b\, dV\rg] \haf u_e\,,\label{eq:finte}
\end{align}
where $\hfR s^h(\mathbf{x},\ome)$ and $\haR p^h(\mathbf{x},\ome)$ are the spatially discretized counterparts of $\hfR s(\mathbf{x},\ome)$ and $\haR p(\mathbf{x},\ome)$, respectively, with $\hfR s^h(\mathbf{x},\ome)$ written in Voigt notation; $\fR B_e$ is the discrete strain-displacement differential operator (a $4\times 6$ array); $\fR m = [1\,1\, 1\,0\,]\Tra$; $\fR B_D = \f\Pi_D\fR B_e$ with $\f\Pi_D = \mathbf{I}_4 -\fr 1 3  \fR m\, \fR m\Tra$ (where $\mathbf{I}_4$ is the $4\times 4$ unit matrix); $\fR b = \fR m\Tra \fR B_e$;
\bEq
\fR D = 
\lf[ 
\begin{array}{c c c c}
2 & 0 & 0 & 0\\
0 & 2 & 0 & 0\\
0 & 0 & 2 & 0\\
0 & 0 & 0 & 1\\
\end{array}
\rg],
\eEq
 and
$\haf u_e$ is the $6\times 1$ vector of the element nodal displacements in the frequency domain, $\haf u_e(\ome) = \baf u_e(\ome)\exp(i\phi)$, where $\baf u_e(\ome)$ is the modulus and $\phi$ the phase shift at each frequency $\ome$. 
From \eqref{eq:finte}, the (complex) element  stiffness matrix is thus obtained as
\bEq
\f k_e = \lf[ G^\ast(i\ome)\!\int_{\C B_e} \fR B_e\Tra \fR D\fR B_D\, dV + K^\ast(i\ome)\!\int_{\C B_e} \fR b\Tra \fR b\, dV\rg] \,.\label{eq:ke}
\eEq

\section{EUCLID for identification of linear viscoelastic constitutive laws}
As follows, we describe the four fundamental ingredients of EUCLID \citep{Flaschel_etal2021,joshi_bayesian-euclid_2022,Flaschel_etal2022,thakolkaran_nn-euclid_2022}: i. a wide and versatile material model library; ii. the data; iii. the linear momentum balance constraint; iv. the sparsity constraint\footnote{Conceptually, one would expect the data to be described first. However, since in the present investigation the data are generated numerically, for the clarity of the presentation it is more convenient to start from the material model library, part of which is then used for finite element data generation.}. 

\subsection{Material model library}
\label{library}
As anticipated earlier, we describe linear viscoelastic behavior with
the generalized Maxwell model, a highly versatile ansatz in which the relaxation functions are expressed as the following Prony series
\begin{align}
G(t) & = G_\infty + \sum_{\alp = 1}^{N_G} G_\alp \exp(-\fr t {\tau_{G_\alp}})\,,\label{eq:Prony_G_td}\\
K(t) & = K_\infty + \sum_{\alp = 1}^{N_K} K_\alp \exp(-\fr t {\tau_{K_\alp}})\,.\label{eq:Prony_K_td}
\end{align} 
Here $G_\infty, G_\alp, \tau_{G_\alp}$ with $\alp = 1,\ldots,N_G$ are the material parameters related to the deviatoric response, whereas $K_\infty, K_\alp, \tau_{K_\alp}$ with $\alp = 1,\ldots,N_K$ are those of the volumetric response, and $N_G$ and $N_K$ are the numbers of Maxwell elements for the deviatoric and volumetric series, respectively. The generalized Maxwell model is known to be able to describe viscoelastic materials of arbitrary complexity if the number of rheological elements is sufficiently large. Hence, we intend to adopt very large values of $N_G$ and $N_K$ to obtain a very flexible model ansatz able to reproduce highly complex material behavior.

Through the Fourier transform of Eqs.~\eqref{eq:Prony_G_td} and \eqref{eq:Prony_K_td}, we obtain $G^\ast(i\ome)$ and $K^\ast(i\ome)$ as follows 
\begin{align}
G^\ast(i\ome) & = G_\infty + \sum_{\alp = 1}^{N_G} G_\alp \fr{\ome^2 \tau_{G_\alp}^2}{1+\ome^2 \tau_{G_\alp}^2} + i \sum_{\alp = 1}^{N_G} G_\alp \fr{\ome\tau_{G_\alp}}{1+\ome^2 \tau_{G_\alp}^2} \,,\label{eq:PronyG_fd}\\
K^\ast(i\ome) & = K_\infty + \sum_{\alp = 1}^{N_K}  K_\alp \fr{\ome^2 \tau_{K_\alp}^2}{1+\ome^2 \tau_{K_\alp}^2} + i \sum_{\alp = 1}^{N_K} K_\alp \fr{\ome\tau_{K_\alp}}{1+\ome^2 \tau_{K_\alp}^2}\,,\label{eq:PronyK_fd}
\end{align}
which can be conveniently written in a more compact form as
\begin{align}
G^\ast(i\ome) &= \f G\Tra \f B_G^s (\ome;\tau_{G_1}...\tau_{G_{N_G}}) + i\,\f G\Tra \f B_G^l(\ome;\tau_{G_1}...\tau_{G_{N_G}})\,, \label{eq:Prony_G_fd_scalarpr}\\
K^\ast(i\ome) &= \f K\Tra \f B_K^s (\ome;\tau_{K_1}...\tau_{K_{N_K}}) + i\,\f K\Tra \f B_K^l(\ome;\tau_{K_1}...\tau_{K_{N_K}})\,, \label{eq:Prony_K_fd_scalarpr}
\end{align}
where
\begin{align*} 
\f G &= \lf[ G_\infty, G_1,\ldots,G_{N_G} \rg]\Tra\,,\\
\f B_G^s &= \lf[1, \fr{\ome^2\tau_{G_1}^2}{1+\ome^2\tau_{G_1}^2},\ldots, \fr{\ome^2\tau_{G_{N_G}}^2}{1+\ome^2\tau_{G_{N_G}}^2} \rg]\Tra\,,\\  
\f B_G^l &= \lf[0, \fr{\ome \tau_{G_1}}{1+\ome^2\tau_{G_1}^2},\ldots, \fr{\ome\tau_{G_{N_G}}}{1+\ome^2\tau_{G_{N_G}}^2} \rg]\Tra\,,\\
\f K &= \lf[ K_\infty, K_1,\ldots,K_{N_K} \rg]\Tra\,,\\ 
\f B_K^s &= \lf[1, \fr{\ome^2\tau_{K_1}^2}{1+\ome^2\tau_{K_1}^2},\ldots, \fr{\ome^2\tau_{K_{N_K}}^2}{1+\ome^2\tau_{K_{N_K}}^2} \rg]\Tra\,,\\   
\f B_K^l &= \lf[0, \fr{\ome \tau_{K_1}}{1+\ome^2\tau_{K_1}^2},\ldots, \fr{\ome\tau_{K_{N_K}}}{1+\ome^2\tau_{K_{N_K}}^2} \rg]\Tra.
\end{align*}

The objective of the identification is thus to select, out of the very wide initial library (containing $N_K+N_G$ Maxwell elements), the minimum number of terms which accurately describes the material behavior, and to calibrate the corresponding values of the bulk and shear moduli.

\subsection{Input data}
\label{data}
In the spirit of  EUCLID, we rely on the availability of experimental full-field displacements (e.g. from DIC/DVC) and net force data, and do not use labeled stress-strain data pairs. Possible data from DMA testing, if available, could be exploited additionally but are not considered here.

In this study, we employ data generated numerically by solving the forward discretized problem of linear viscoelasticity in the frequency domain outlined in Section \ref{weak_form}. For each frequency, the global algebraic system of equations in the unknown nodal displacements is obtained by assembling the element stiffness matrices and internal force vectors in Eqs.~\eqref{eq:ke} and \eqref{eq:finte} and applying Dirichlet boundary conditions.
To set these, we  choose first the frequency range $[\ome_{\min}, \ome_{\max}]$ relevant for the material at hand, and then the number $N_\ome$ of frequencies to be excited during the test. Then, we assign the modulus (and possibly the phase angle) of the applied displacements.
Note that the lowest frequency $\ome_{\min}$ directly affects the capability of the model to discover very large relaxation times. Since we need to identify the moduli at ``infinite" time, a sufficiently small value for $\ome_{\min}$ must be used. Being the problem formulated in the frequency domain, this is not detrimental for the efficiency of data generation.

We mimic experimental displacements from DIC or DVC, which are inevitably affected by noise, by adding to the numerical data a spatially uncorrelated Gaussian white noise with zero mean and standard deviation $\sigma$. The noise is generated in the time domain by choosing, at each node of the mesh, a constant amplitude and a different seed for the generation of random phases. The noise in the frequency domain is then obtained through Fast Fourier Transform and added to the noise-free solution of the forward problem at the chosen $N_\ome$ frequencies.

\subsection{Enforcing balance of linear momentum on the data}

\label{sec:inverse_problem}

With data at hand, and having formulated a wide material model library, we now seek to identify which terms in the library are relevant to describe the material response as observed in the data, and to simultaneously compute the corresponding unknown material parameters.
As full-field displacements in the bulk and net force data for the loaded portions of the boundary are known at different frequencies, the material parameters remain the only unknowns in the weak linear momentum balance equations.
Hence, these equations can serve as the basis to formulate a physics-driven inverse problem for material parameter identification.
To set this problem, it is convenient to rearrange Eq.~\eqref{eq:finte} to express the elemental internal force vector as a linear function of the unknown shear and bulk moduli, i.e. as 
$\f f^{int}_e = \f a_e\f\tht$. Here $\f\tht = [\f G;\f K] = \lf[G_\infty, G_1,\ldots,G_{N_G},K_\infty, K_1,\ldots,K_{N_K} \rg]\Tra$ is a $N_f\times 1$ array collecting all the unknown shear and bulk moduli\footnote{We denote with $[(\cdot)\, , (\cdot)]$ and with $[(\cdot)\, ; (\cdot)]$ horizontal and vertical concatenations of arrays, respectively.}, with $N_f = (N_G+N_K+2)$, and $\f a_e = \lf[\f a^G_e,\f a^K_e\rg] $ is a $6\car N_f$ matrix, with the sub-matrices $\f a^G_e$ and $\f a^K_e$ given by 
\begin{align}
\f a^G_e & = \lf[ \int_{\C B_e}            \lf( \fR B\Tra \fR D\fR B_D\rg)  dV  \haf u_e \rg] \lf[ \f B^s_G + i \f B^l_G \rg]\Tra \label{eq:aGe}\\
 \f a^K_e & = \lf[ \int_{\C B_e}             \lf(\fR b\Tra \fR b\rg) dV  \haf u_e \rg] \lf[ \f B^s_K + i \f B^l_K \rg]\Tra\,.\label{eq:aKe}
\end{align} 

Let $D =\{ (a,i) : a = 1,\ldots,N; i = 1, 2\}$ denote the set of all nodal degrees of freedom. 
$D$ is split in two subsets of internal and boundary degrees of freedom, namely
$D^{int} \sub D$ and $D^{bnd} = D \setminus D^{int}$, respectively. Due to the assumption of  displacement-controlled experiments, we have on the domain boundary either homogeneous Neumann or Dirichlet boundary conditions. Let $D^{bnd,\alp} \subseteq D^{bnd}$ with $\alp = 1,\ldots,N_{bnd}$, such that $D^{bnd,\alp} \cap D^{bnd,\bet} =  \emptyset$ for $\alp\neq\bet$, be the boundary subsets where Dirichlet boundary conditions are enforced and reaction forces can be measured. 
Note that, on the generic subset $D^{bnd,\alp}$, only the net reaction force $\haf r^\alp$  (given by the sum of the reaction forces at all degrees of freedom in the subset) is assumed to be known, as it is realistically measurable through load cells. 

Through the assembly of the element matrices $\f a_e$, the global nodal force vector is obtained as a linear function of the unknown moduli $\f A(\ome) \f\tht = \haf f(\ome)$, where the system matrix $\f A$ has dimensions $2N\car N_f$. 
Let $\f A^{int} = \f A_{|D^{int}}$ be the restriction of the matrix $\f A$ to the internal nodal degrees of freedom $D^{int}$ and 
$\f A^{bnd} = [ \f A_{|D^{bnd,1}}; \ldots ; \f A_{|D^{bnd,N_{bnd} }}  ]$ be the vertical concatenation of the restrictions of the matrix $\f A$ to the boundary degrees of freedom $D^{bnd,\alp}$ with $\alp = 1,\ldots,N_{bnd}$.
The corresponding restrictions on the vector $\haf f$ are $\haf f^{int} = \haf f_{| D^{int}} = \f 0$ (since we do not have body forces) and $\haf f^{bnd} = [\haf r^1; \ldots;\haf r^{N_{bnd}}]$. 

The above matrices $\f A^{int} (\ome),\, \f A^{bnd} (\ome)$ and reaction force vector $\hat {\f f}^{bnd}(\ome)$ are frequency dependent. As mentioned in Section \ref{data}, we assume to perform an experiment in which $N_\ome$ frequencies are excited. By vertically concatenating the contribution of each frequency $\ome_h$ with $h = 1,\ldots,N_\ome$, we build the system $\fR A\f\tht = \fR f$, where
\begin{align}
\fR A & = \lf[\f A^{int}(\ome_1);\ldots ; \f A^{int}(\ome_{N_\ome}); \f A^{bnd}(\ome_1); \ldots; \f A^{bnd}(\ome_{N_\ome}) \rg]\,, \\
\fR f & = \lf[\f f^{int}(\ome_1);\ldots ; \f f^{int}(\ome_{N_\ome}); \f f^{bnd}(\ome_1); \ldots; \f f^{bnd}(\ome_{N_\ome}) \rg]\,, 
\end{align}
which condenses all the measured information on the material response (combining full-field displacements and reaction forces) corresponding to the excited frequencies.

Finally we note that the obtained complex system $\fR A\f\tht = \fR f$ is equivalent to the two real systems of equations involving its real and imaginary parts. Hence, the final linear system of real equations $\mathbbm{A}\f\tht = \mathbbm{b}$ is obtained by vertically concatenating the real and imaginary parts of $\fR A$ and $\fR f$, i.e., $\mathbbm{A} = [ \Re(\fR A); \Im(\fR A) ]$ and  $\mathbbm{b} = [\Re(\fR f); \Im(\fR f) ]$.

\subsection{Sparsity promotion through Lasso regularization} 
The linear system obtained in the previous section is overdetermined and can be solved in a least square sense as follows
\bEq
\f\tht^{opt} = \arg \min_{\f\tht} || \mathbbm A  \f{\tht} - \mathbbm b ||^2\,. \label{eq:opt_pb}
\eEq
However, due to the ill-posed nature of the problem, ordinary least square estimates obtained by \eqref{eq:opt_pb} are often not satisfactory. 
Further, recall that we do not know the relaxation times upfront and thus start from a highly flexible model ansatz containing a very large number of Maxwell elements (see Section \ref{library}). This implies that solving \eqref{eq:opt_pb} would in general result in a highly complicated material model with a very large number of material parameters (as large as in the assumed library). 
In the spirit of EUCLID, we seek to promote sparsity, i.e. to automatically select only a small subset of the material parameters contained in the model library to obtain a parsimonious model. To this end, we exploit the Lasso (least absolute shrinkage and selection operator) regularization technique \cite{Tibshirani1996} (see also the preliminary work in \cite{frank_statistical_1993}) and rewrite the optimization problem in Eq.~\eqref{eq:opt_pb} as follows
\bEq
\f\tht^{opt} = \arg \min_{\f\tht}\lf( || \mathbbm A  \f{\tht} - \mathbbm b ||^2  +\lam \sum_{i = 1}^{N_f} |\tht_i| \rg)\,. \label{eq:opt_pb_lasso}
\eEq
The regularization term added to the loss function penalizes solution vectors with many non-zero entries and hence promotes sparsity in $\f\tht$. The penalty parameter $\lam$ controls the importance of the regularization term relative to the linear momentum balance term. The higher $\lam$, the larger the number of features which are set to zero in the final solution vector, i.e. removed from our material model library.
An intuitive rule for selecting $\lam$ is discussed in Section \ref{sec:2stage_strategy}.
To solve \eqref{eq:opt_pb_lasso}, we use the built-in Matlab function \texttt{lasso}, which employs the coordinate descent technique \cite{Friedman_etal2010}.

\section{Two-stage identification strategy}\label{sec:2stage_strategy}
To promote parsimony in the most effective way, we propose an identification strategy based on two subsequent stages.
In the first stage, the Lasso-regularized optimization problem \eqref{eq:opt_pb_lasso} is solved to obtain a sparse solution with a small number of Maxwell elements; in the second stage, the sparsity of the solution is further enhanced by merging Maxwell elements with similar relaxation times.

\subsection{Stage 1: sparse regression}
The first stage of the discovery process consist in the following steps: 
\begin{enumerate}
\item Set the size of the material library, i.e. choose $N_ G$ and $N_K$, from which the total number of unknown model features results as $N_f = N_G+N_K +2$.
\item Choose the range of relaxation times  $[\tau_{\min}, \tau_{\max}]$ relevant for the material at hand and a discrete set of relaxation times in the range. We assume the $N_G$ relaxation times for the shear deformation and the $N_K$ relaxation times for the bulk deformation to be both equally spaced on a logarithmic scale in the chosen range. 
\item Build the arrays $\mathbbm A$ and $\mathbbm b$ in~\eqref{eq:opt_pb_lasso} as described in \cref{sec:inverse_problem}.
\item Define a discrete set of $\lam$ values from very small (corresponding to almost no regularization) to very large (causing all features to be set to zero) and solve \eqref{eq:opt_pb_lasso} for each of these values.
\item Set a threshold $e_{\lam}$ for the Mean Squared Error (MSE), defined as $|| \mathbbm A  \f{\tht} - \mathbbm b ||^2/ N_f$, and identify the largest value of $\lam$ (i.e. the one leading to the most parsimonious model) which corresponds to an MSE below the threshold, $\lam^{opt}$. Further clarification will follow in Section \ref{Noise-free case}.
\end{enumerate}
The solution of \eqref{eq:opt_pb_lasso} obtained for $\lam=\lam^{opt}$ is the outcome of stage 1. It delivers a material model characterized by a drastically reduced number of features $\f\tht^{opt,(1)}\sub\f\tht^{opt}$ with respect to the initially chosen number $N_f$. The model is thus at the same time parsimonious and accurate, whereby the accuracy (in the satisfaction of linear momentum balance on the data) is dictated by the user-defined choice of $e_{\lam}$.

\subsection{Stage 2: clustering}
\label{clustering}
The standard EUCLID strategy proposed in \cite{Flaschel_etal2021,Flaschel_etal2022,flaschel_automated_2022} is limited to stage 1. With respect to the previous investigations, the present linear viscoelastic case displays two unique features which motivate the introduction of a second stage, namely, 
\begin{itemize}
\item the \textit{a priori} choice of a discrete set of relaxation times stemming from the fine discretization (in the logarithmic scale) of a relaxation time interval, and
\item  the equivalence of two (or more) Maxwell elements with the same relaxation times and different shear (bulk) moduli to a single Maxwell element with the same relaxation time and shear (bulk) modulus equal to the sum of the two (or more).
\end{itemize}

This equivalence naturally calls for a clustering procedure able to condensate Maxwell elements with close relaxation times and thus to further reduce the number of features, reaching the highest level of parsimony in the material model. The ensuing second stage compensates for the difficulties of Lasso regression in choosing among almost linearly dependent features, such as in the case of Maxwell elements with very close relaxation times. 
Moreover, the clustering algorithm in stage 2 is extremely efficient, since it operates on the results of stage 1, for which the number of features is already reduced by orders of magnitude with respect to the initial catalogue size.

Let $N^{(1)}_{G} < N_G$ and $N^{(1)}_{K} < N_K$ be the number of Maxwell elements with non-zero moduli selected in stage 1 for shear and bulk deformations, respectively, such that $N^{(1)}_f = N^{(1)}_G +N^{(1)}_K + 2$ is the number of non-zero model features (dimension of $\f\tht^{opt,(1)}$) after stage 1. Our objective is to find the \emph{minimum} number of clusters $N^{(2)} = N^{(2)}_{G} = N^{(2)}_{K}$ such that the material model obtained by condensing the Maxwell elements belonging to each cluster accurately describes the material response\footnote{We assume here that the optimal final number of Maxwell elements is the same for shear and bulk deformations. This assumption, however, could be easily removed by formulating two separate clustering algorithms, one for the shear and one for the bulk elements.}.

To this end, we gradually increase the number of clusters, $i_{cls}$, starting from 1. For each $i_{cls}$ we deploy a k-means clustering algorithm, based on \cite{Lloyd1982} and implemented in the built-in Matlab function \texttt{kmeans}, which partitions the active relaxation times selected in stage 1 into $i_{cls}$ clusters by minimizing the sum of the distances of the relaxation times within a cluster to the centroid relaxation time. After clustering, the centroid of a cluster represents the corresponding relaxation time, whereas the sum of the moduli belonging to that cluster is the corresponding modulus. For each $i_{cls}$ we compute the associated MSE, and the optimal number of clusters is automatically identified as the value of $i_{cls}$ at which the MSE decreases abruptly. More details will follow in Section \ref{Results}.

\section{Numerical results}
\label{Results}
In this section we test the performance of EUCLID by applying the identification strategy described in Section~\ref{sec:2stage_strategy} to numerically generated data augmented with artificial noise. 

\subsection{Data generation}
Data are generated numerically by adopting a generalized Maxwell model characterized by a total of $N_f = 12$ rheological components with $N_G = N_K = 5$. These ``true" parameters, selected from \cite{Kim_etal2010}, are reported in Table~\ref{tb:true_params}. 

The specimen has a rectangular shape with dimensions $L_x = \SI{100}{mm}$ and $L_y = \SI{500}{mm}$. The two vertical sides are free, the bottom side is fixed in both directions ($u_x = u_y = 0$), whereas on the top side $u_x = 0$ and $u_y$ is given by a periodic function $\bar u_y(\ome_j)\exp(i\phi_j)$ with $j = 1,\ldots,N_\ome$, such that $N_\ome$ frequencies equally spaced over a logarithmic scale spanning from $\ome_{\min} = \SI{0.0009}{rad/s}$ to $\ome_{\max} = \SI{907.5291}{rad/s}$ are excited. In this study we have set $N_\ome= 15$.
\begin{table}
	\centering
	\begin{tabular}{r r r r}
	$\f G$			 & $\f \tau_G$	 	&  $\f K$			& $\f \tau_K$	\\[-3pt]
	$[\SI{}{N/mm^2}]$	 & $[\SI{}{s}]$ 	&  $[\SI{}{N/mm^2}]$	& $[\SI{}{s}]$	\\
	\hline
  	  500				 &	  -				&	2000				&	-\\	
  	  779				 & 0.0728	  			&	2242				& 0.007693\\
  	1019				 & 0.4824	  			&	2712				& 0.063440\\
  	  529				 & 3.9150	  			&	2366				& 0.457000\\
  	 201.1				 & 30.2100	  			&	1097				& 4.197000\\
              96				 & 629.4000	  			&	460.1				& 35.120000\\
              \hline
	\end{tabular}
	\caption{True material parameters selected from \cite{Kim_etal2010}.}\label{tb:true_params}
\end{table}

We performed several tests considering different levels of noise. In the following, we report the results for the two most representative cases, namely the one with no noise and the one with a magnitude of the noise which starts influencing results to a non-negligible extent. This noise level is quite high, indicating a low sensitivity of the proposed strategy to noisy data (at least for the type of noise considered here). All the other tested cases with intermediate levels of noise are not discussed, as their results are nearly indistinguishable from those of the noise-free case. 

\subsection{Inverse problem settings}
For the material library we choose $N_G = N_K = 300$, hence our identification procedure starts with a total number of features $N_f = 602$. Additionally, we choose a relaxation time range spanning seven orders of magnitude between $\tau_{\min} = \SI{e-3}{s}$ and $\tau_{\max} = \SI{e4}{s}$. The 300 candidate relaxation times for both  shear and bulk moduli are taken equally spaced on a logarithmic scale ranging from $\tau_{\min}$ to $\tau_{\max}$. For the regularization parameter $\lam$, we consider 1000 values equally spaced in a logarithmic scale between $\SI{E-12}{}$ and $\SI{E-1}{}$.

\subsection{Noise-free case}
\label{Noise-free case}
\begin{figure}
\centering
\includegraphics[width=0.5\linewidth]{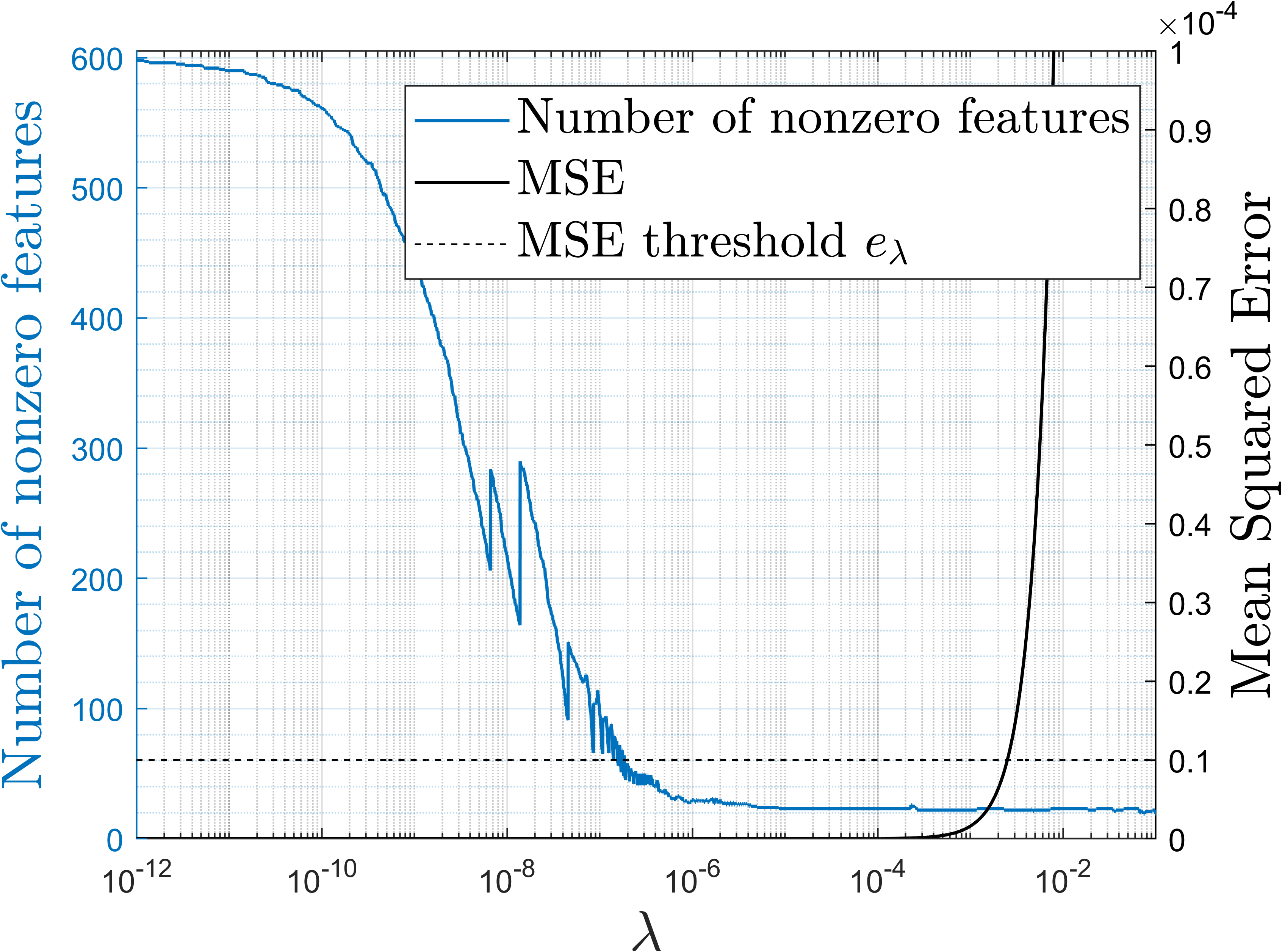}
\caption{Mean Squared Error and number of non-zero features vs. $\lam$ for the noise-free case ($\sig = 0$).}\label{fig:MSE+threshold_vs_lam_sig0}
\end{figure}

Figure~\ref{fig:MSE+threshold_vs_lam_sig0} shows the MSE obtained for the range of tested values of the regularization parameter $\lam$.  As $\lam$ increases, the MSE increases accordingly, indicating that linear momentum balance is satisfied with decreasing accuracy; at the same time, the number of active (non-zero) features in the identified model decreases. Thus the model becomes increasingly simple and decreasingly accurate, confirming a trend that was consistently encountered in the previous investigations on EUCLID \cite{Flaschel_etal2021,Flaschel_etal2022,flaschel_automated_2022}. The horizontal line corresponds to the MSE threshold (here $e_\lam = \SI{1e-5}{}$) selected by the user as the accepted level of error in the satisfaction of linear momentum balance. The largest value of $\lam$ leading to an MSE below this threshold, $\lam^{opt}$, is chosen as the best compromise between complexity and accuracy (in our case, $\lam^{opt} = 0.0025$). Note, however, than any value of $\lam$ larger than about $10^{-6}$ leads to approximately the same  number of non-zero features, hence the specific choice of $\lam^{opt}$ (or of $e_\lam$) is not crucial for the success of the method. 

\begin{figure}[h!]
\centering
\begin{subfigure}{0.47\textwidth}
\centering
\includegraphics[width=1\linewidth]{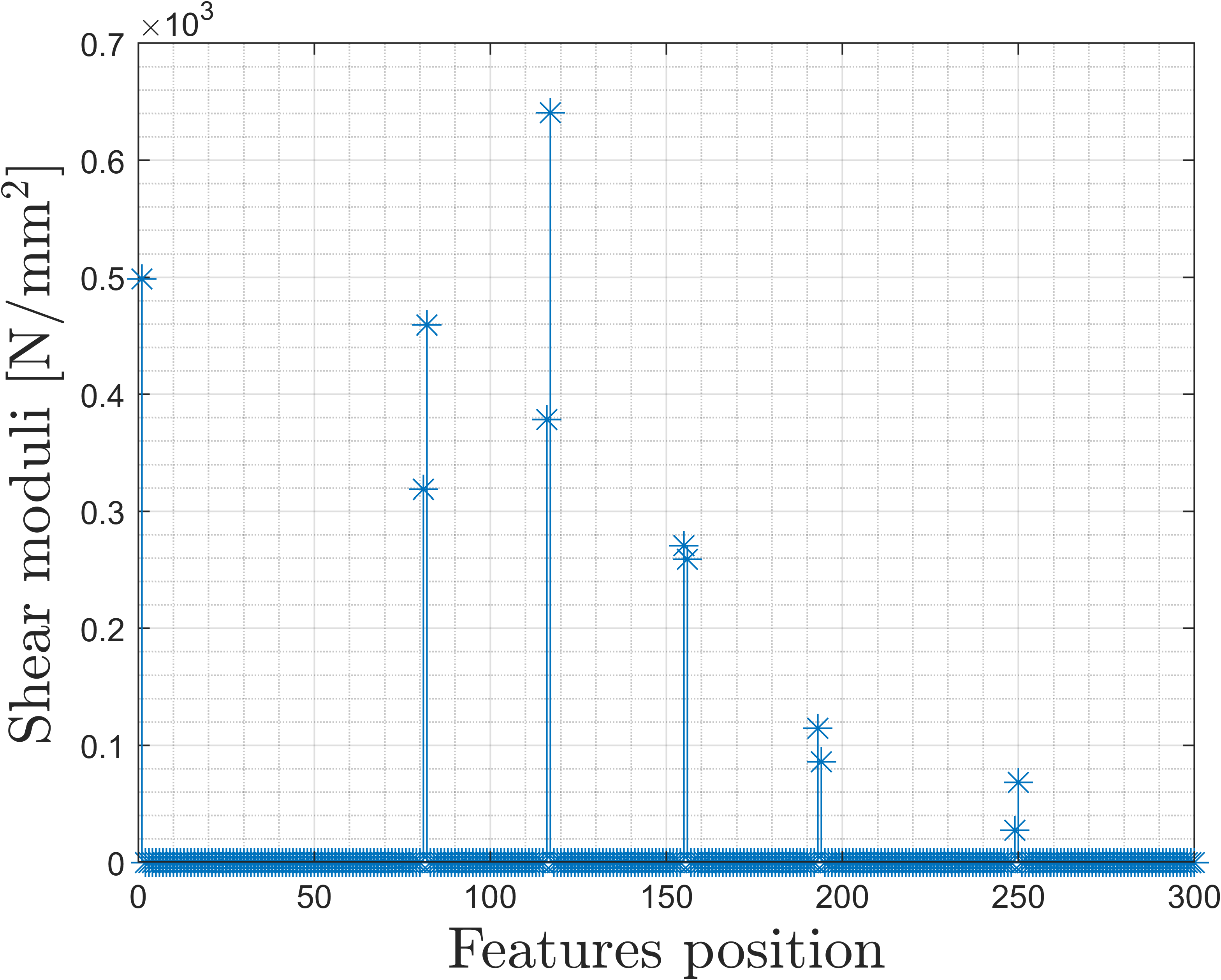} 
\end{subfigure}
\hspace{9pt}
\begin{subfigure}{0.47\textwidth}
\centering
\includegraphics[width=1\linewidth]{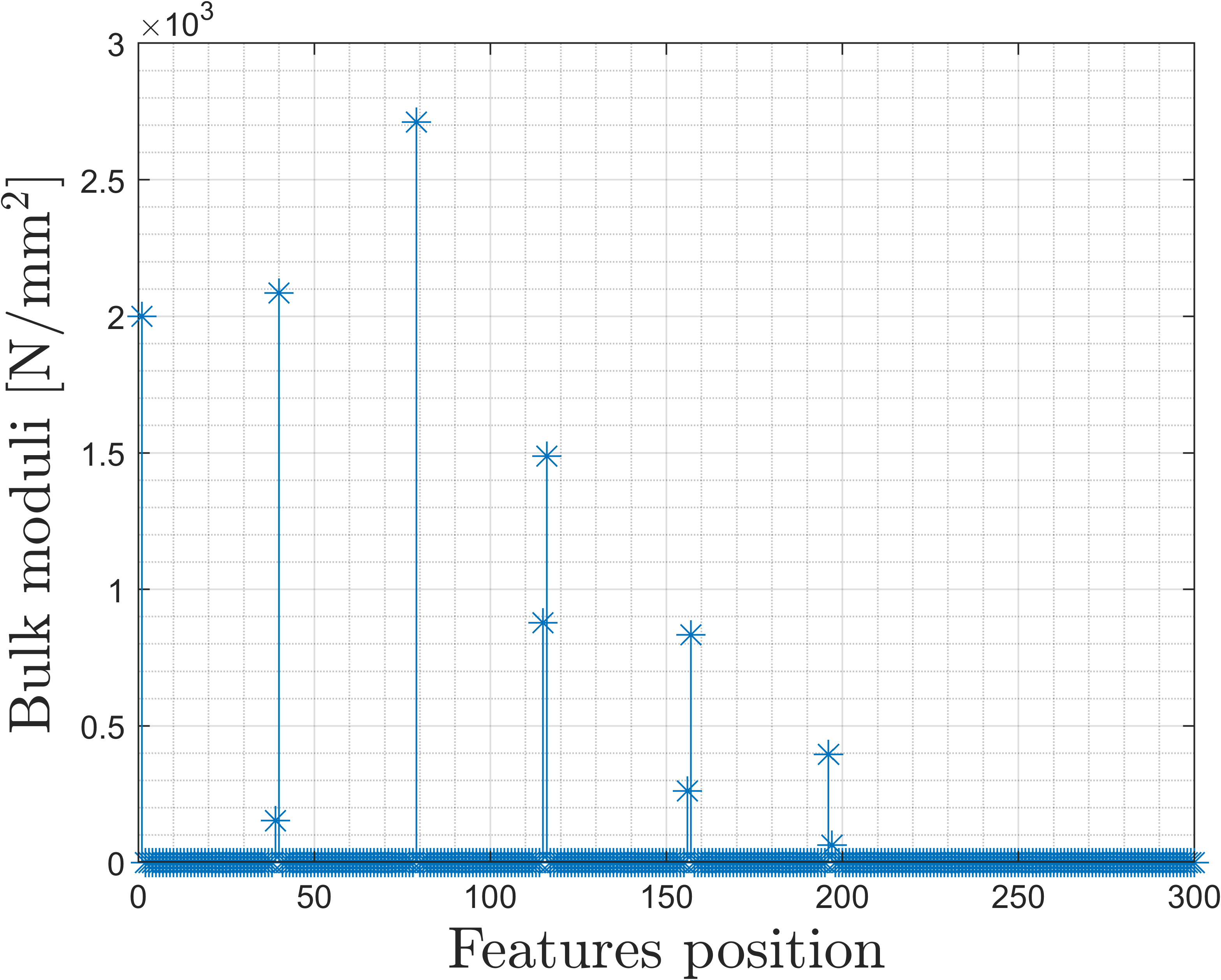} 
\end{subfigure}
\caption{Activated shear (a) and bulk (b) moduli over the entire library of rheological components after stage 1 for the noise-free case ($\sig = 0$).}\label{fig:activated_GK_vs_position}
\end{figure}

\begin{figure}
\centering
\begin{subfigure}{0.47\textwidth}
\centering
\includegraphics[width=1\linewidth]{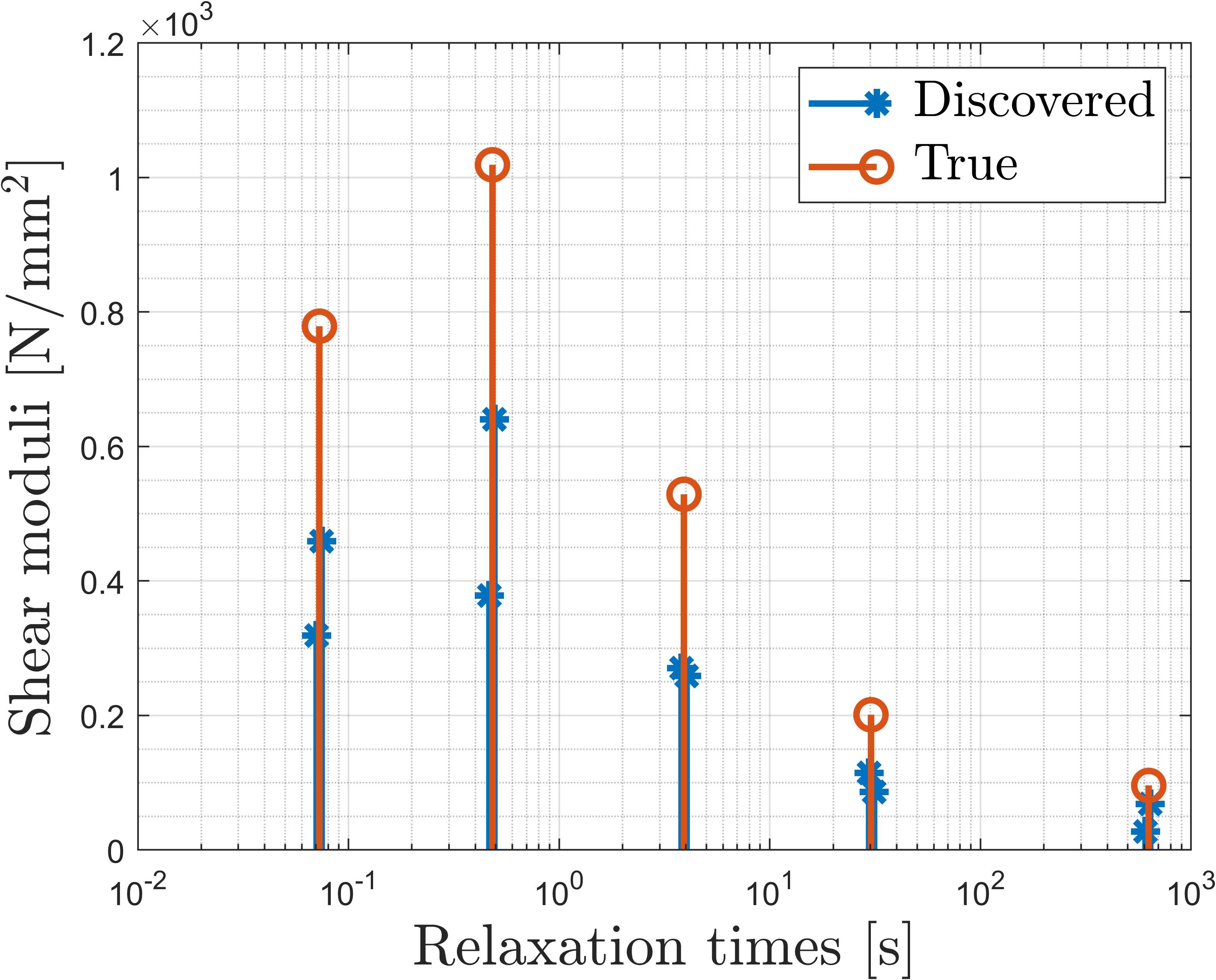} 
\end{subfigure}
\hspace{9pt}
\begin{subfigure}{0.47\textwidth}
\centering
\includegraphics[width=1\linewidth]{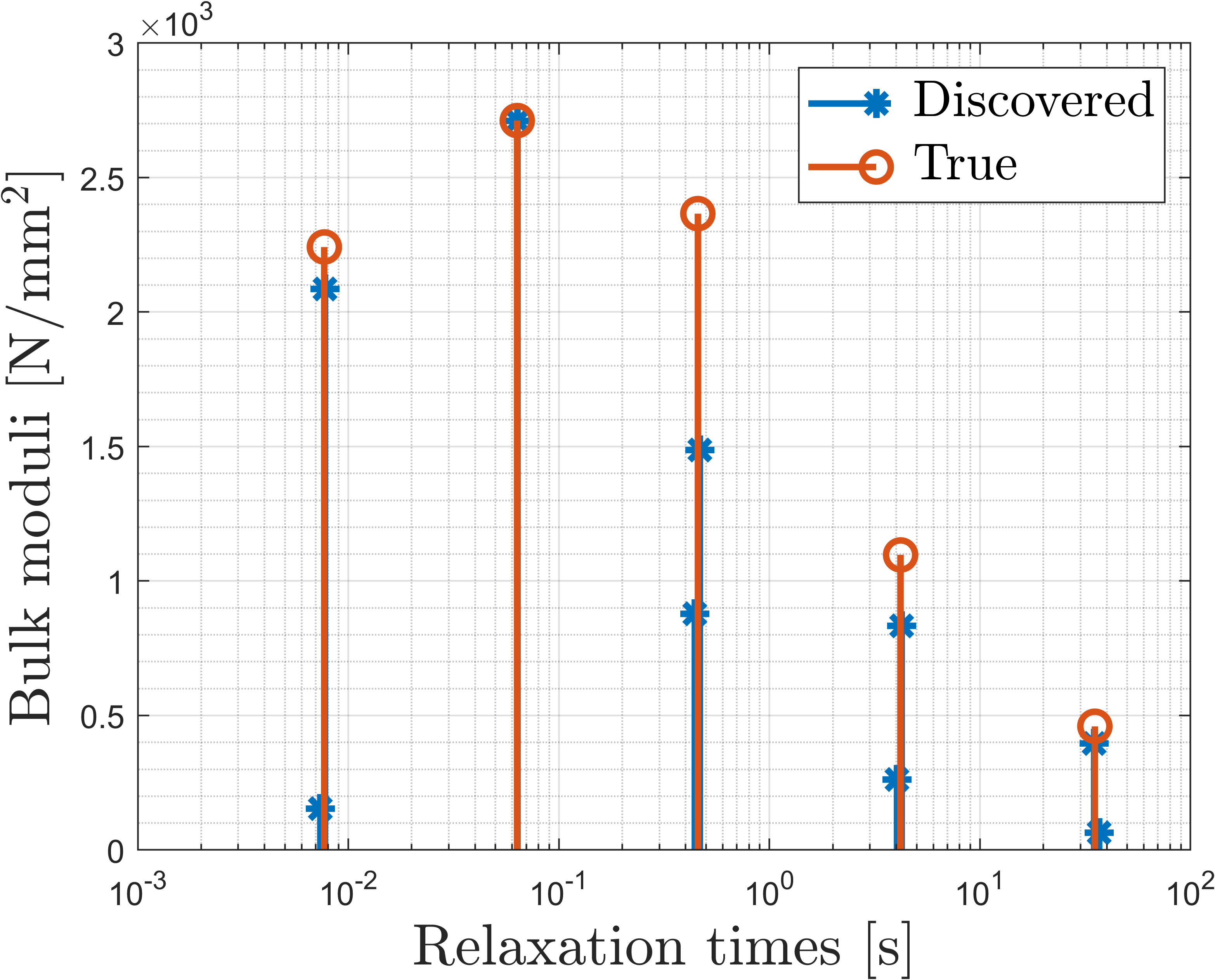} 
\end{subfigure}
\caption{Activated shear (a) and bulk (b) moduli and corresponding relaxation times after stage 1 for the noise-free case ($\sig = 0$).}\label{fig:activated_GK_vs_relaxtimes}
\end{figure}

With this value of $\lam$, the active parameters automatically selected by sparse regression are illustrated in Figure~\ref{fig:activated_GK_vs_position}, which gives the  magnitude of the identified moduli vs. their position in the features vector. 
Out of the original 602 features, the vast majority are automatically set to zero and only 22 are retained (11 for the shear and 11 for the bulk response). 
Figure~\ref{fig:activated_GK_vs_relaxtimes} provides further details by illustrating the selected moduli (excluding $G_\infty$ and $K_\infty$, so that we now show 10 shear and 10 bulk moduli) with the corresponding relaxation times in comparison with the true moduli hidden in the input data. It is evident that the active moduli correspond to quite accurate relaxation times; however, the number of active moduli is the double of the true number, as in the neighborhood of each true relaxation time two moduli are active after stage 1. A closer look reveals that, for each couple of moduli corresponding to the neighborhood of a given relaxation time, the sum of the moduli is very close to the true modulus. Indeed, two Maxwell elements with equal (in our case, very similar) relaxation times are equivalent to a single Maxwell element with the same relaxation time, and modulus given by the sum of the moduli. Thus, the results in Figure~\ref{fig:activated_GK_vs_relaxtimes} motivate the need for stage 2 of the identification procedure (see Section \ref{clustering}). 

\begin{figure}
\centering
\includegraphics[width=0.5\linewidth]{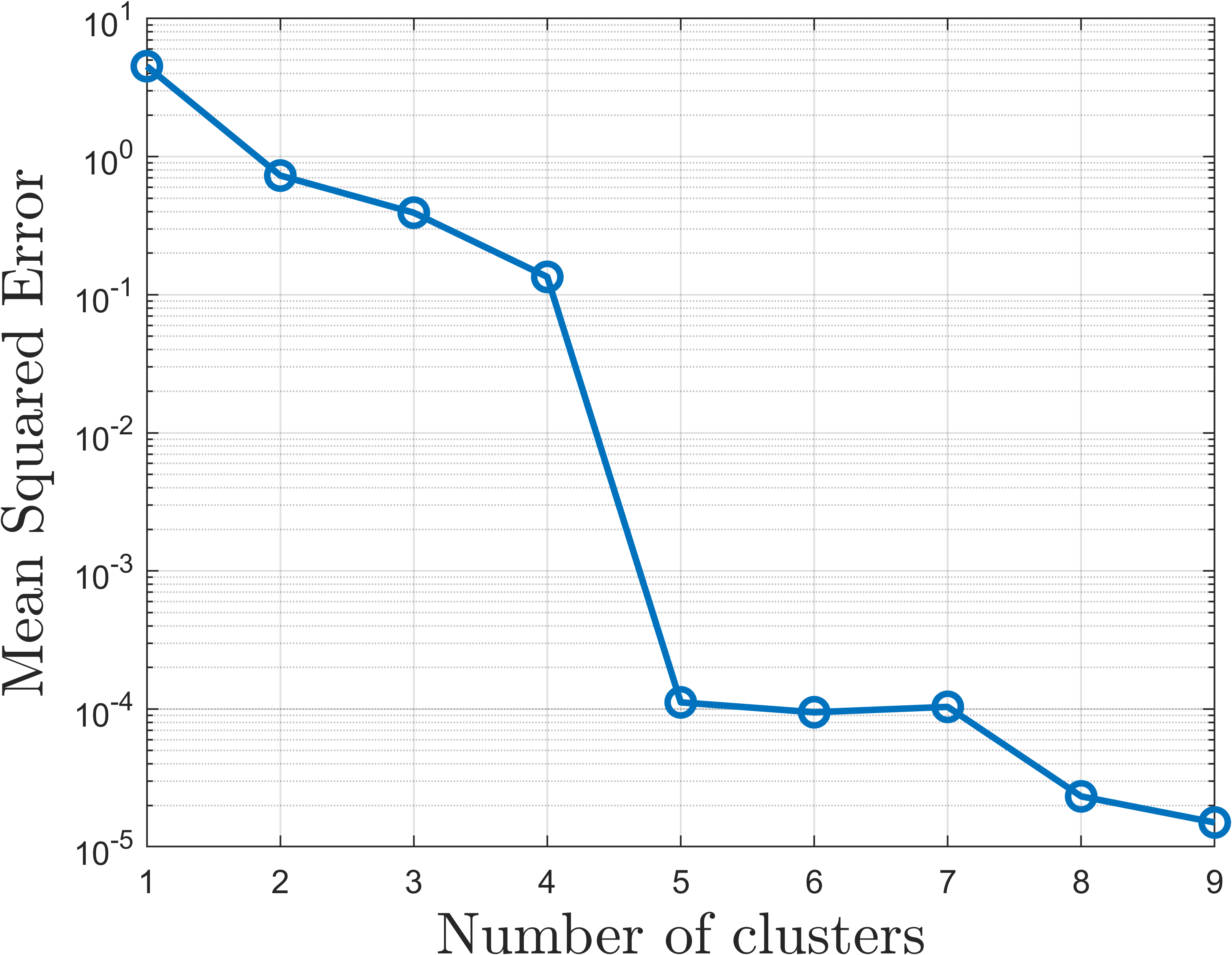}
\caption{Mean Squared Error vs. number of clusters for the noise-free case ($\sig = 0$).}\label{fig:MSElog_vs_clsts_sig0}
\end{figure}

\r{To set an automated procedure to group Maxwell elements,} we loop over the number of clusters, $i_{cls}$, starting from 1 and ending at the unclustered number of moduli (in the present case, 11 for shear and 11 for bulk). For each $i_{cls}$ we perform clustering using the k-means algorithm implemented in Matlab and compute the MSE. Results are shown in Figure~\ref{fig:MSElog_vs_clsts_sig0}, and indicate that a sudden decrease in the MSE (three orders of magnitude) is obtained as the optimal number of clusters is reached (5 in our case).  
For a number of clusters between 5 and 7 the MSE does not vary appreciably, whereas for a number larger than 7 it  slowly decreases further. The observed trend indicates that the choice of 5 as the optimal number of clusters can be easily automatized introducing a suitable criterion on the MSE drop. 

\begin{figure}
\centering
\begin{subfigure}{1\textwidth}
\centering
\hspace{0.05\textwidth}
\includegraphics[width=.2\linewidth]{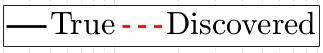} 
\end{subfigure}\\
\begin{subfigure}{0.02\textwidth}
\begin{turn}{90} 
1 Cluster
\end{turn}
\end{subfigure}
\hspace{6pt}
\begin{subfigure}{0.2\textwidth}
\centering
\includegraphics[width=1\linewidth]{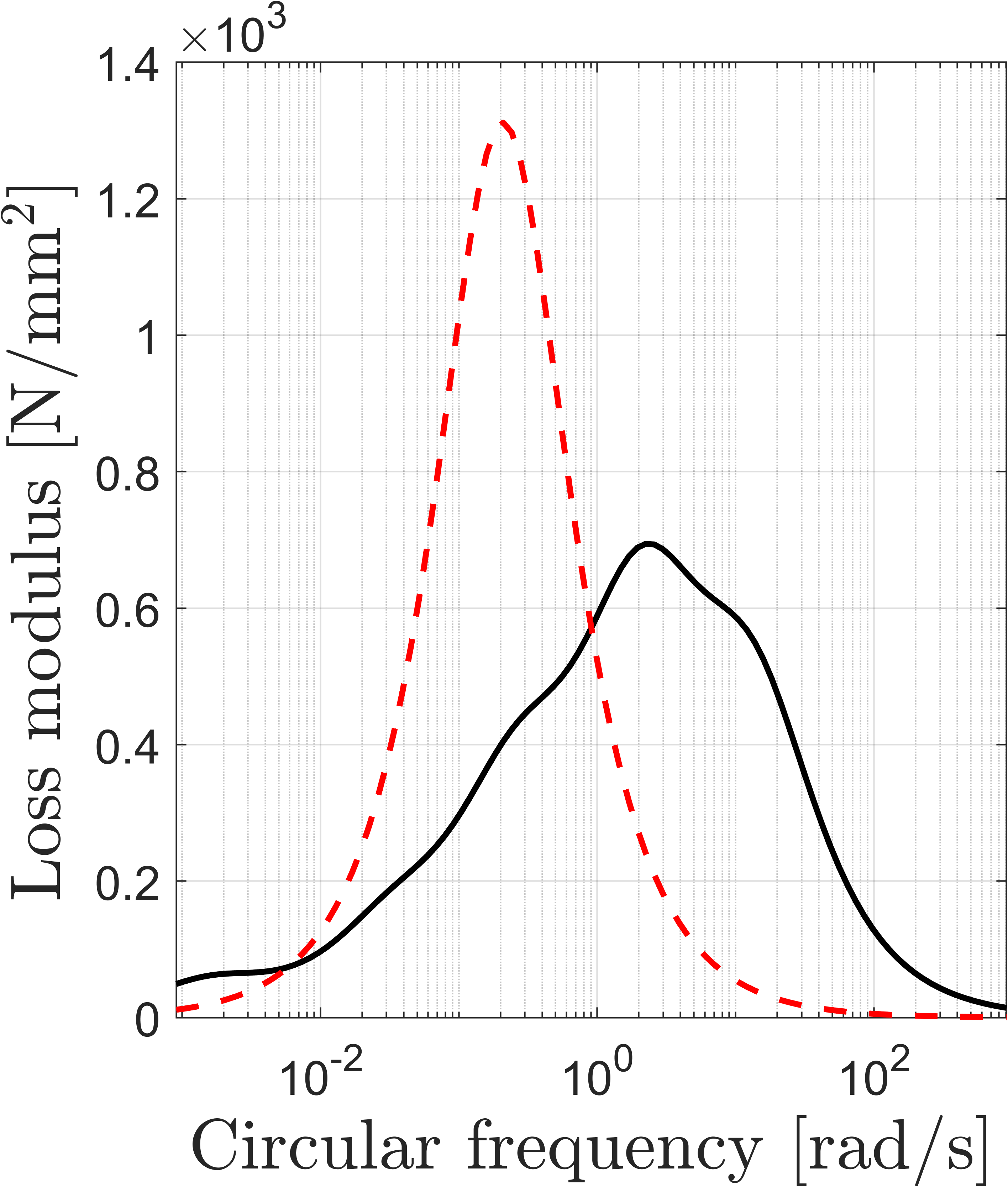} 
\end{subfigure}
\hspace{6pt}
\begin{subfigure}{0.2\textwidth}
\centering
\includegraphics[width=1\linewidth]{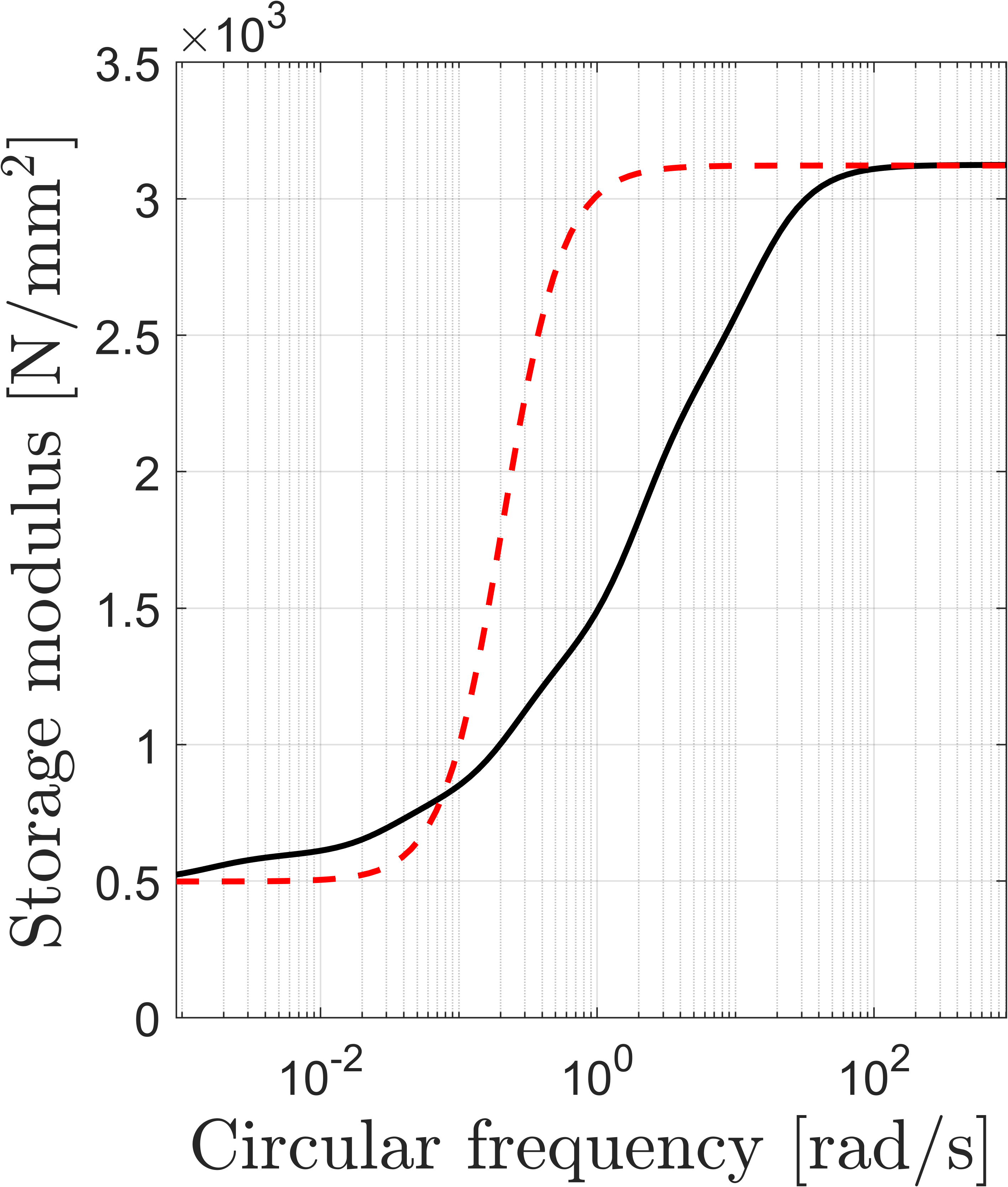} 
\end{subfigure}
\hspace{6pt}
\begin{subfigure}{0.2\textwidth}
\centering
\includegraphics[width=1\linewidth]{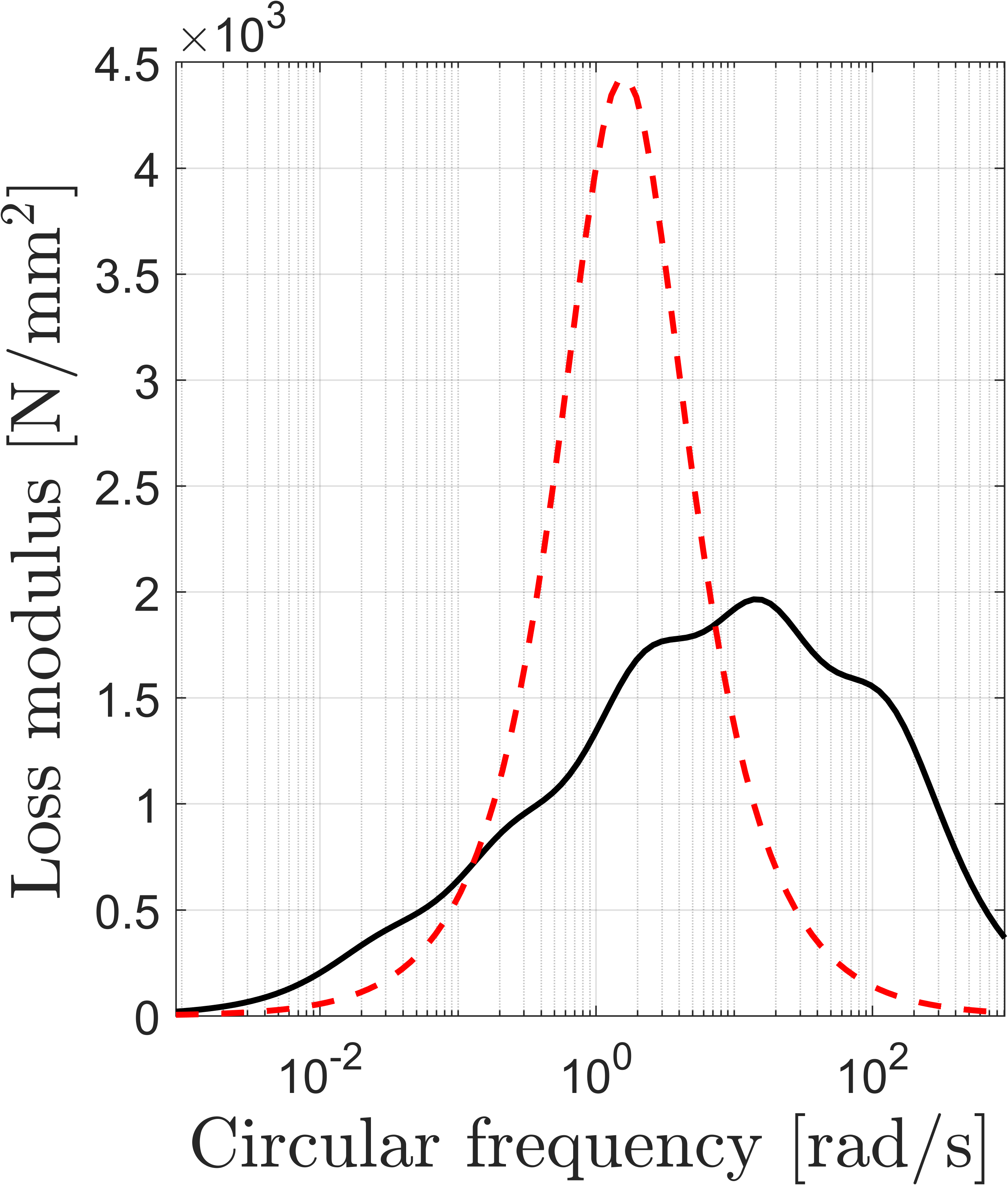} 
\end{subfigure}
\hspace{6pt}
\begin{subfigure}{0.2\textwidth}
\centering
\includegraphics[width=1\linewidth]{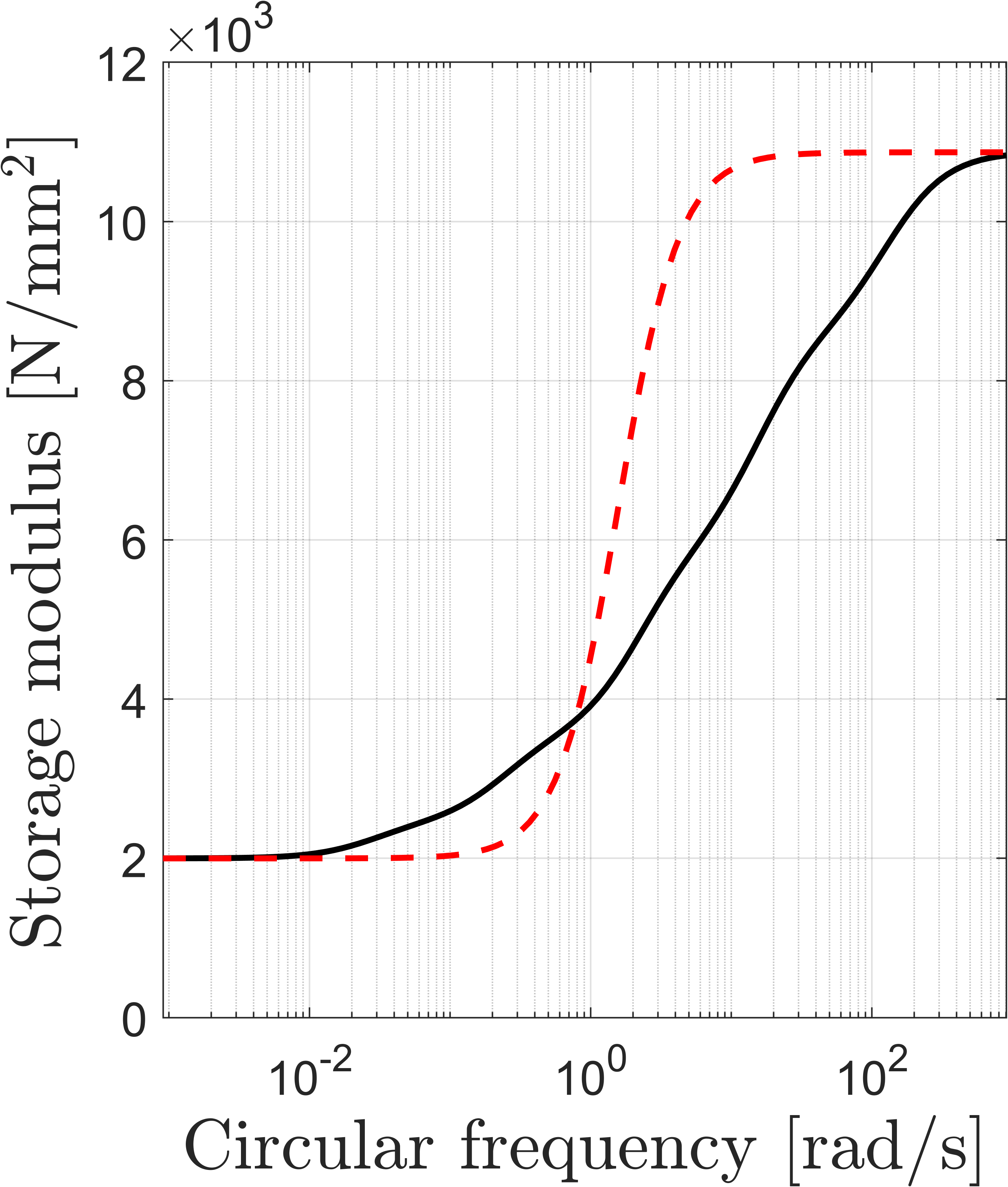} 
\end{subfigure}\\[12pt]
\begin{subfigure}{0.02\textwidth}
\begin{turn}{90} 
2 Clusters
\end{turn}
\end{subfigure}
\hspace{6pt}
\begin{subfigure}{0.2\textwidth}
\centering
\includegraphics[width=1\linewidth]{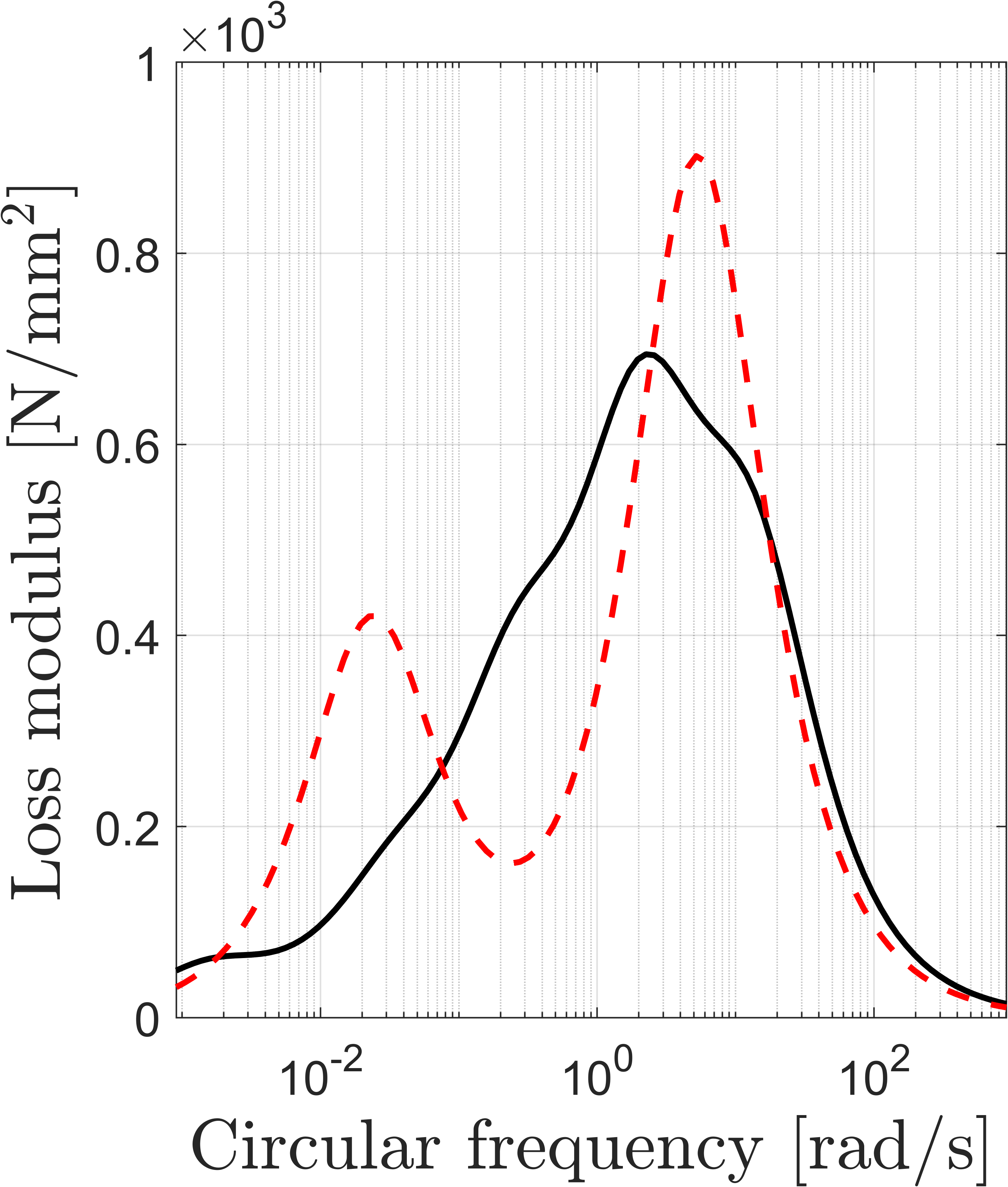} 
\end{subfigure}
\hspace{6pt}
\begin{subfigure}{0.2\textwidth}
\centering
\includegraphics[width=1\linewidth]{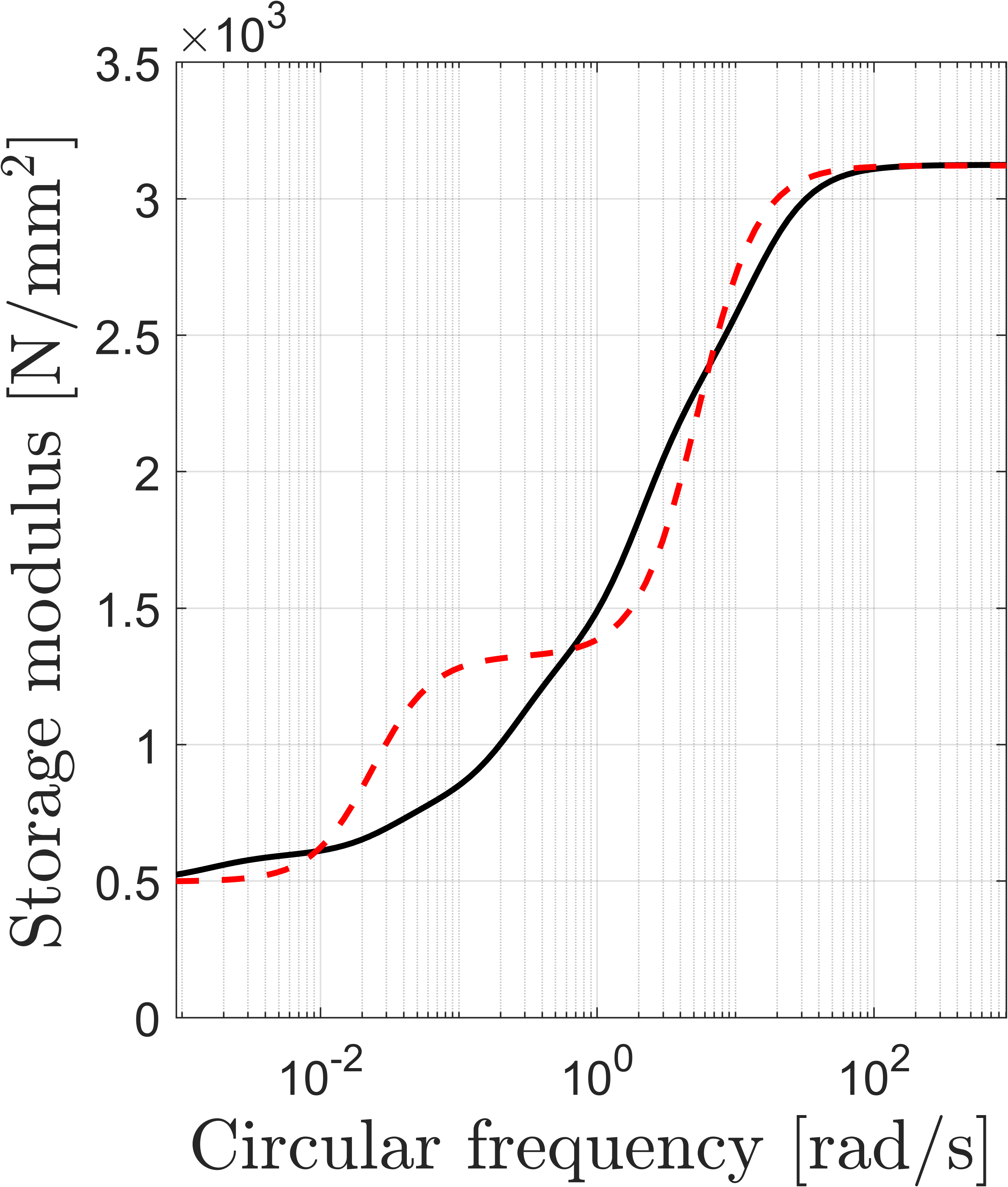} 
\end{subfigure}
\hspace{6pt}
\begin{subfigure}{0.2\textwidth}
\centering
\includegraphics[width=1\linewidth]{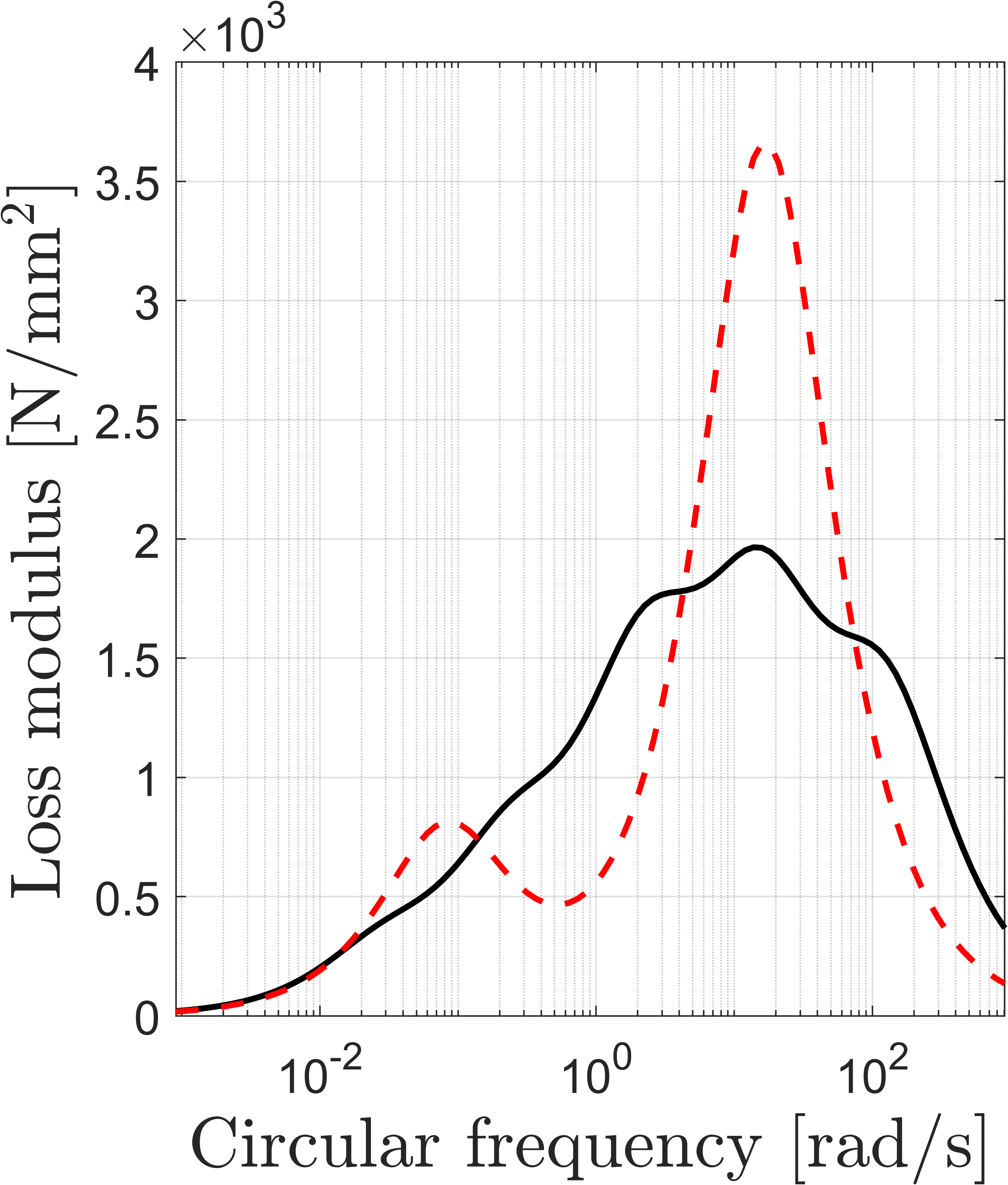} 
\end{subfigure}
\hspace{6pt}
\begin{subfigure}{0.2\textwidth}
\centering
\includegraphics[width=1\linewidth]{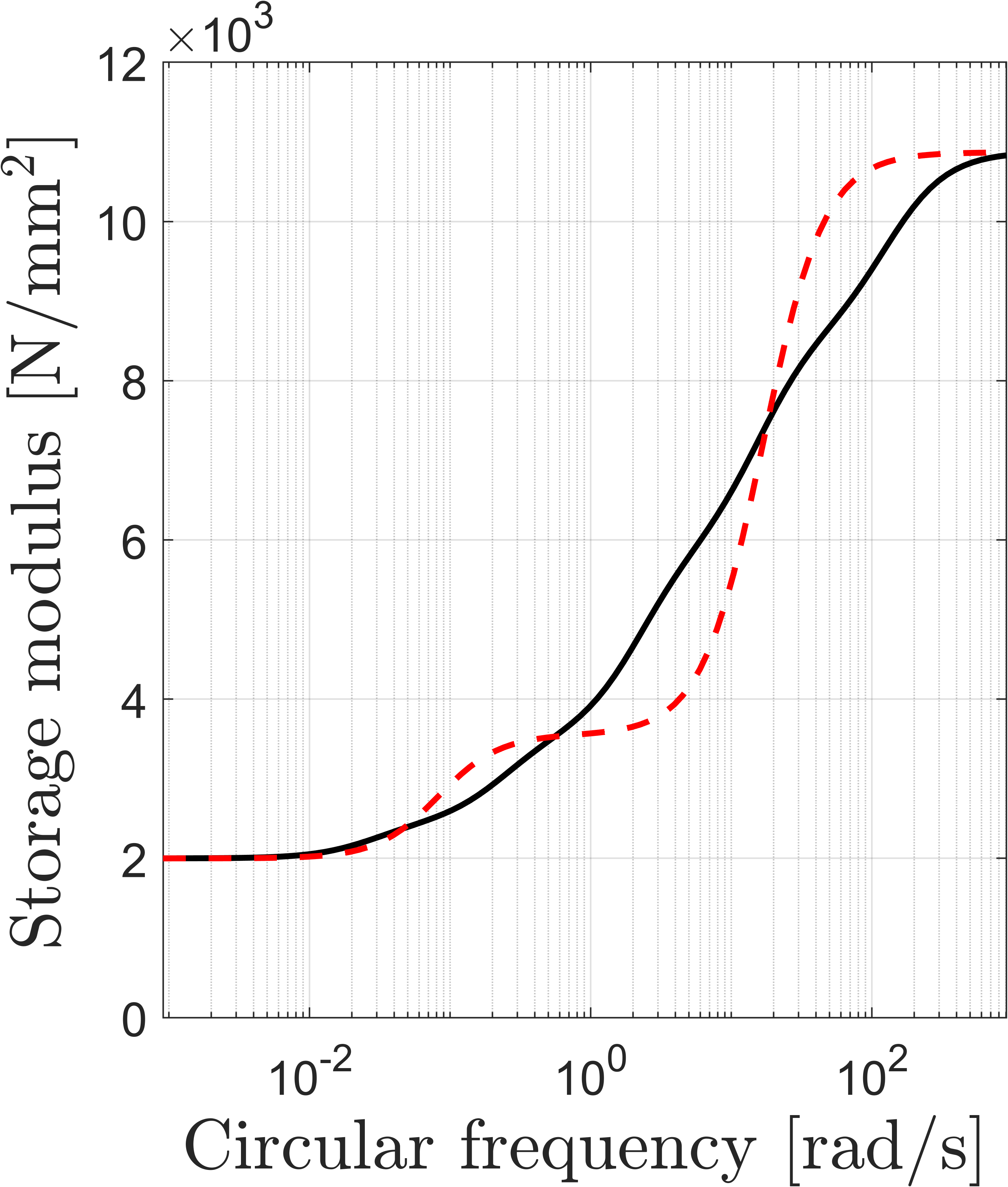} 
\end{subfigure}\\[12pt]
\begin{subfigure}{0.02\textwidth}
\begin{turn}{90} 
3 Clusters
\end{turn}
\end{subfigure}
\hspace{6pt}
\begin{subfigure}{0.2\textwidth}
\centering
\includegraphics[width=1\linewidth]{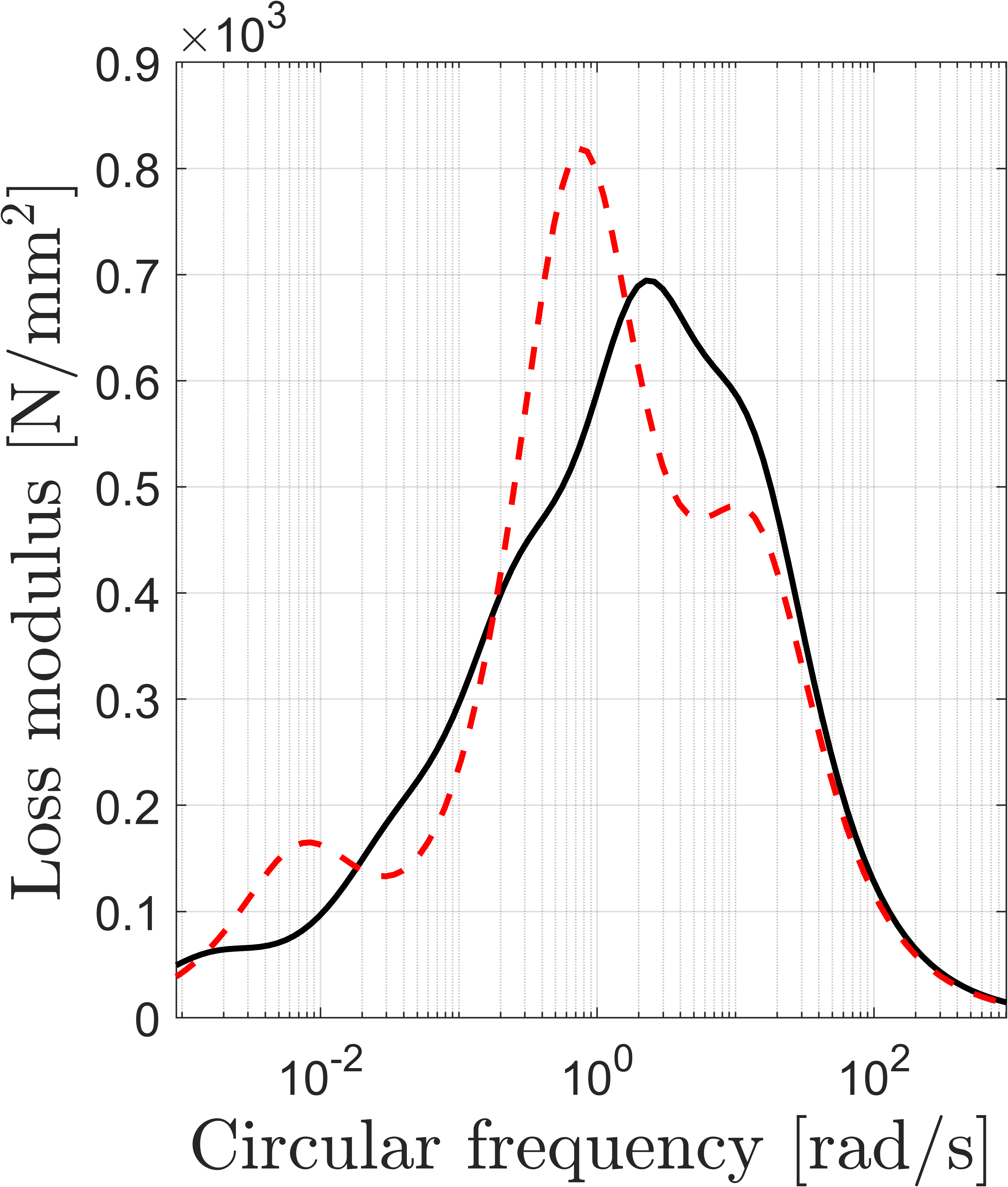} 
\end{subfigure}
\hspace{6pt}
\begin{subfigure}{0.2\textwidth}
\centering
\includegraphics[width=1\linewidth]{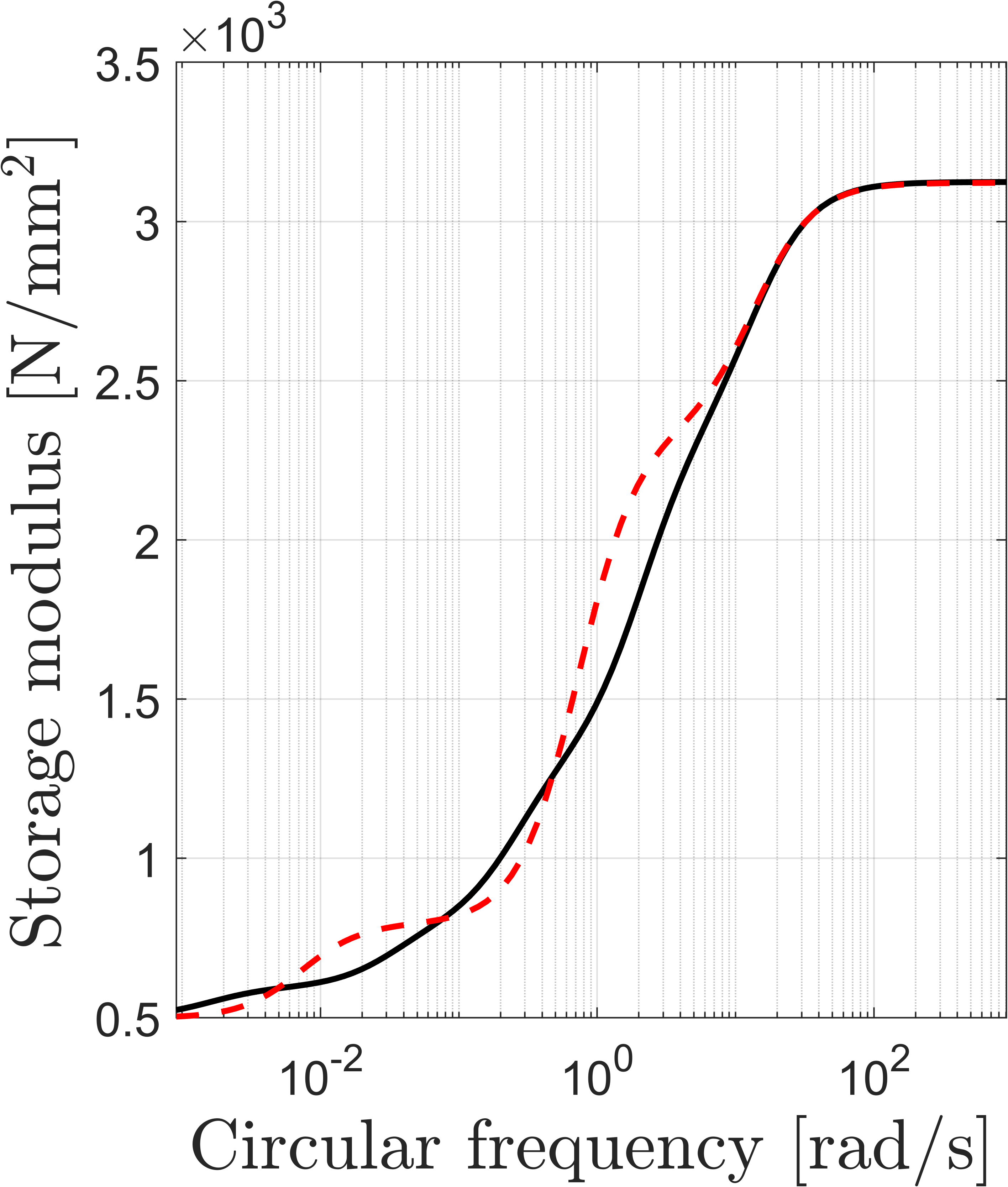} 
\end{subfigure}
\hspace{6pt}
\begin{subfigure}{0.2\textwidth}
\centering
\includegraphics[width=1\linewidth]{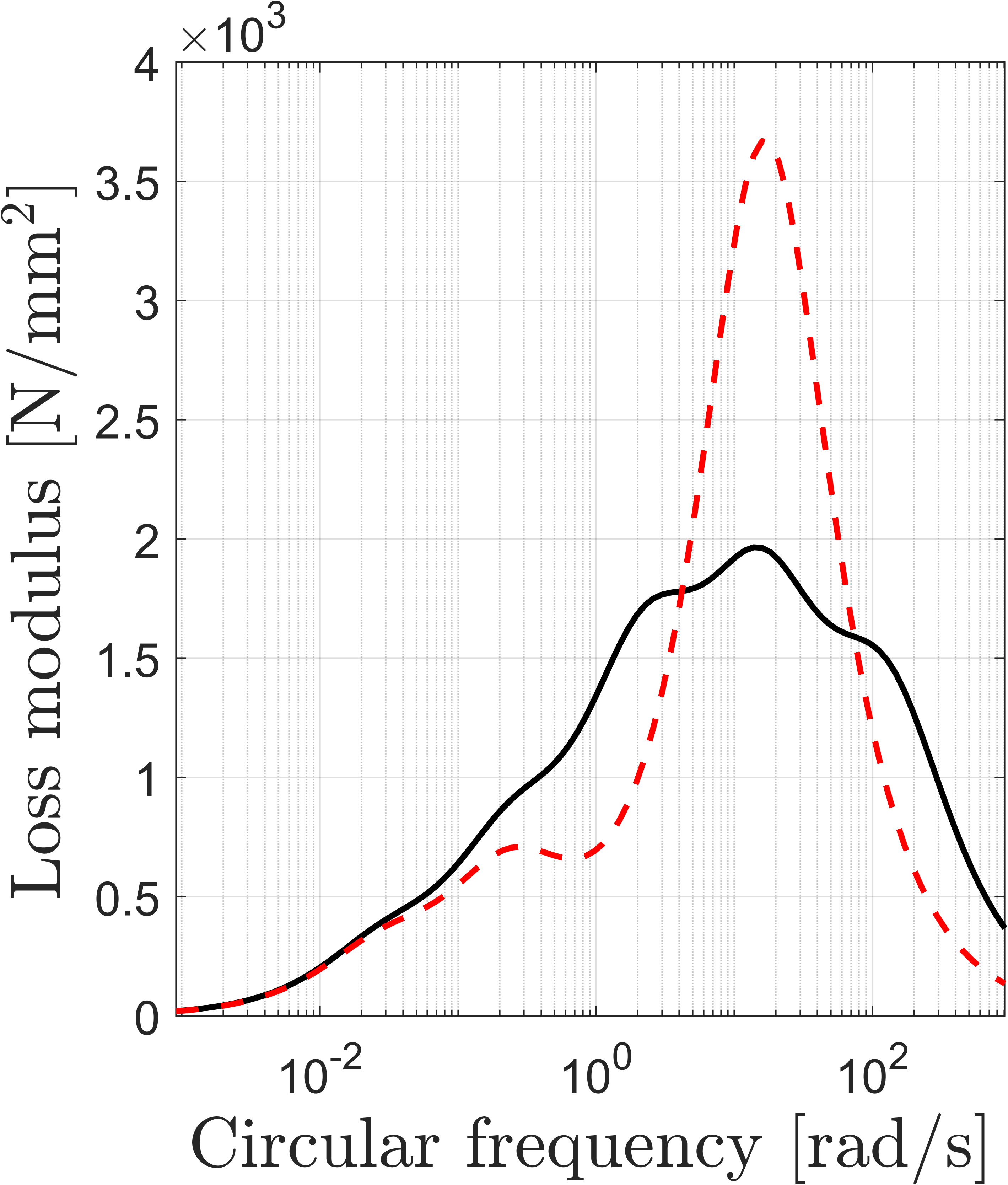} 
\end{subfigure}
\hspace{6pt}
\begin{subfigure}{0.2\textwidth}
\centering
\includegraphics[width=1\linewidth]{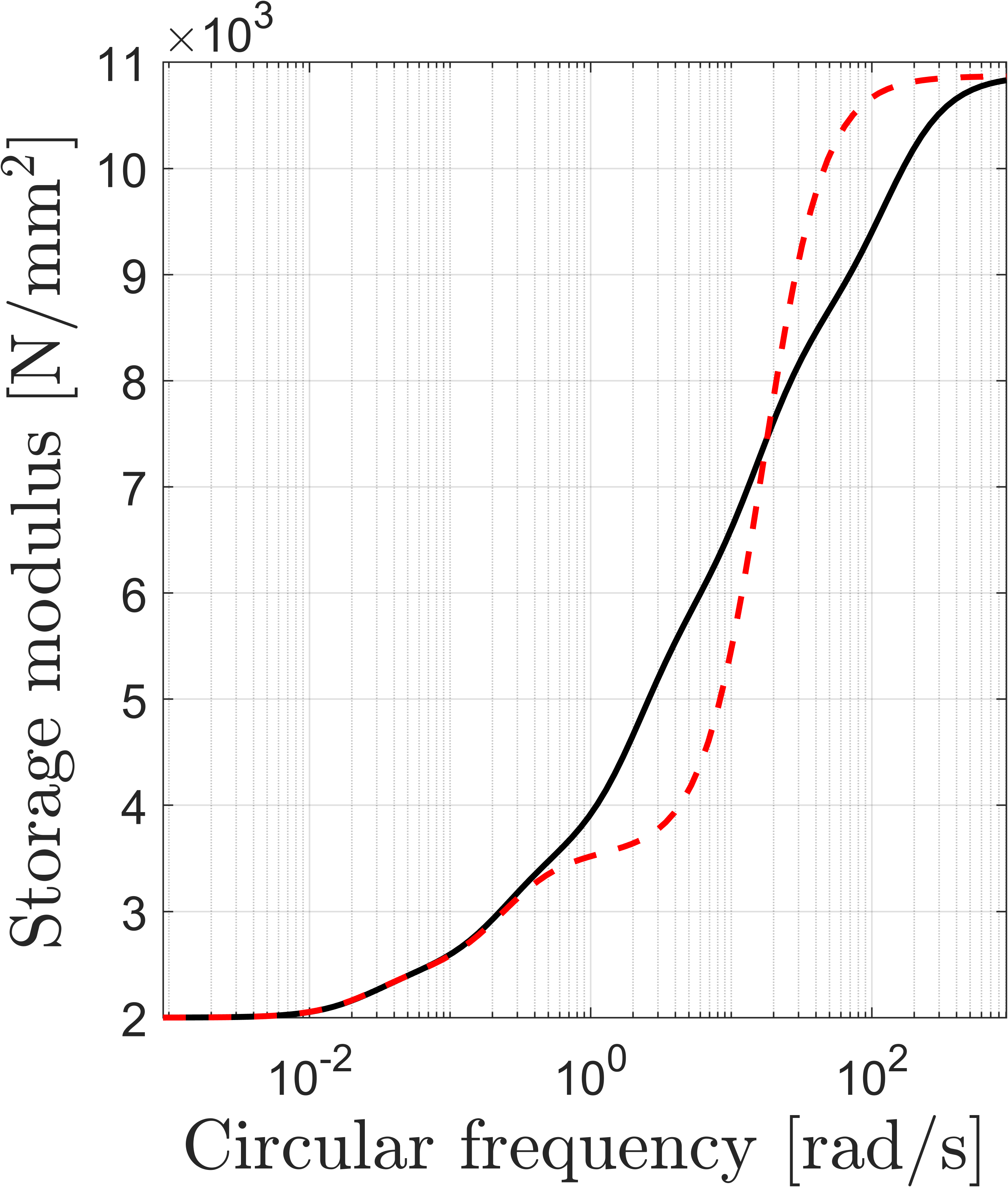} 
\end{subfigure}\\[12pt]
\begin{subfigure}{0.02\textwidth}
\begin{turn}{90} 
4 Clusters
\end{turn}
\end{subfigure}
\hspace{6pt}
\begin{subfigure}{0.2\textwidth}
\centering
\includegraphics[width=1\linewidth]{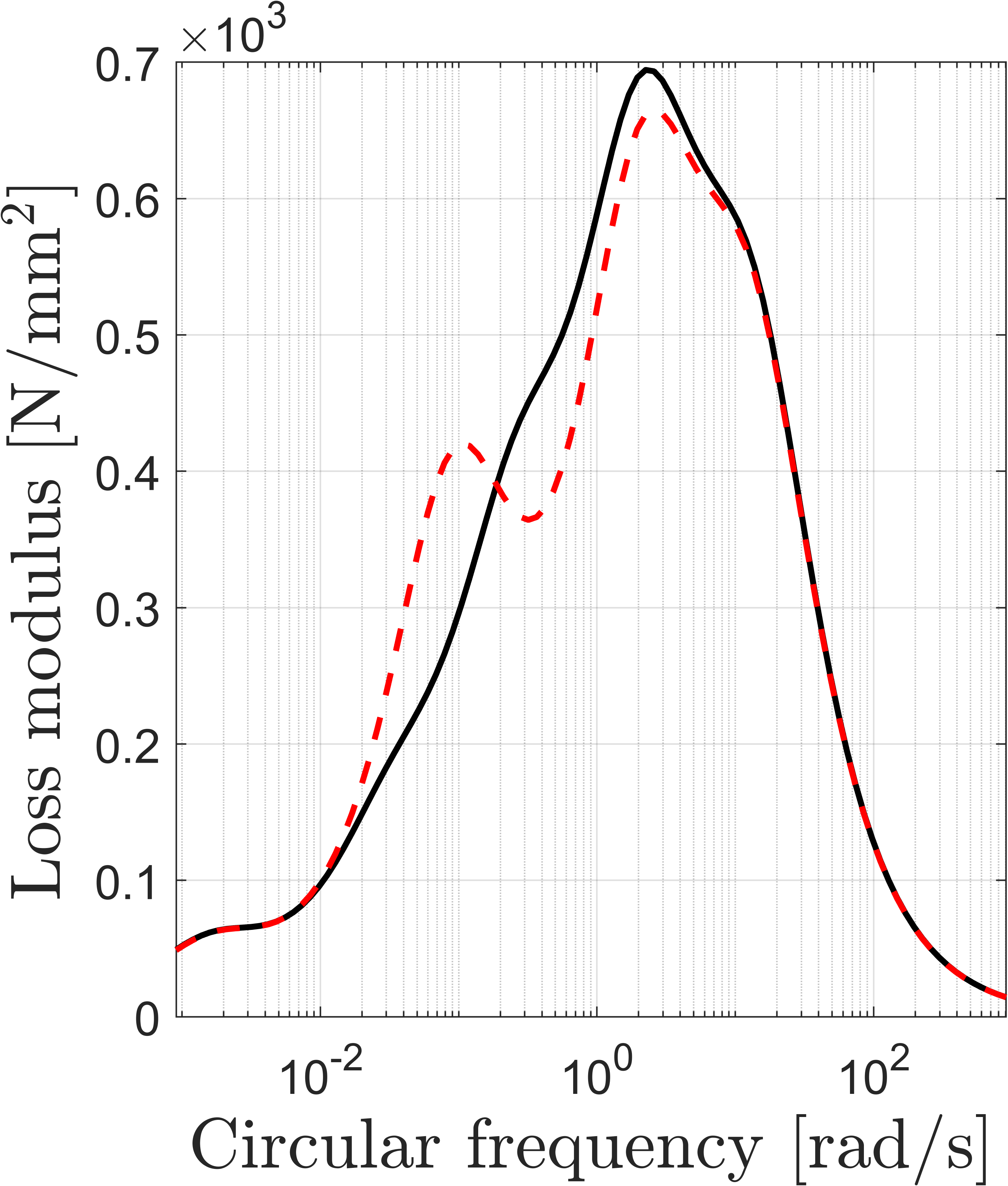} 
\end{subfigure}
\hspace{6pt}
\begin{subfigure}{0.2\textwidth}
\centering
\includegraphics[width=1\linewidth]{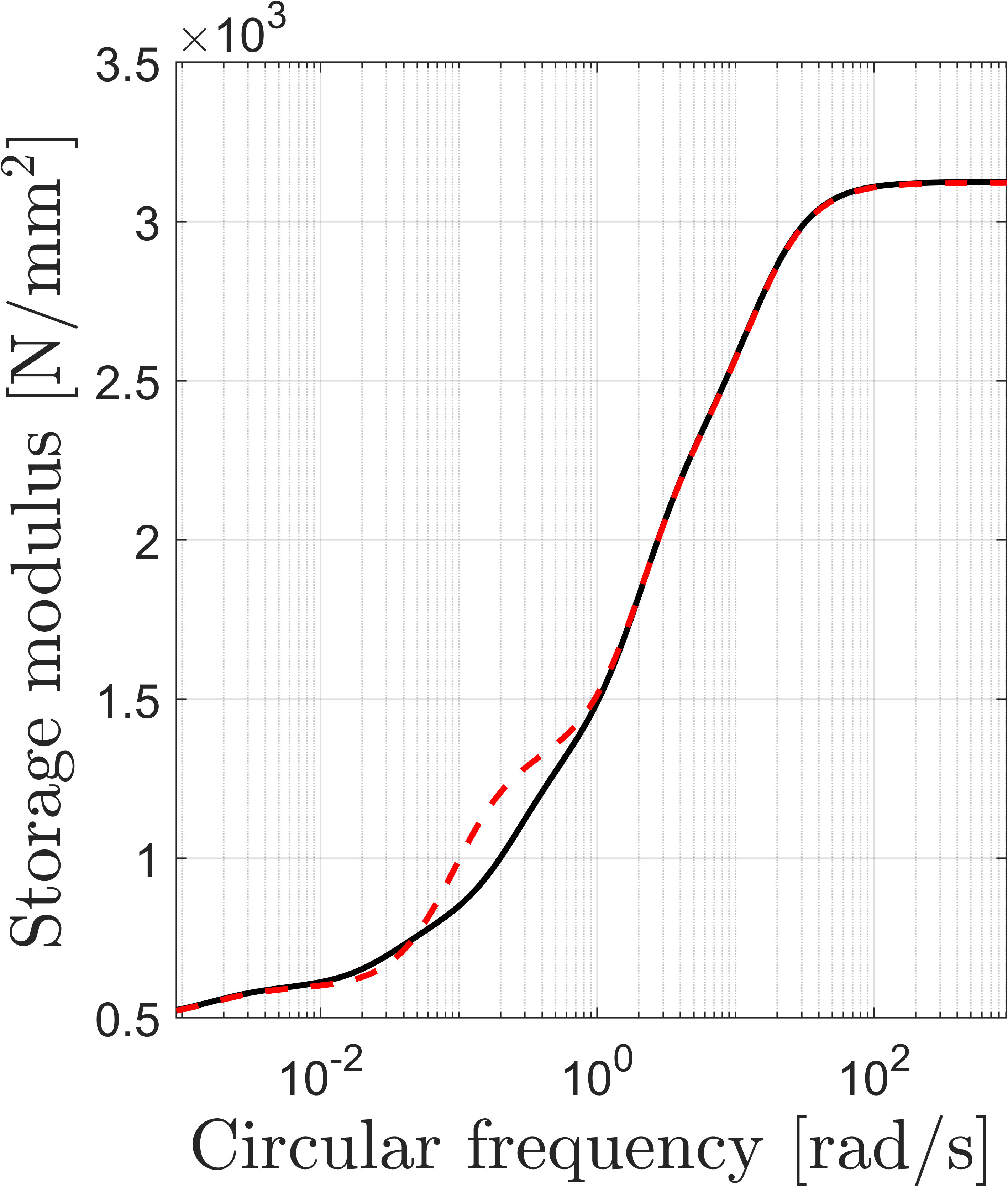} 
\end{subfigure}
\hspace{6pt}
\begin{subfigure}{0.2\textwidth}
\centering
\includegraphics[width=1\linewidth]{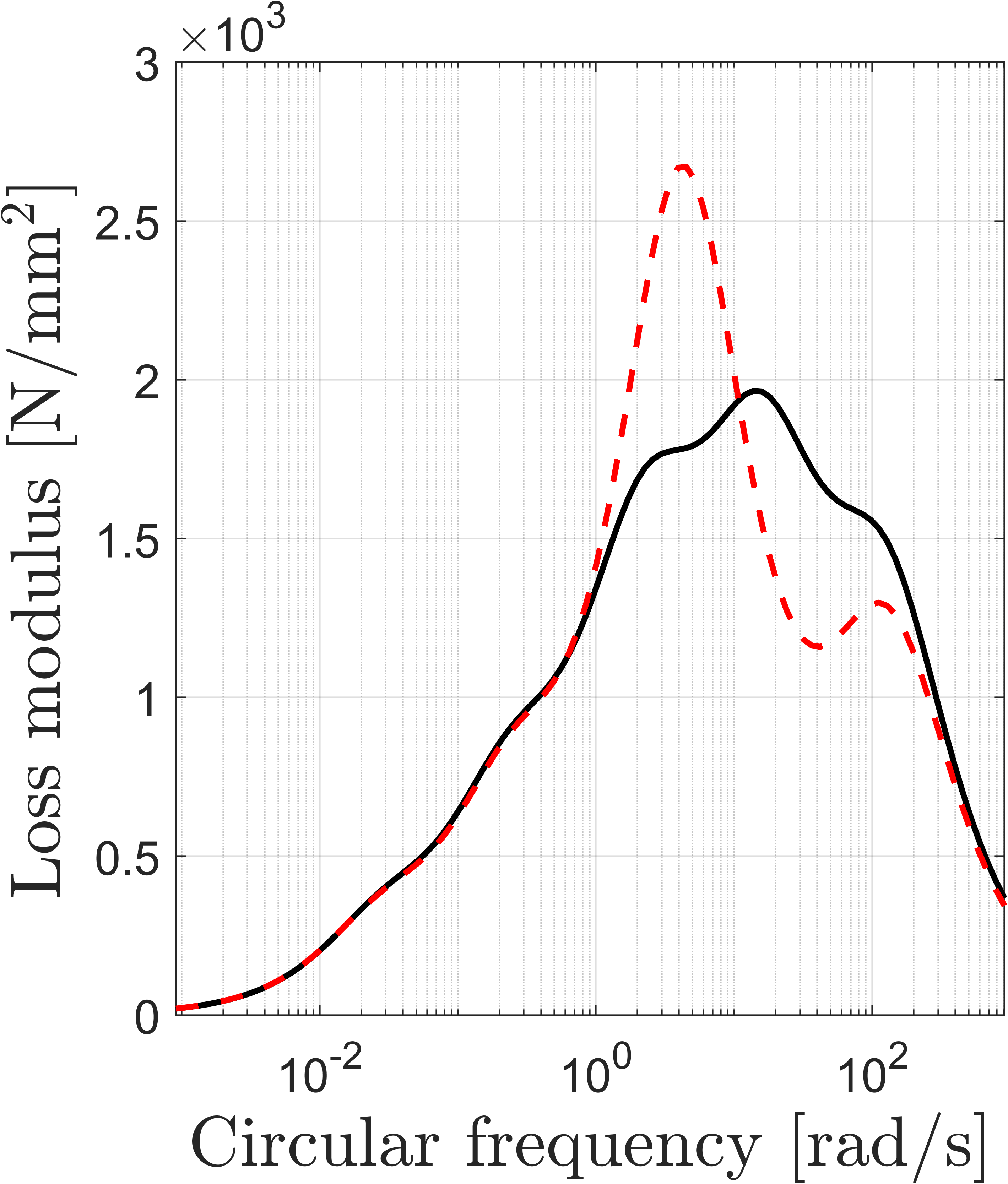} 
\end{subfigure}
\hspace{6pt}
\begin{subfigure}{0.2\textwidth}
\centering
\includegraphics[width=1\linewidth]{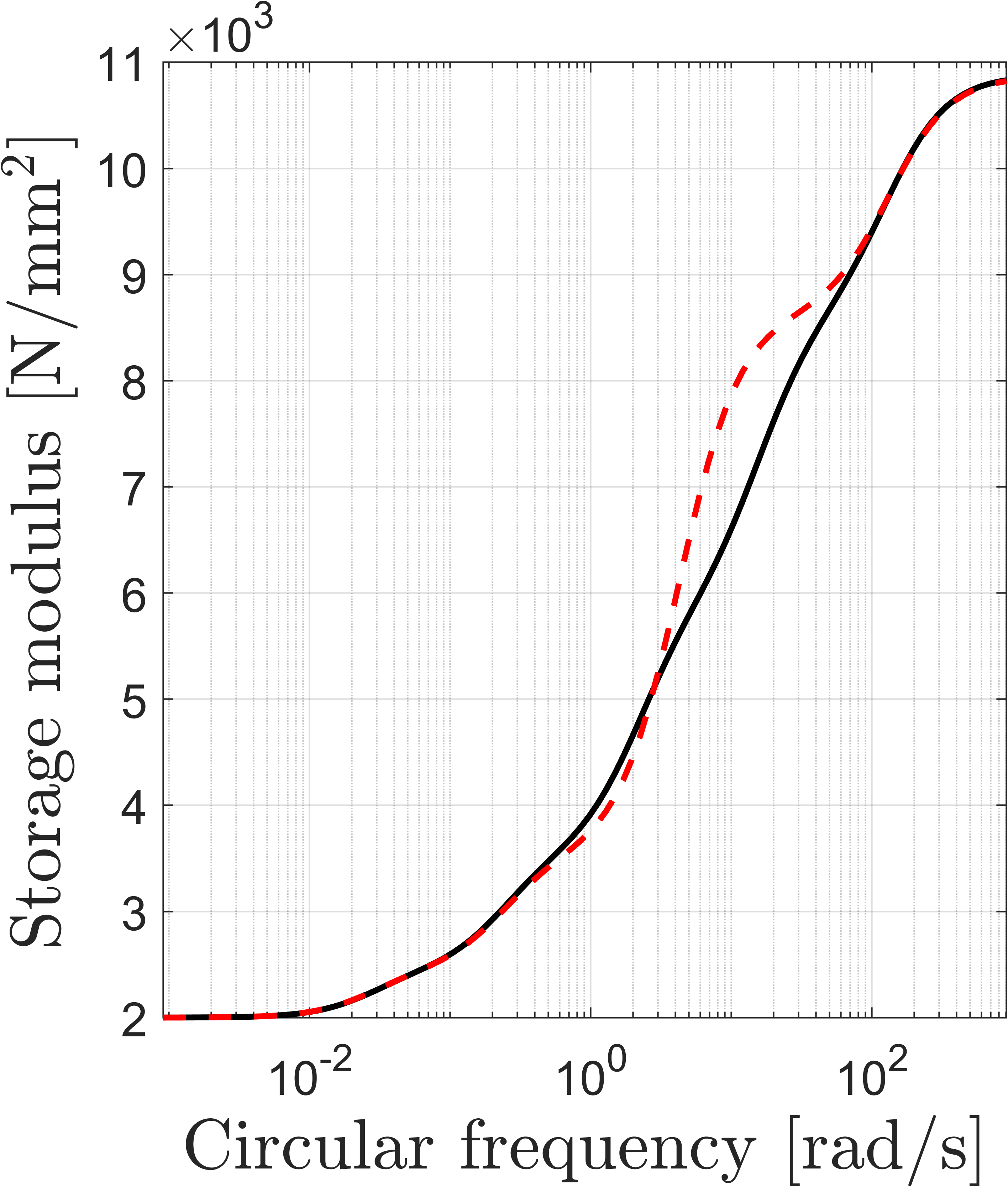} 
\end{subfigure}\\[12pt]
\begin{subfigure}{0.02\textwidth}
\begin{turn}{90} 
5 Clusters
\end{turn}
\end{subfigure}
\hspace{6pt}
\begin{subfigure}{0.2\textwidth}
\centering
\includegraphics[width=1\linewidth]{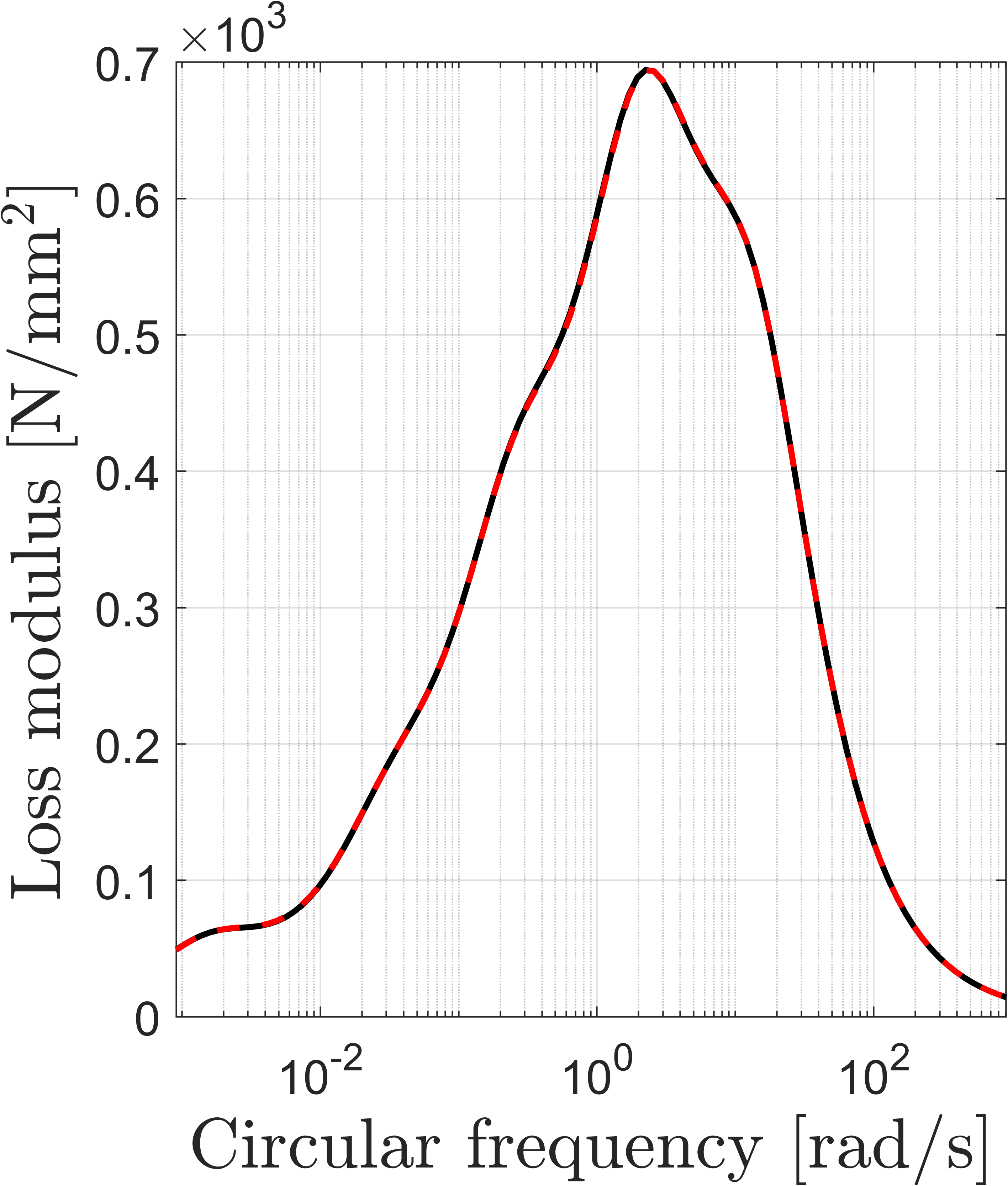} 
\end{subfigure}
\hspace{6pt}
\begin{subfigure}{0.2\textwidth}
\centering
\includegraphics[width=1\linewidth]{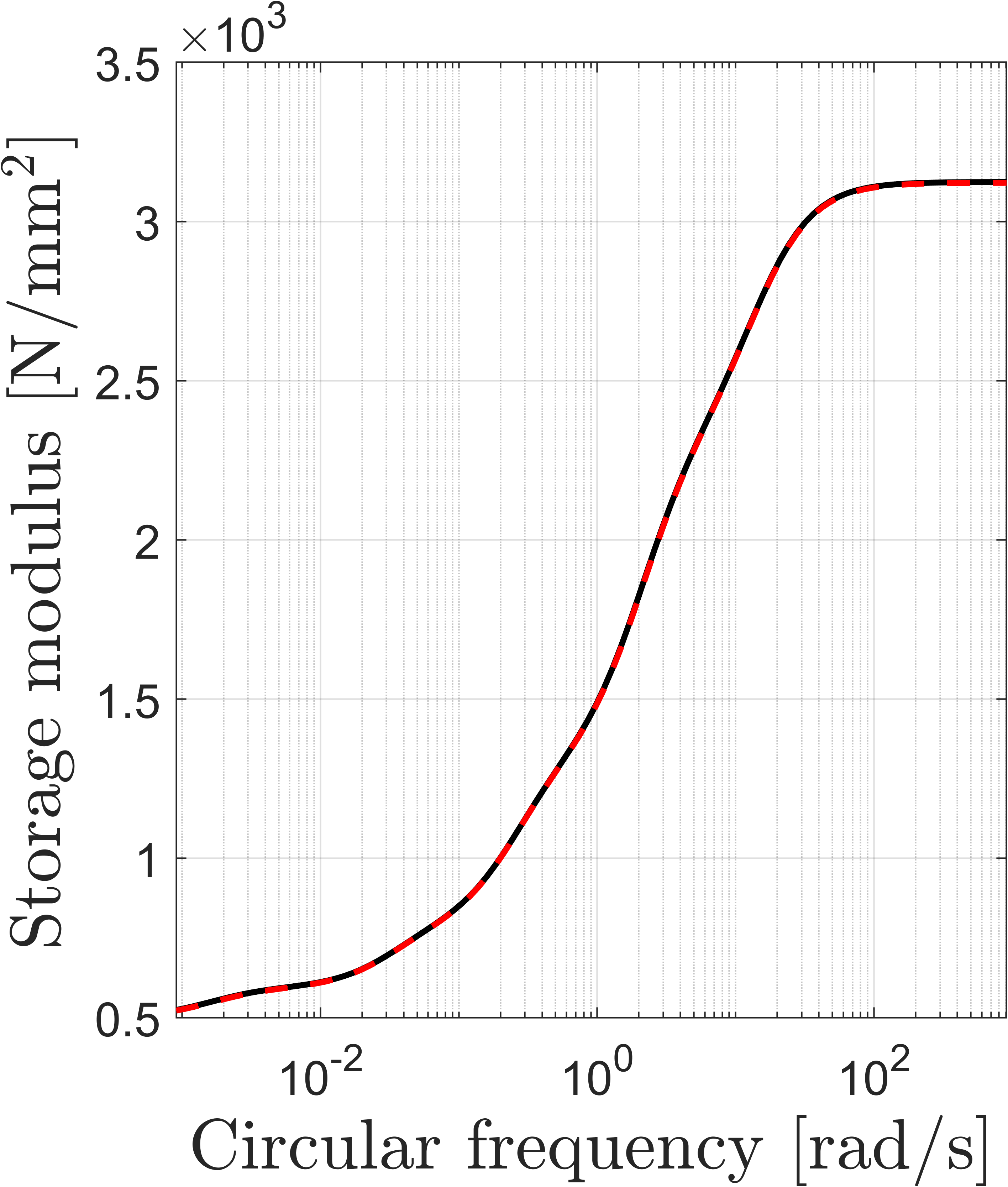} 
\end{subfigure}
\hspace{6pt}
\begin{subfigure}{0.2\textwidth}
\centering
\includegraphics[width=1\linewidth]{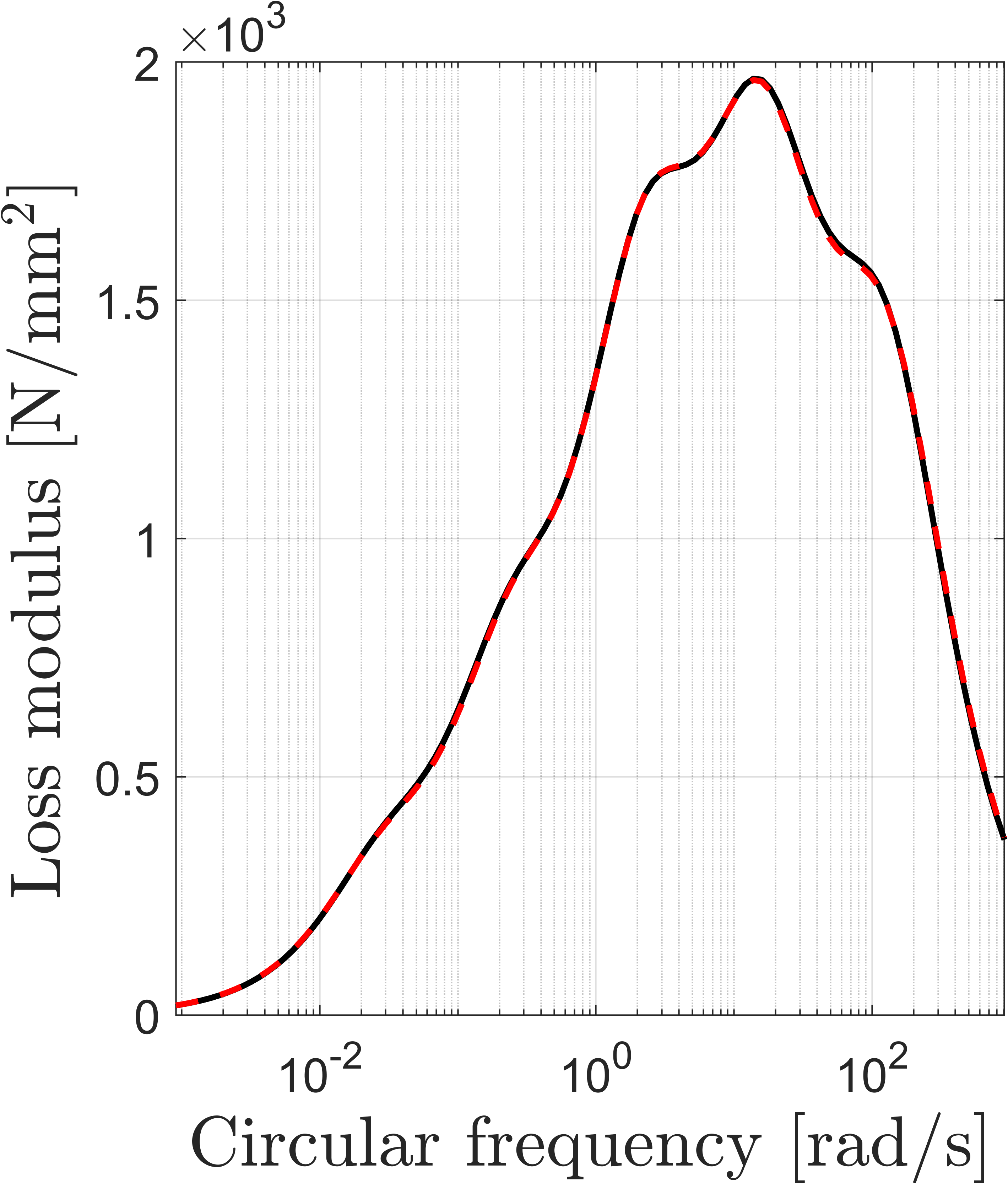} 
\end{subfigure}
\hspace{6pt}
\begin{subfigure}{0.2\textwidth}
\centering
\includegraphics[width=1\linewidth]{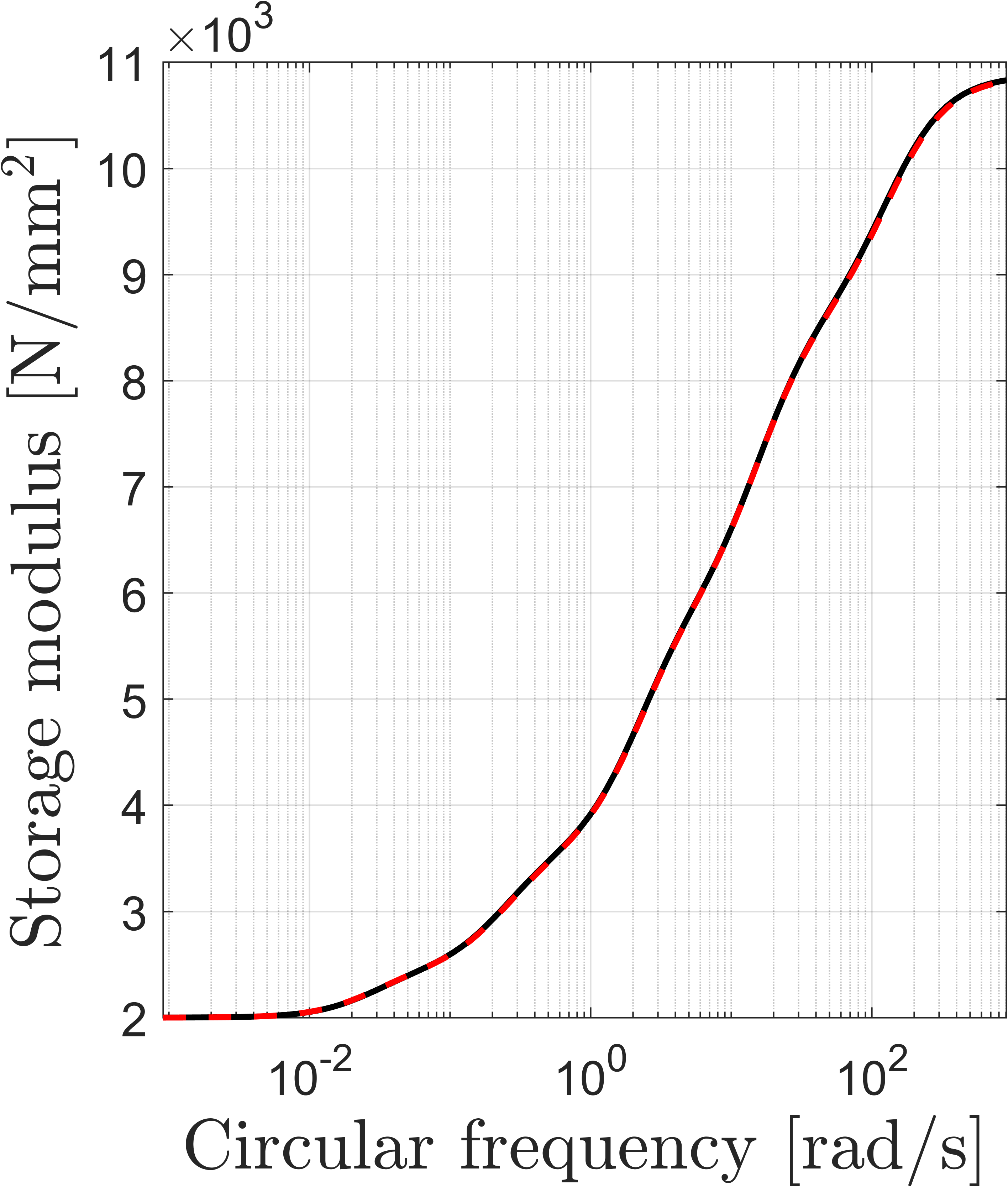} 
\end{subfigure}
\caption{Comparison of true and identified response functions ordered as: shear loss, shear storage, bulk loss, bulk storage (row-wise from left to right) and with increasing number of clusters from 1 to 5 (column-wise from top to bottom) for the noise-free case ($\sig = 0$).}
\label{fig:comparison_moduli_sig0} 
\end{figure}

The final identified material parameters after stage 2 are reported in Table~\ref{tb:discovered_params_sig0}. 
To assess their quality, Figure~\ref{fig:comparison_moduli_sig0} reports a comparison of identified (red dashed line) vs. true (solid black line) loss and storage functions, both for shear and bulk deformations, and for different number of clusters. Column-wise, the figure shows the improvement of the agreement as the number of clusters increases until an excellent matching is achieved with 5 clusters. Plots with higher number of clusters are not reported since they would be indistinguishable from those with 5 clusters.  

\begin{table}
	\centering
	\begin{tabular}{r r r r}
	$\f G$			 & $\f \tau_G$	 	&  $\f K$			& $\f \tau_K$	\\[-3pt]
	$[\SI{}{N/mm^2}]$	 & $[\SI{}{s}]$ 	&  $[\SI{}{N/mm^2}]$	& $[\SI{}{s}]$	\\
	\hline
  	  499			 &	  -				&	2000				&	-\\	
  	  778				 & 0.0726	  			&	2239				&0.0075\\
  	 1019			 & 0.4793	  			&	2711				&0.0635\\
  	  527				 &3.9232	  			&	2365				&0.4541\\
  	  201				 &30.4283  				&	1095				&4.1405\\
              96				 &622.7509	  			&	460				&35.7694\\
              \hline
	\end{tabular}
	\caption{Identified parameters after the two-stage procedure for the noise-free case ($\sig = 0$).}\label{tb:discovered_params_sig0}
\end{table}

\cleardoublepage

\subsection{Noisy data case}
We now test the sensitivity of the proposed strategy to noise by adding to the data an artificial noise with a quite high standard deviation $\sig = \SI{1e-2}{mm}$, corresponding to a noise-excitation standard deviation ratio of about 0.001. In additional tests which we do not show here, all noise levels below this value gave results practically indistinguishable from the noise-free results.

\begin{figure}
\centering
\includegraphics[width=0.5\linewidth]{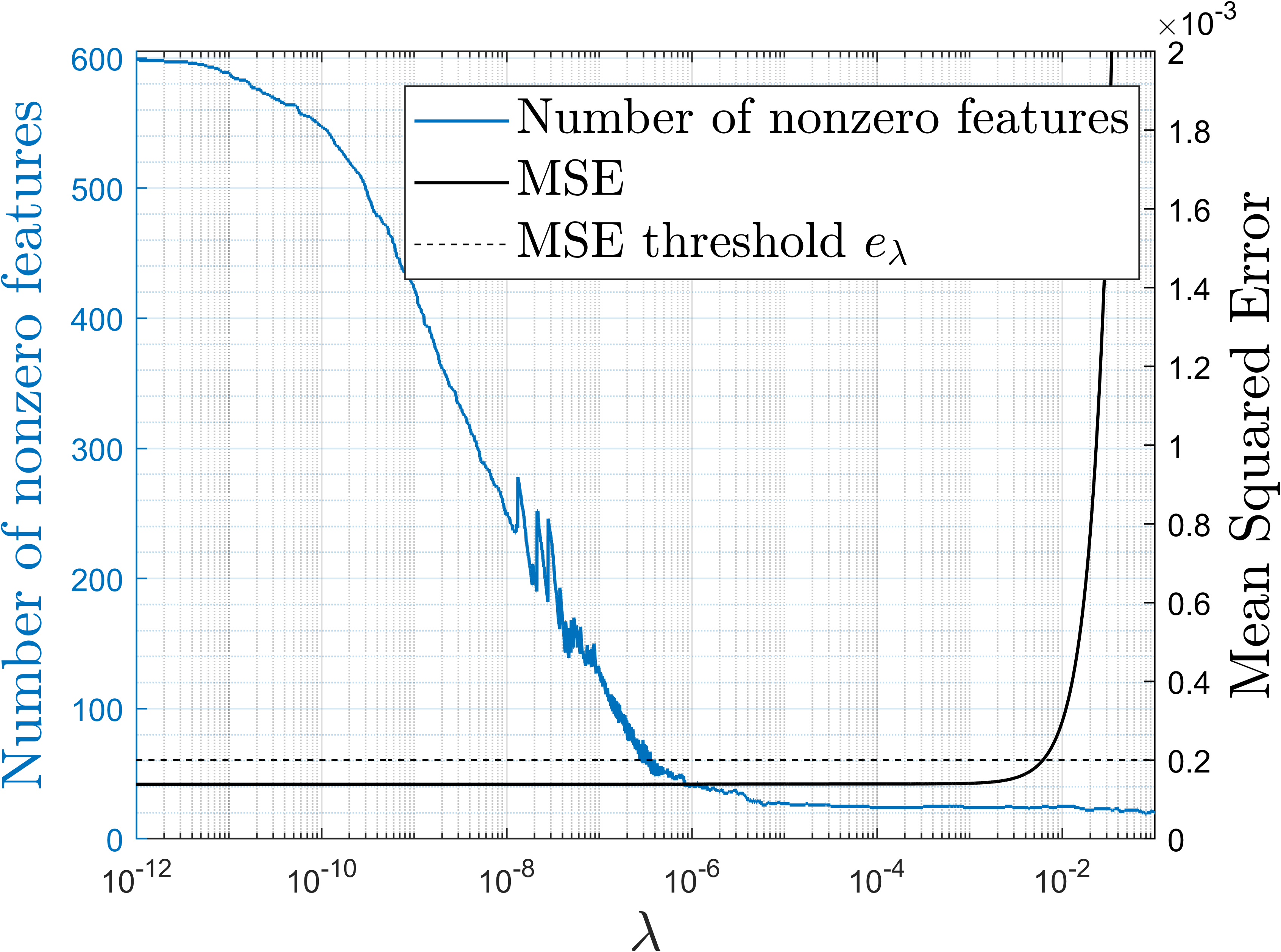}
\caption{Mean Squared Error and number of non-zero features vs. $\lam$ with noise $\sig = \SI{1e-2}{mm}$.}\label{fig:MSE+threshold_vs_lam_sig1em4}
\end{figure}

\begin{figure}
\centering
\begin{subfigure}{0.47\textwidth}
\centering
\includegraphics[width=1\linewidth]{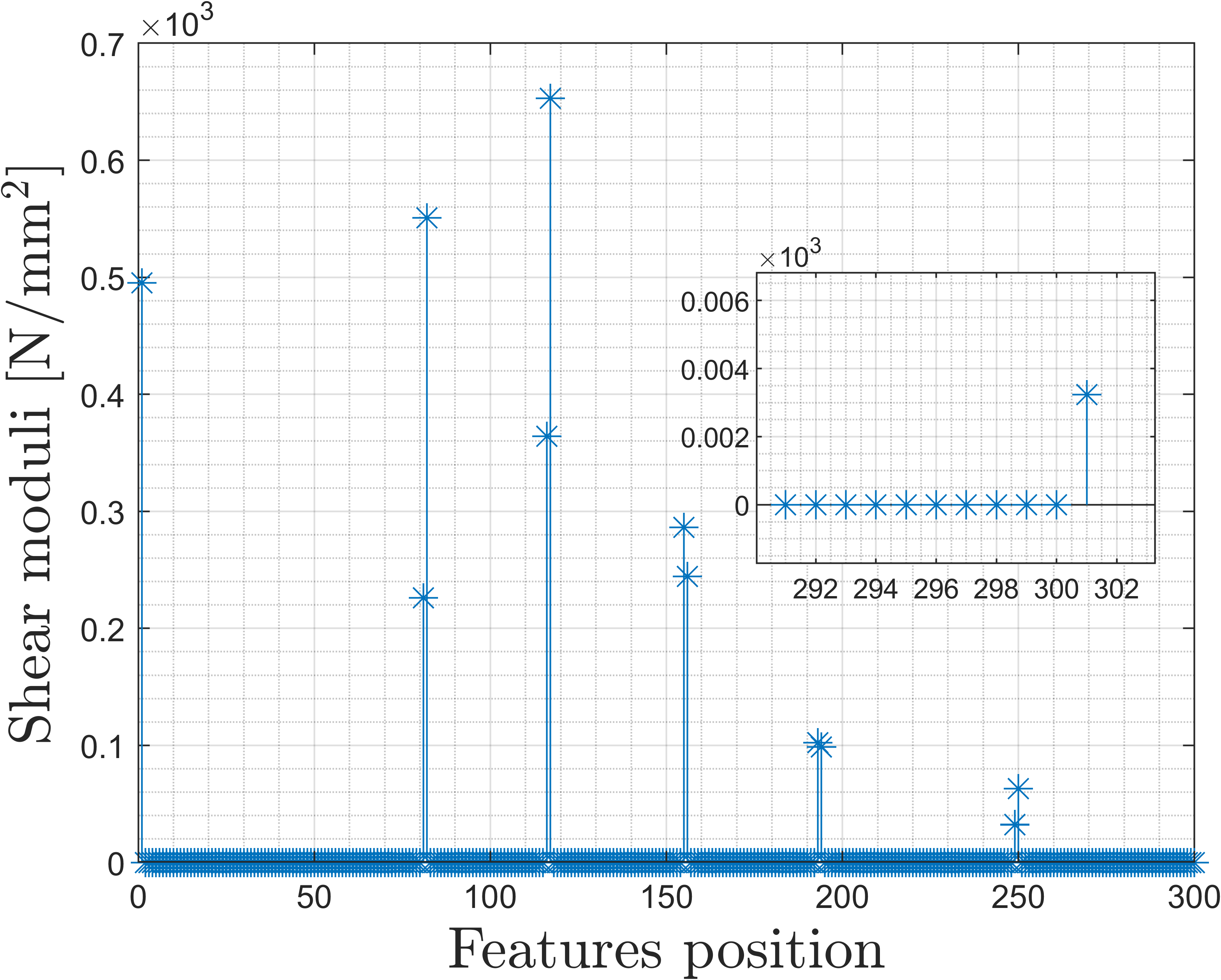} 
\caption{Shear.}\label{fig:activated_GK_vs_position_sig1em4_a}
\end{subfigure}
\hspace{9pt}
\begin{subfigure}{0.47\textwidth}
\centering
\includegraphics[width=1\linewidth]{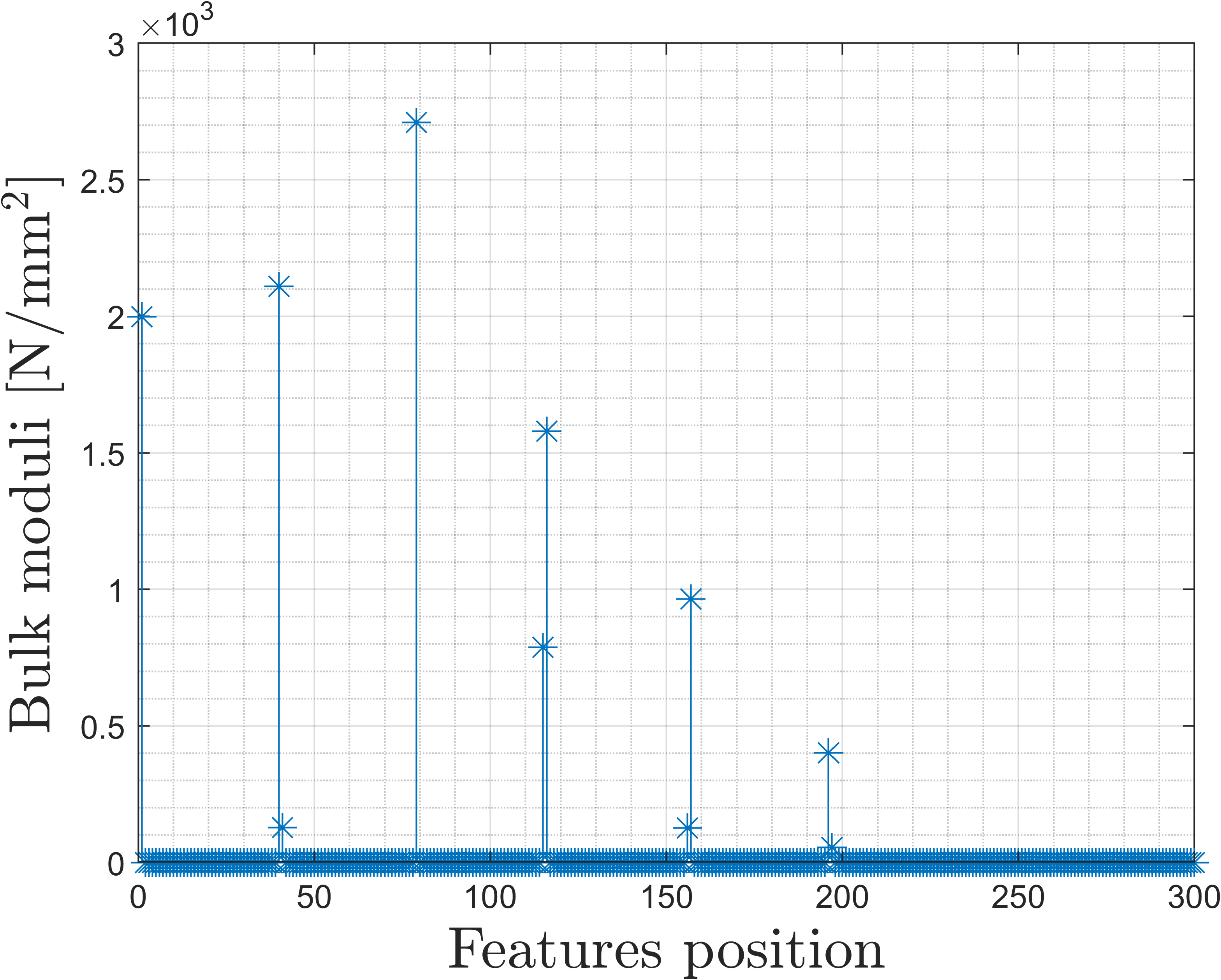} 
\caption{Bulk.}\label{fig:activated_GK_vs_position_sig1em4_b}
\end{subfigure}
\caption{Activated shear (a) and bulk (b) moduli over the entire library of rheological components after stage 1 with noise $\sig = \SI{1e-2}{mm}$.}\label{fig:activated_GK_vs_position_sig1em4}
\end{figure}

\begin{figure}
\centering
\begin{subfigure}{0.47\textwidth}
\centering
\includegraphics[width=1\linewidth]{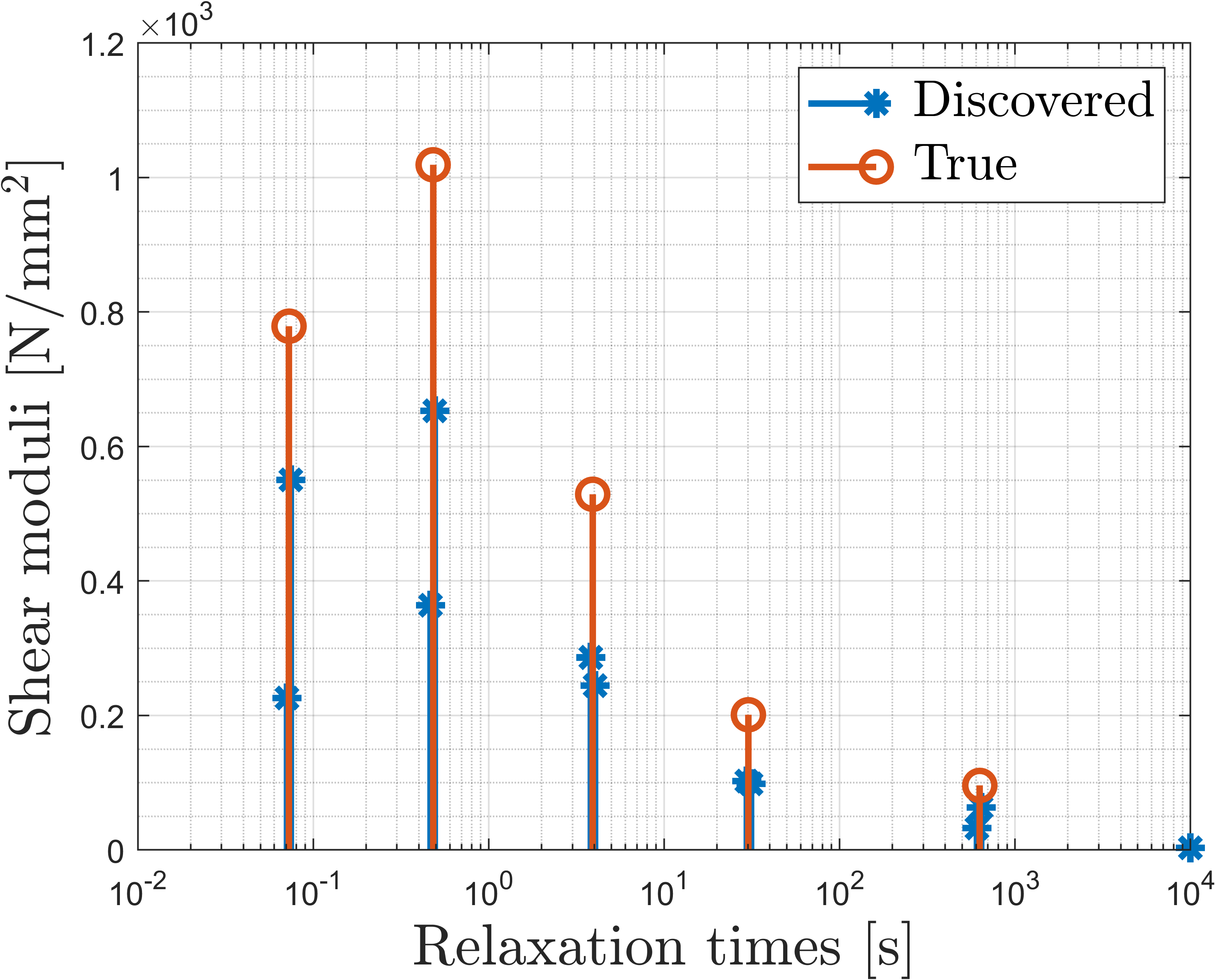} 
\end{subfigure}
\hspace{9pt}
\begin{subfigure}{0.47\textwidth}
\centering
\includegraphics[width=1\linewidth]{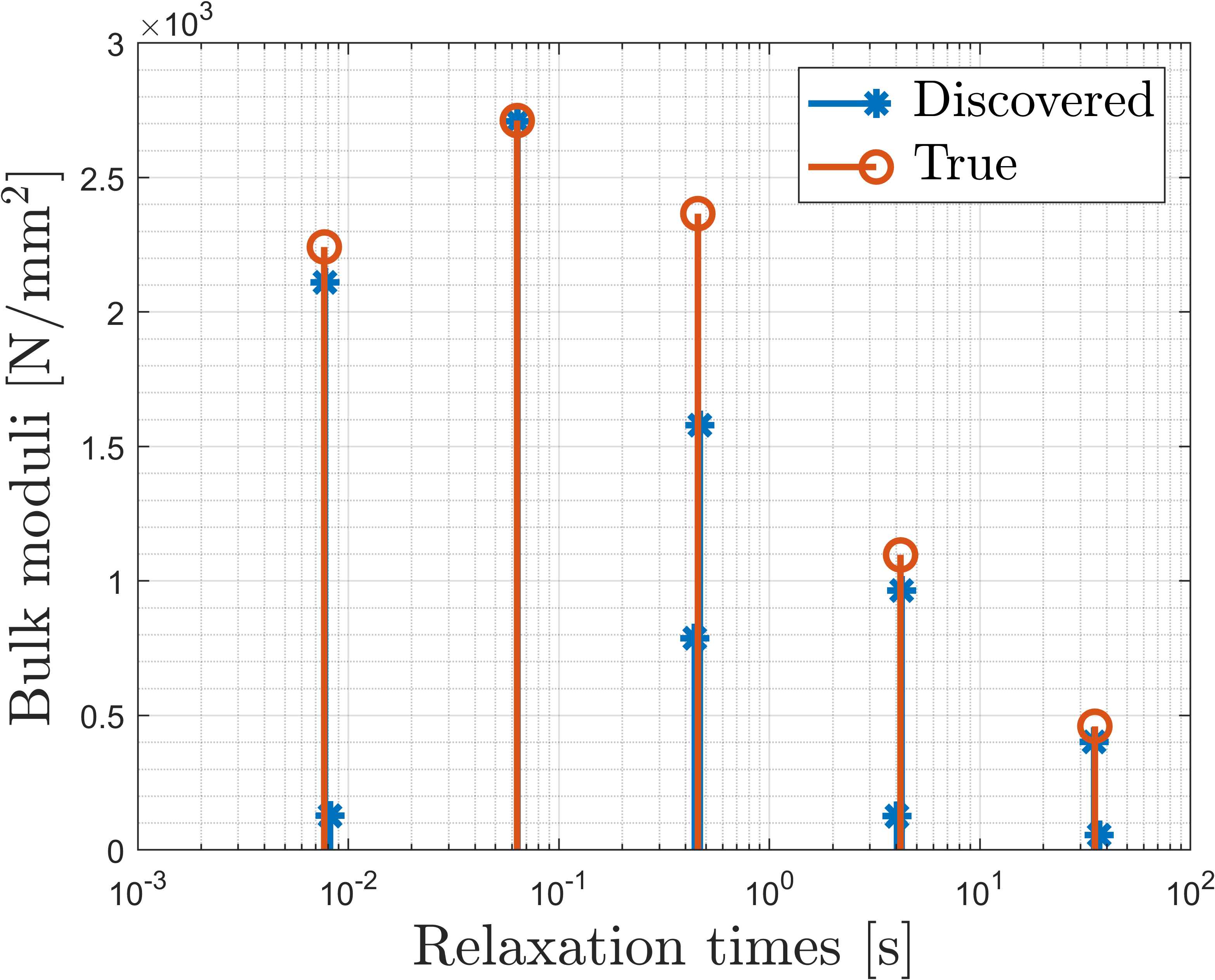} 
\end{subfigure}
\caption{Activated shear (a) and bulk (b) moduli and corresponding relaxation times after stage 1 with noise $\sig = \SI{1e-2}{mm}$.}\label{fig:activated_GK_vs_relaxtimes_sig1em4}
\end{figure}

Figure~\ref{fig:MSE+threshold_vs_lam_sig1em4} shows the MSE vs. the regularization parameter $\lam$, which confirms the trend of Figure \ref{fig:MSE+threshold_vs_lam_sig0} but expectedly with larger MSE values. We now choose an error threshold $e_\lam = \SI{ 2e-4}{}$ and identify $\lam^{opt} = 0.0062$ (however, also in this case any $\lam$ larger than about $10^{-5}$ is equally effective in inducing sparsity). 
The parameters automatically selected by Lasso in stage 1 are shown in Figure~\ref{fig:activated_GK_vs_position_sig1em4}, and Figure~\ref{fig:activated_GK_vs_relaxtimes_sig1em4} illustrates the selected moduli with their corresponding relaxation times in comparison with the true values.
As in the noise-free case, most of the features are suppressed in stage 1; now 12 features remain active for the shear response and 11 for the bulk response.
 As highlighted in the insert of Figure~\ref{fig:activated_GK_vs_position_sig1em4_a}, an extra feature with respect to the noise-free case is now active, corresponding to the upper bound of the relaxation times. 
This indicates that the noise disturbance affects the small frequency response, leading to the identification of $G_{\infty} = \SI{495}{N/mm^2}$ as opposed to the true value of $\SI{500}{N/mm^2}$. 
Interestingly, the identification of $K_{\infty}$ is instead almost unaffected by the presence of noise.

\begin{figure}
\centering
\includegraphics[width=0.5\linewidth]{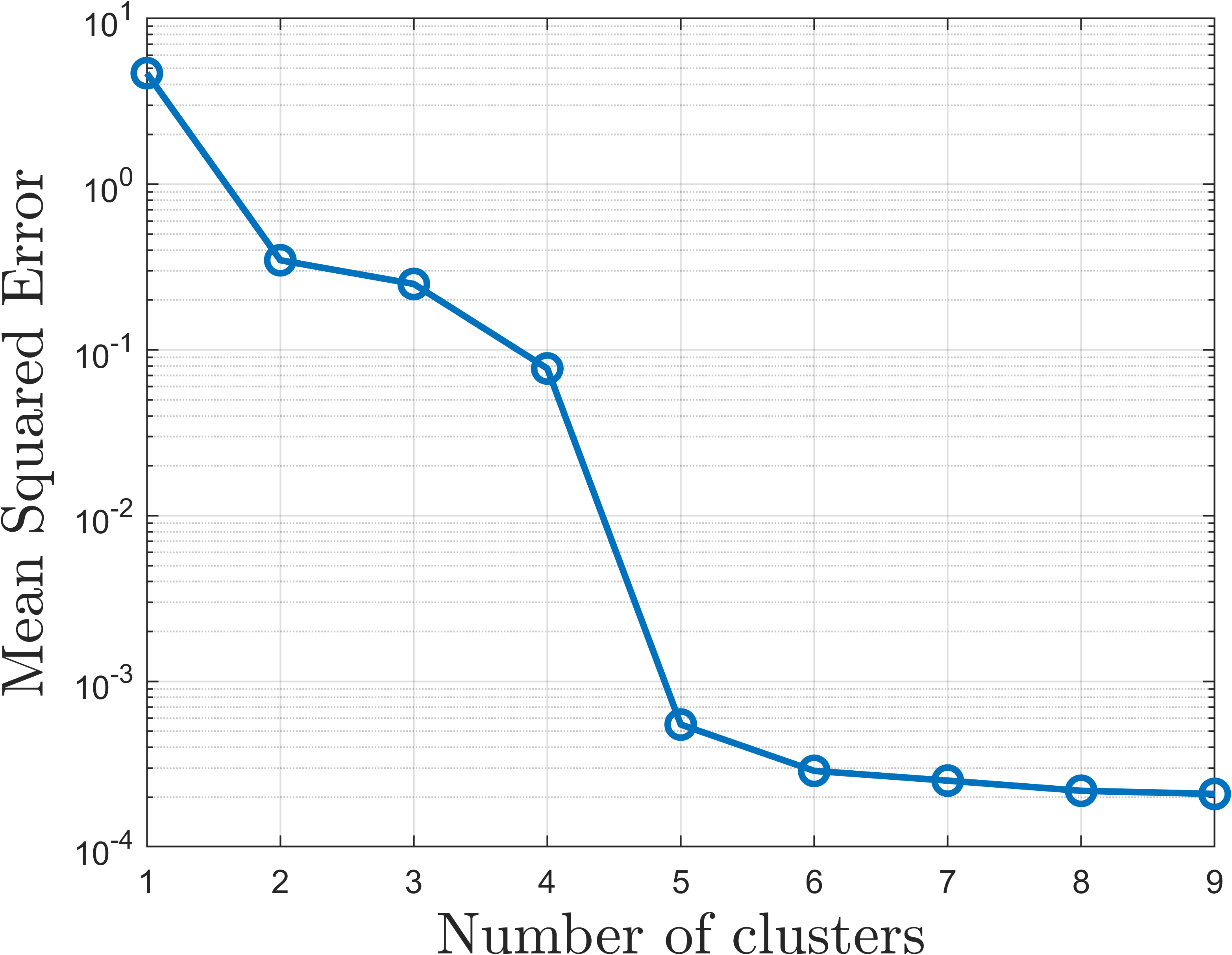}
\caption{Mean Squared Error vs. number of clusters with noise $\sig = \SI{1e-2}{mm}$.}\label{fig:MSElog_vs_clsts_sig5em4+1em5}
\end{figure}

Figure~\ref{fig:MSElog_vs_clsts_sig5em4+1em5} shows the results of the clustering procedure in stage 2 and reveals once again a drastic drop of the MSE at 5 clusters. However, while in the noise-free case the use of 6 or 7 clusters leads to no improvement, here considering 6 clusters does improve results at the small cost of adding one extra rheological component. To understand the reason, we compare in Table~\ref{tb:discovered_params_sig1em4_5-6cls} the final identified material parameters for the choices of 5 and 6 clusters.
It is clear that with 5 clusters the largest relaxation time for the shear deformation is not correctly identified, as no shear modulus should be activated around \SI{1571}{s}. 
Instead, with 6 clusters, a shear modulus much closer to the true value (\SI{95}{N/mm^2} vs. \SI{96}{N/mm^2}) is correctly activated in the neighborhood of the true relaxation time \SI{629.40}{s}. 
Moreover, a shear modulus of \SI{3}{N/mm^2} is activated at the upper limit of the relaxation times (\SI{1E4}{s}), indicating that such a value improves the accuracy of the long-term response. This is confirmed by noting that, summing this value to the identified long-term shear modulus, a better $G_\infty$ is found ($495 + 3 =  \SI{498}{N/mm^2}$, which is much closer to the true value).

\begin{figure}
\centering
\begin{subfigure}{1\textwidth}
\centering
\hspace{0.05\textwidth}
\includegraphics[width=.2\linewidth]{legend.png} 
\end{subfigure}\\
\begin{subfigure}{0.02\textwidth}
\begin{turn}{90} 
2 Cluster
\end{turn}
\end{subfigure}
\hspace{6pt}
\begin{subfigure}{0.2\textwidth}
\centering
\includegraphics[width=1\linewidth]{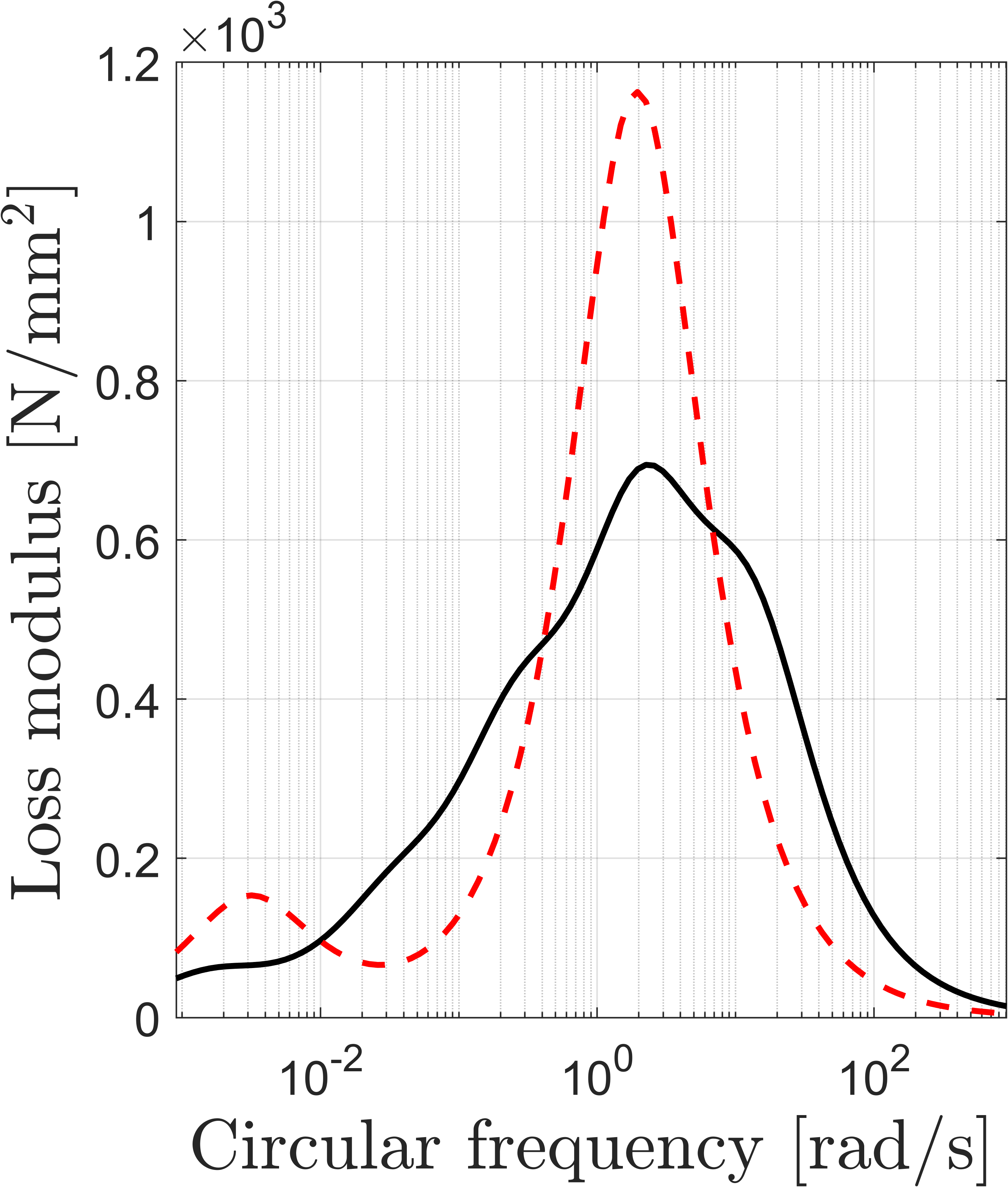} 
\end{subfigure}
\hspace{6pt}
\begin{subfigure}{0.2\textwidth}
\centering
\includegraphics[width=1\linewidth]{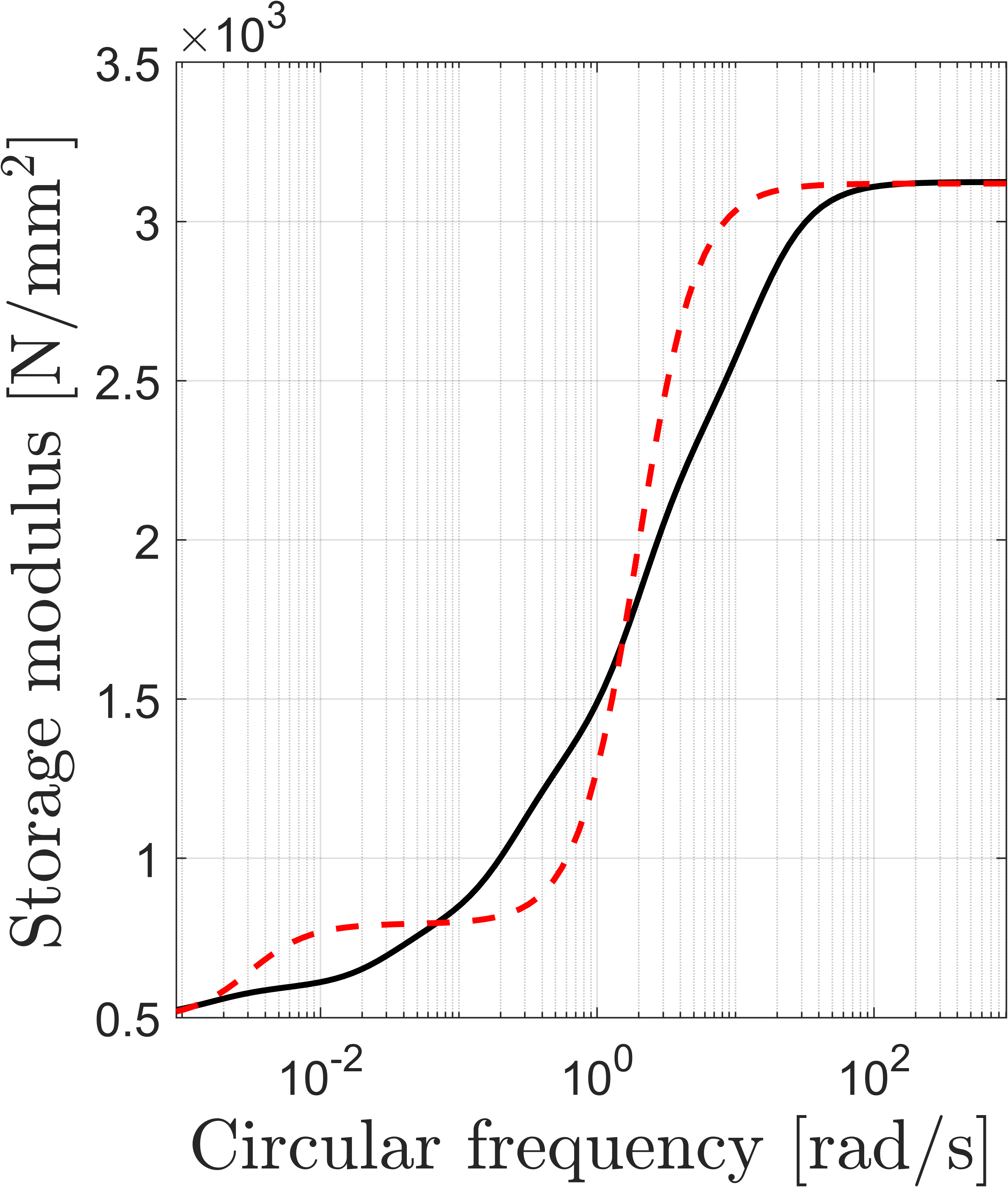} 
\end{subfigure}
\hspace{6pt}
\begin{subfigure}{0.2\textwidth}
\centering
\includegraphics[width=1\linewidth]{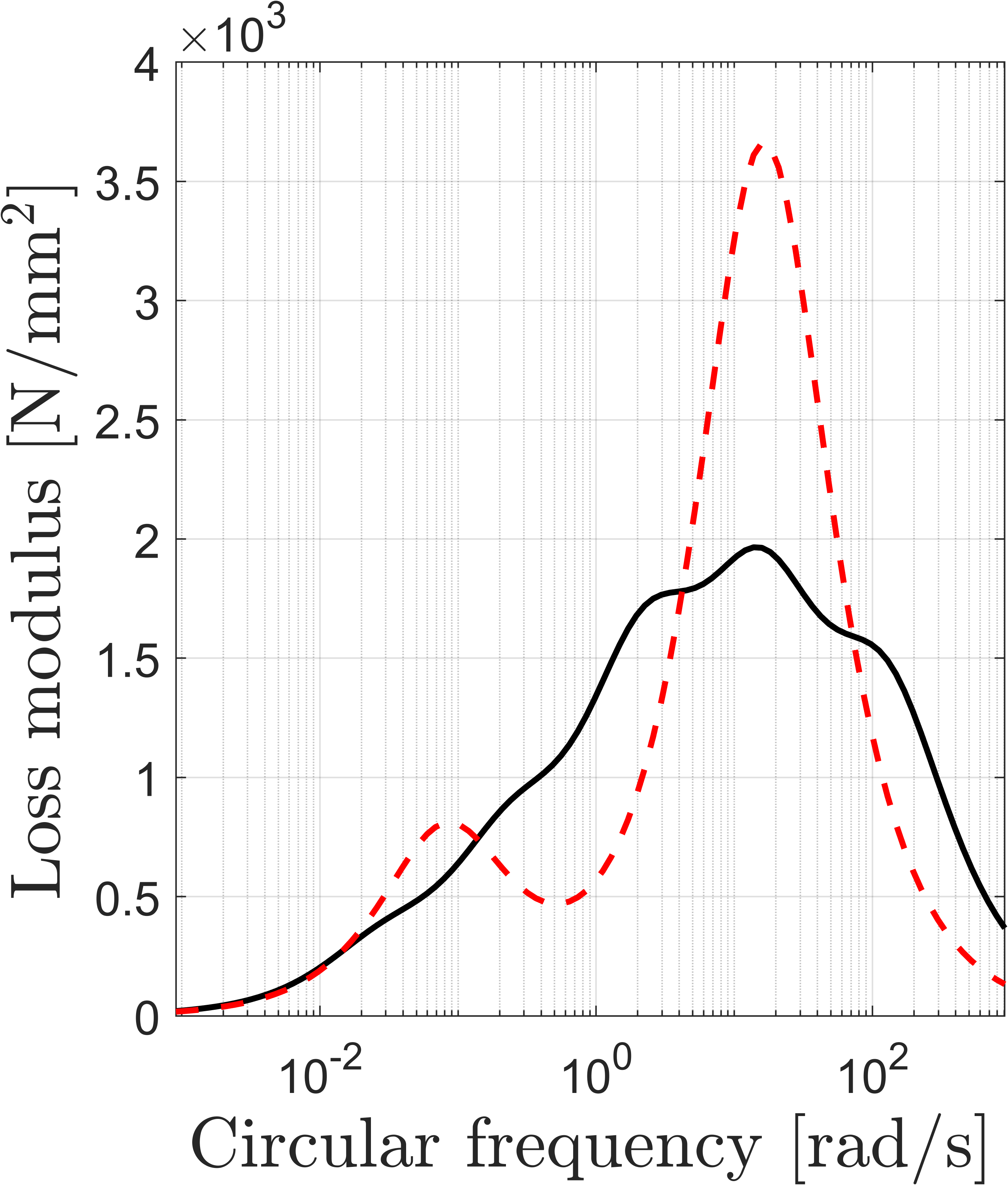} 
\end{subfigure}
\hspace{6pt}
\begin{subfigure}{0.2\textwidth}
\centering
\includegraphics[width=1\linewidth]{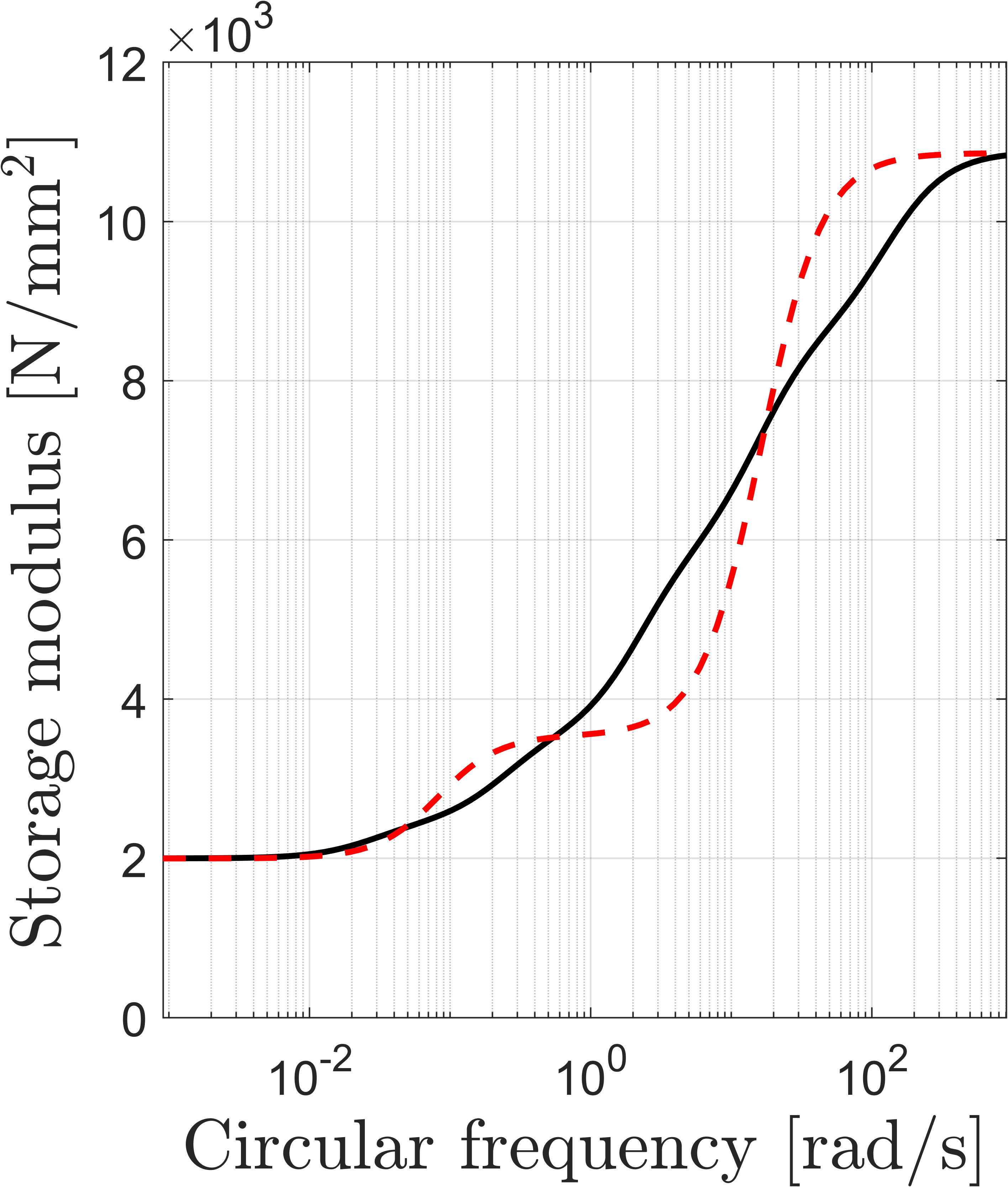} 
\end{subfigure}\\[12pt]
\begin{subfigure}{0.02\textwidth}
\begin{turn}{90} 
3 Clusters
\end{turn}
\end{subfigure}
\hspace{6pt}
\begin{subfigure}{0.2\textwidth}
\centering
\includegraphics[width=1\linewidth]{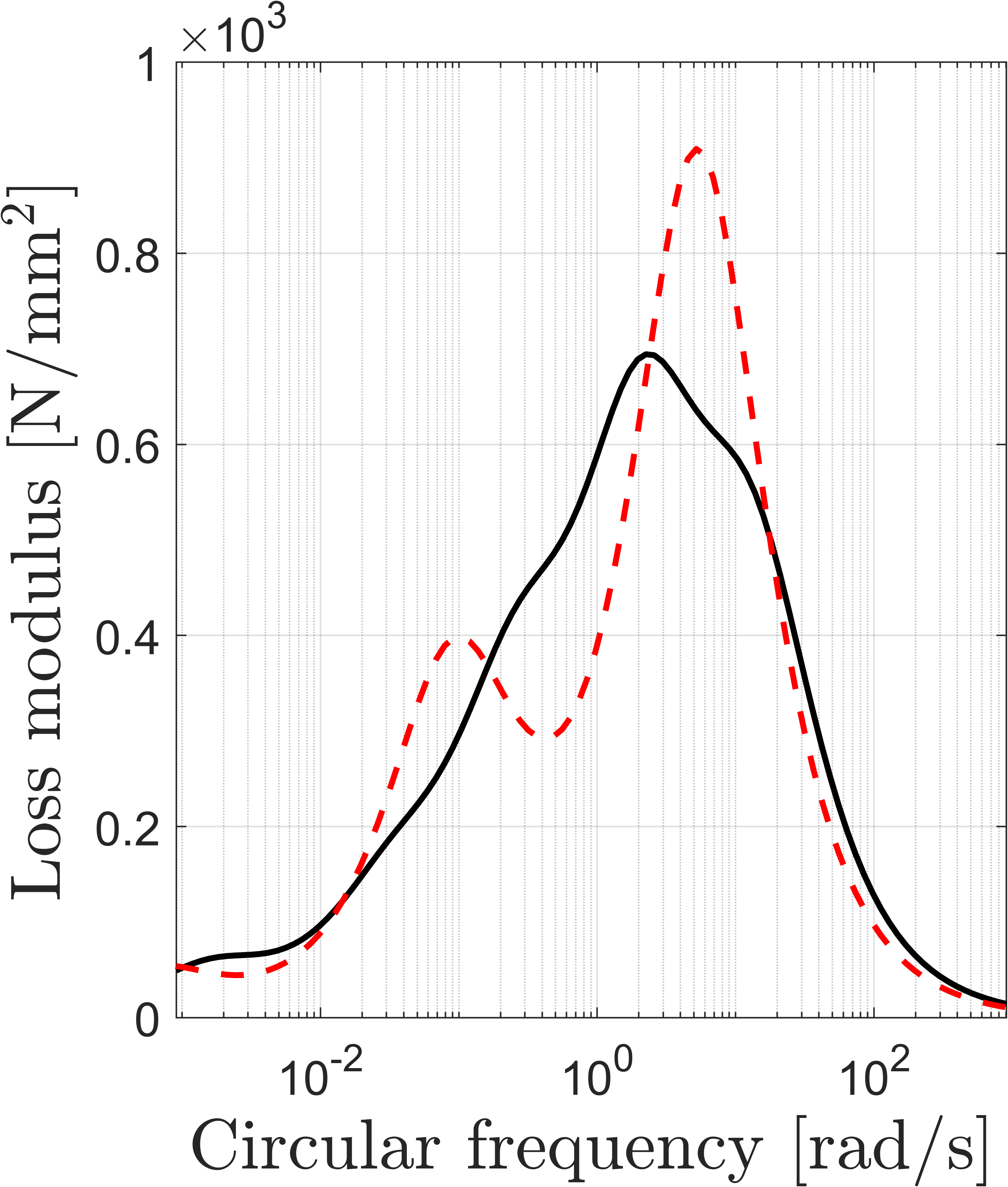} 
\end{subfigure}
\hspace{6pt}
\begin{subfigure}{0.2\textwidth}
\centering
\includegraphics[width=1\linewidth]{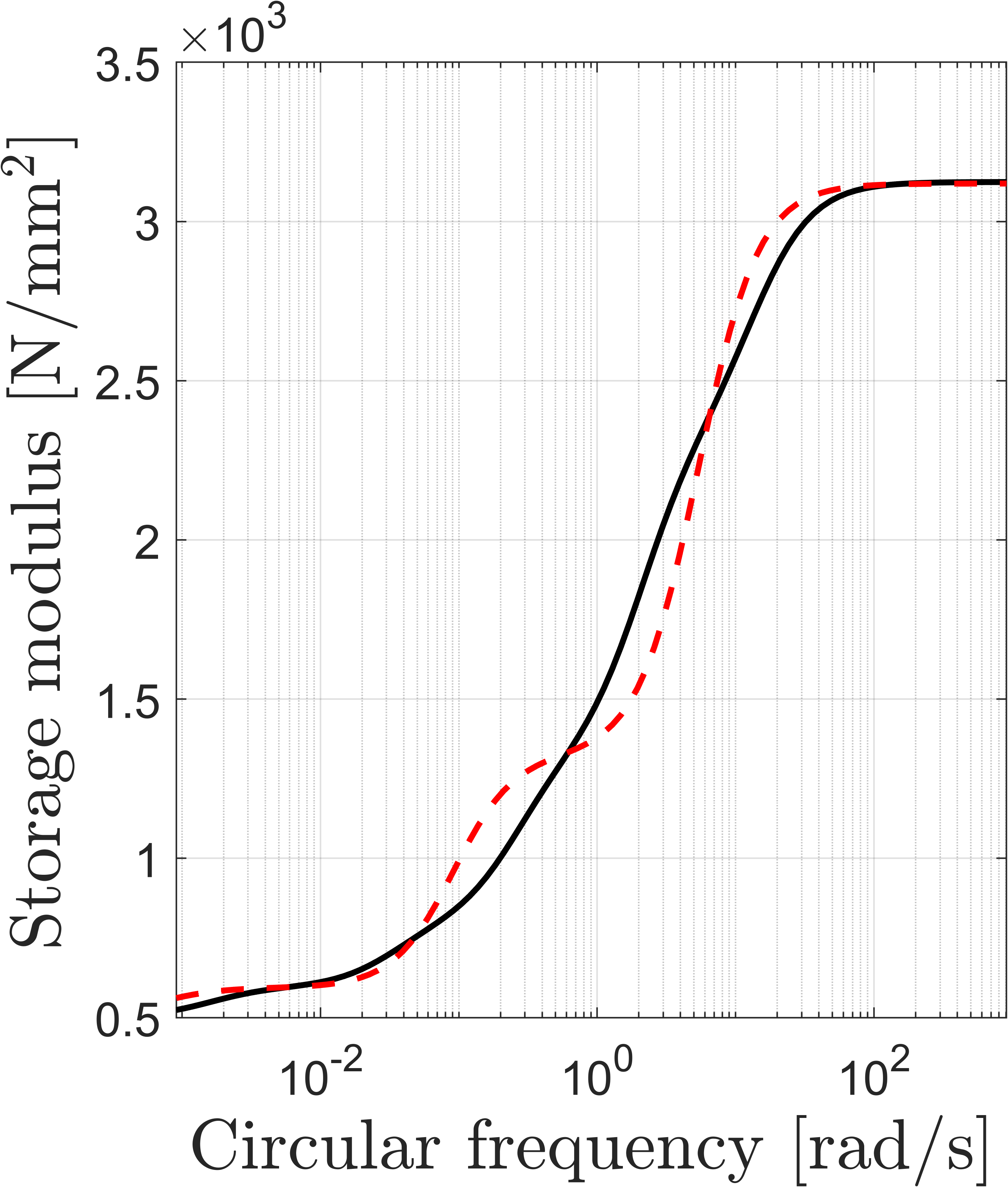} 
\end{subfigure}
\hspace{6pt}
\begin{subfigure}{0.2\textwidth}
\centering
\includegraphics[width=1\linewidth]{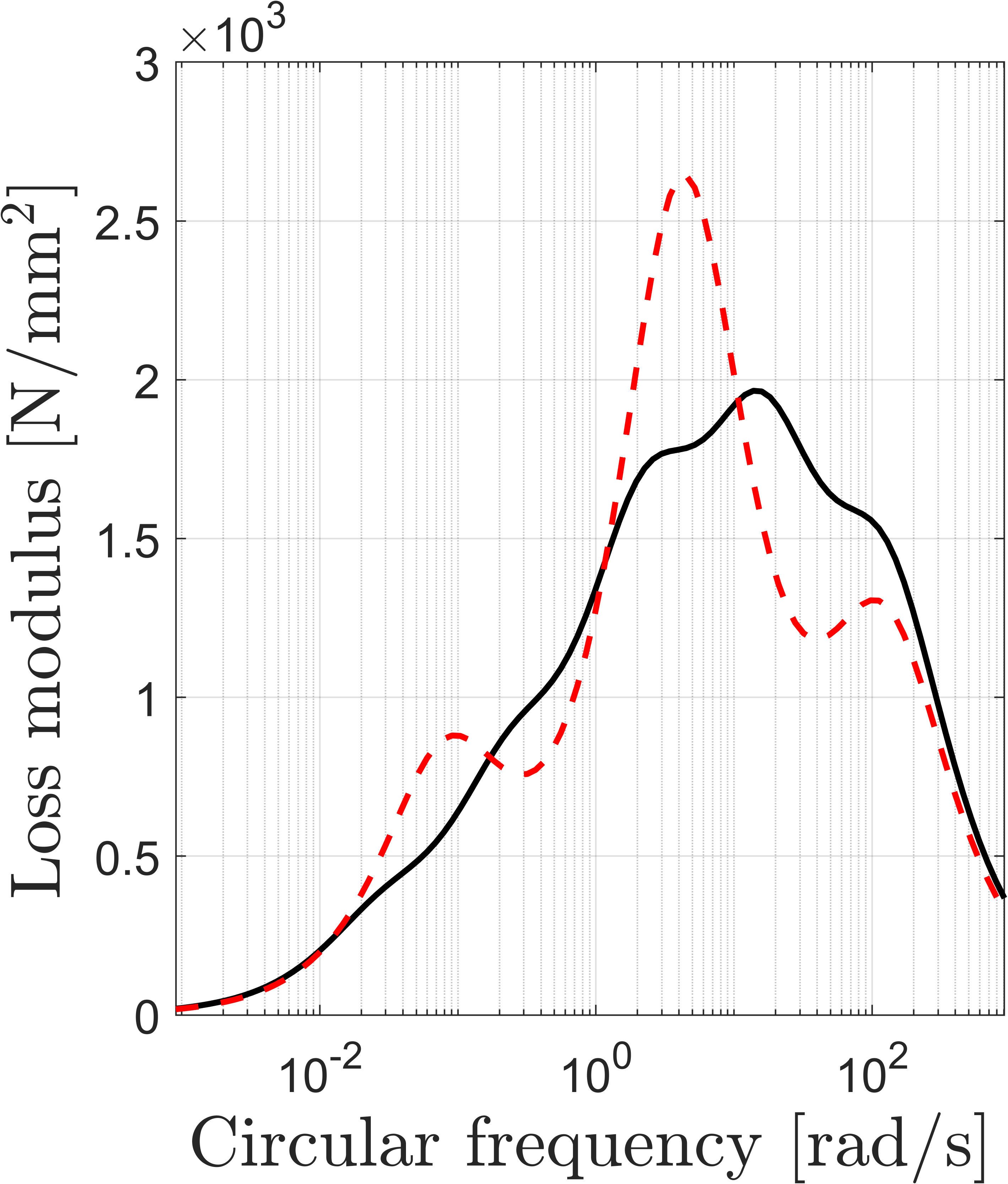} 
\end{subfigure}
\hspace{6pt}
\begin{subfigure}{0.2\textwidth}
\centering
\includegraphics[width=1\linewidth]{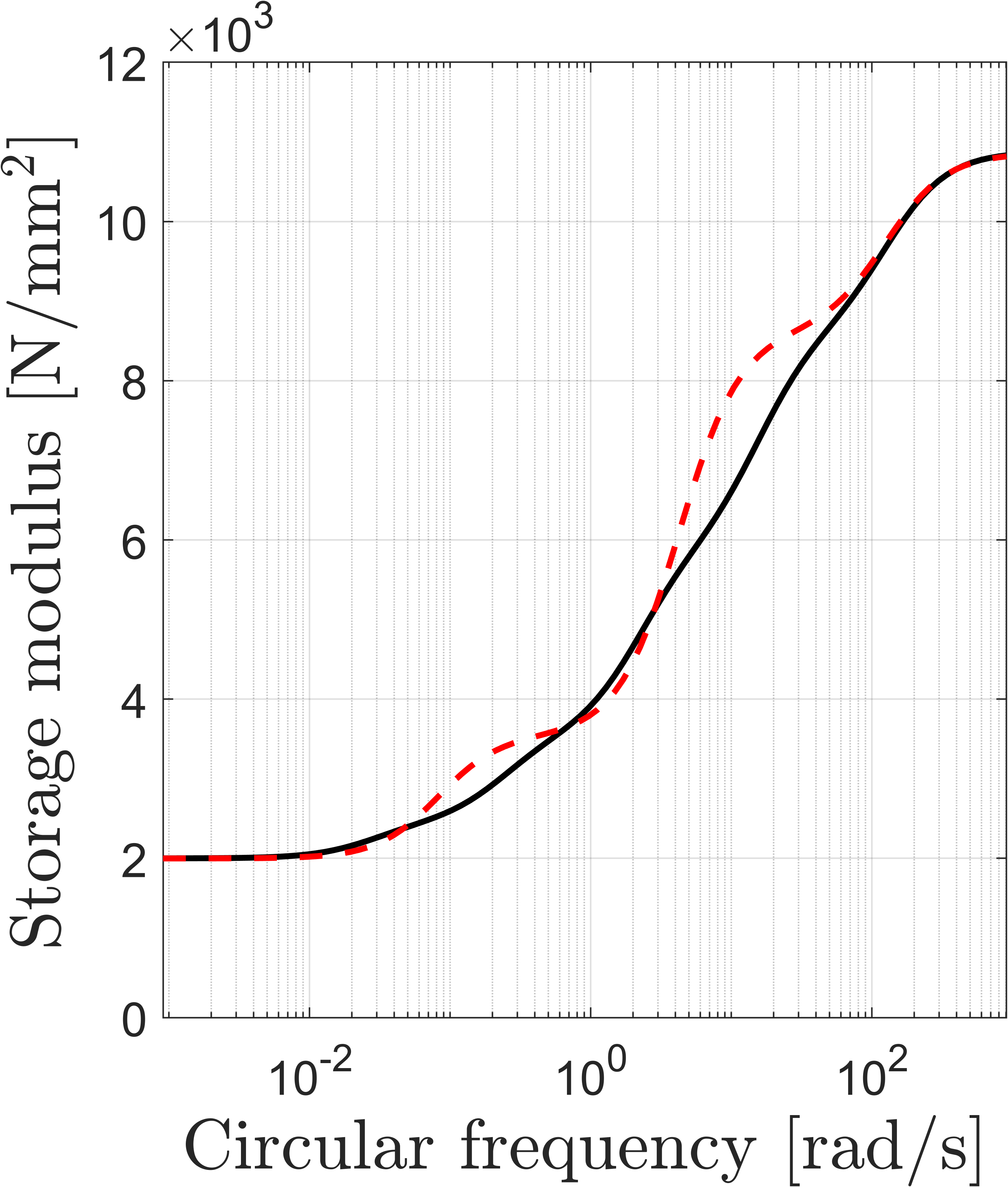} 
\end{subfigure}\\[12pt]
\begin{subfigure}{0.02\textwidth}
\begin{turn}{90} 
4 Clusters
\end{turn}
\end{subfigure}
\hspace{6pt}
\begin{subfigure}{0.2\textwidth}
\centering
\includegraphics[width=1\linewidth]{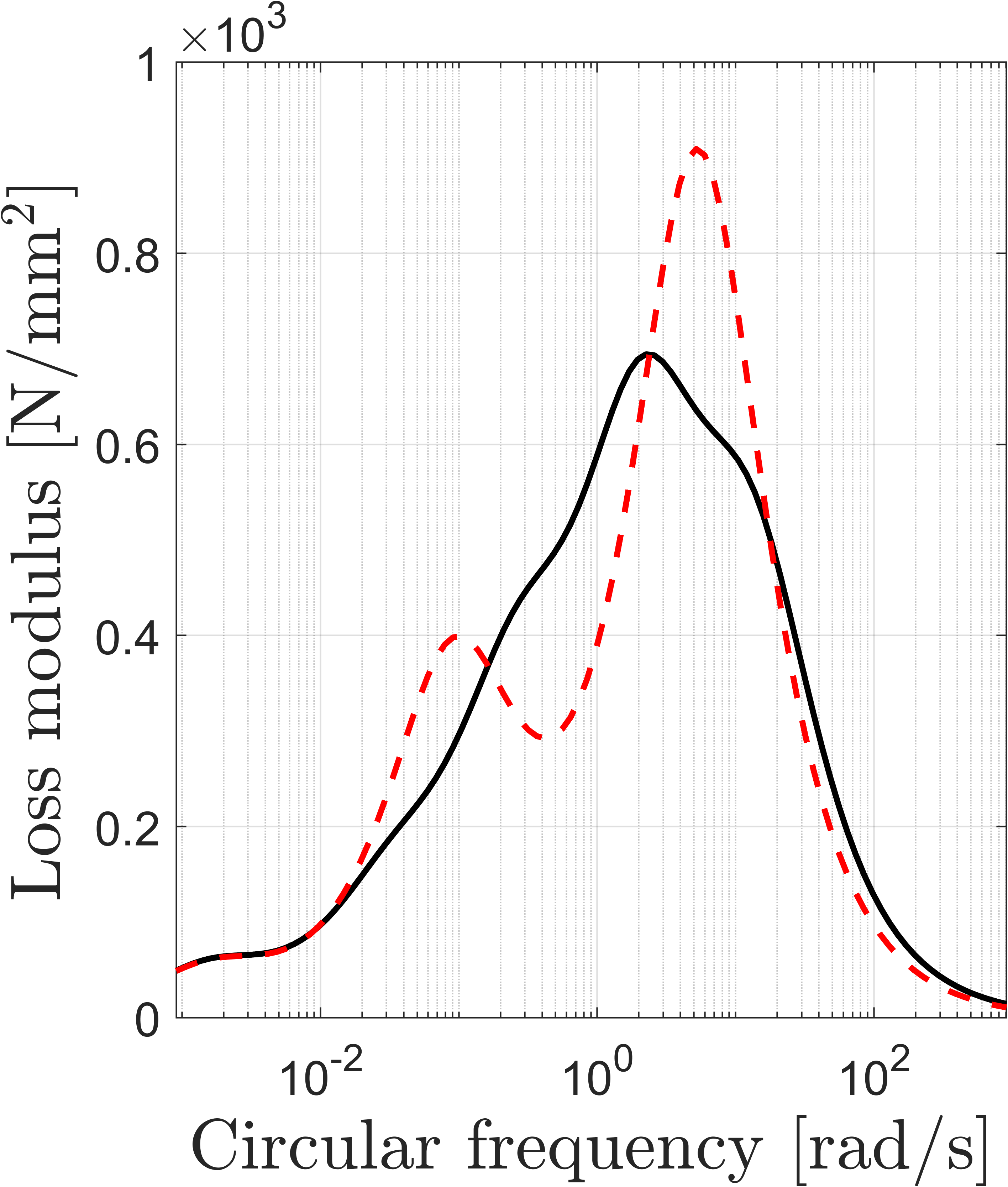} 
\end{subfigure}
\hspace{6pt}
\begin{subfigure}{0.2\textwidth}
\centering
\includegraphics[width=1\linewidth]{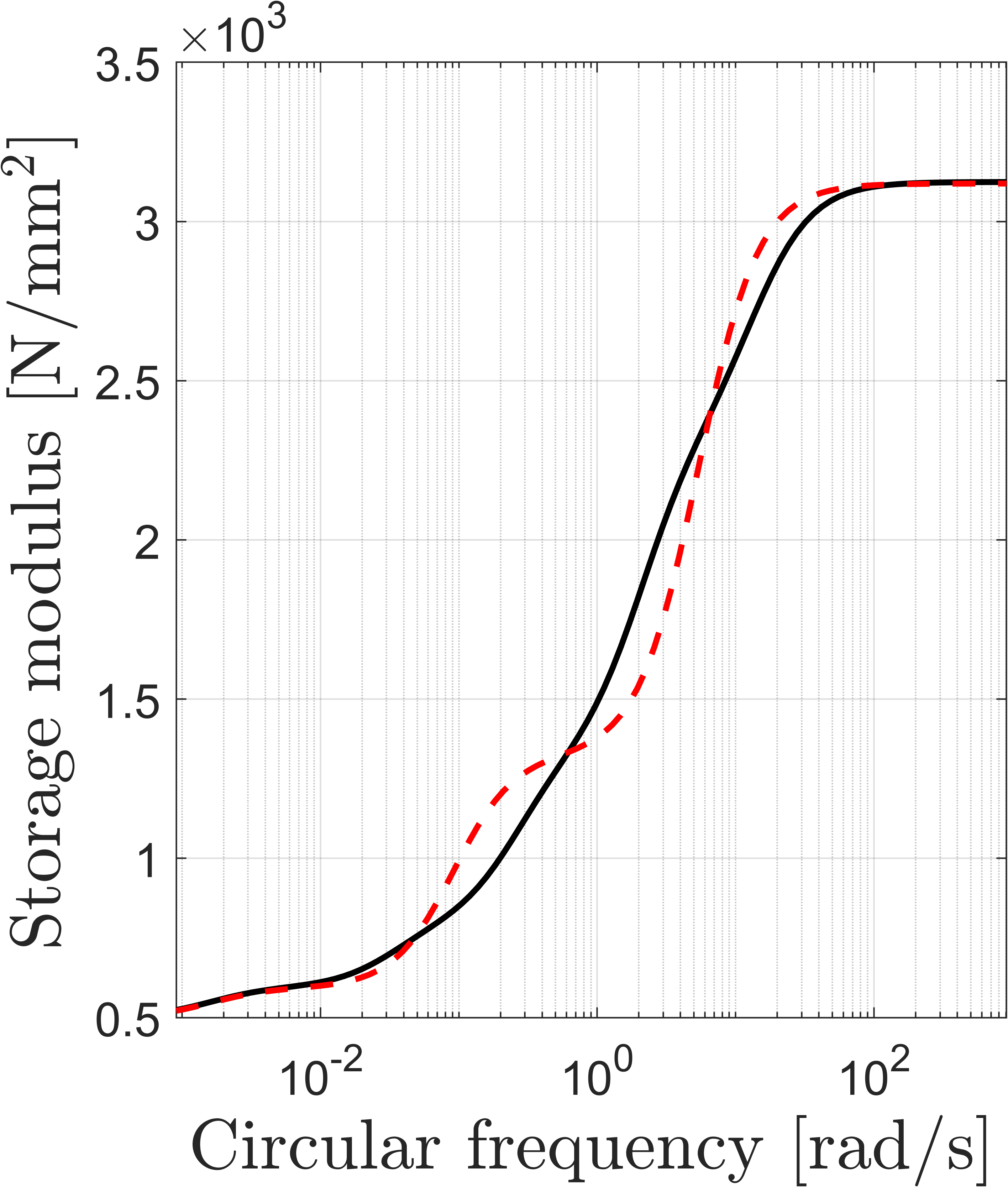} 
\end{subfigure}
\hspace{6pt}
\begin{subfigure}{0.2\textwidth}
\centering
\includegraphics[width=1\linewidth]{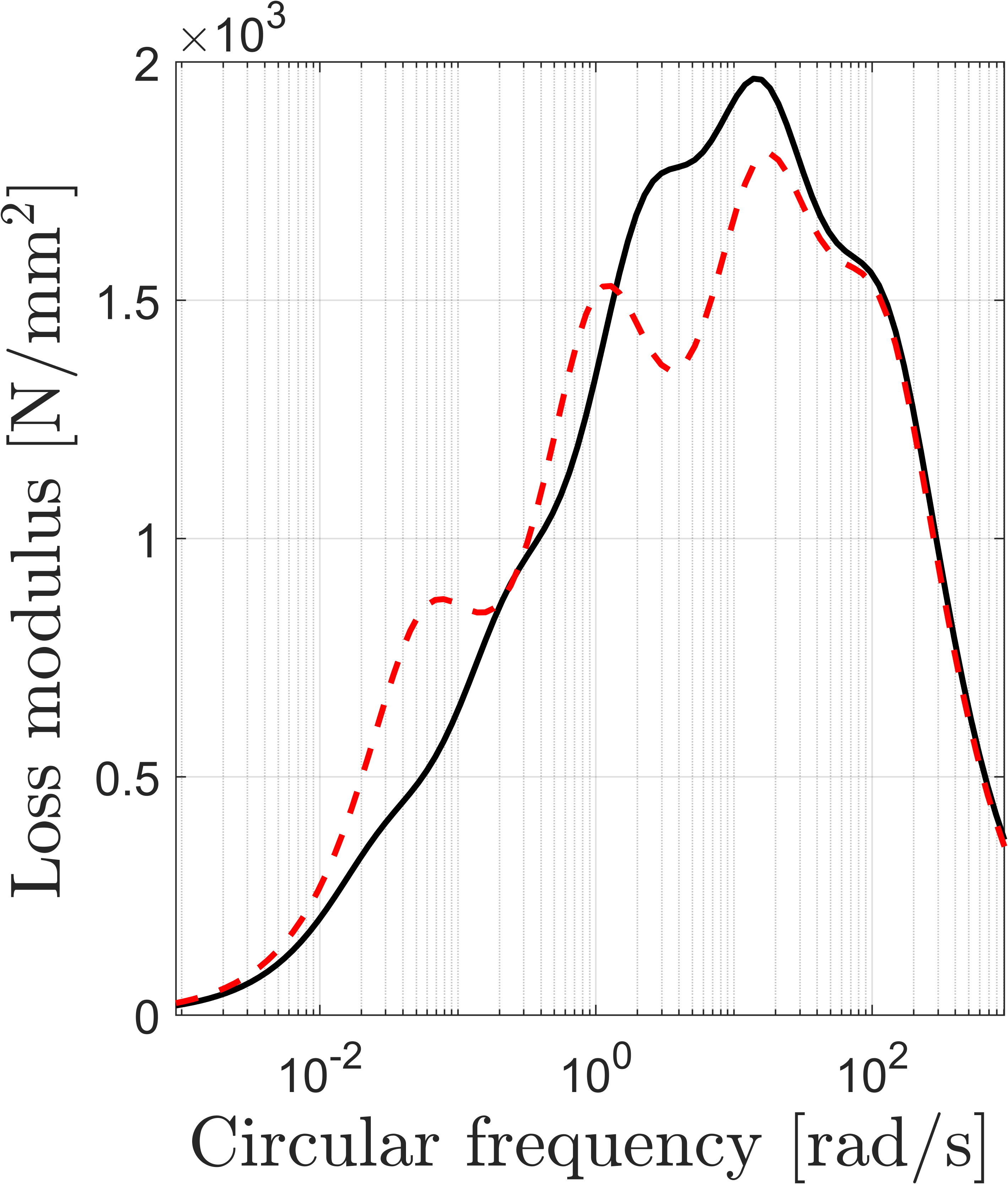} 
\end{subfigure}
\hspace{6pt}
\begin{subfigure}{0.2\textwidth}
\centering
\includegraphics[width=1\linewidth]{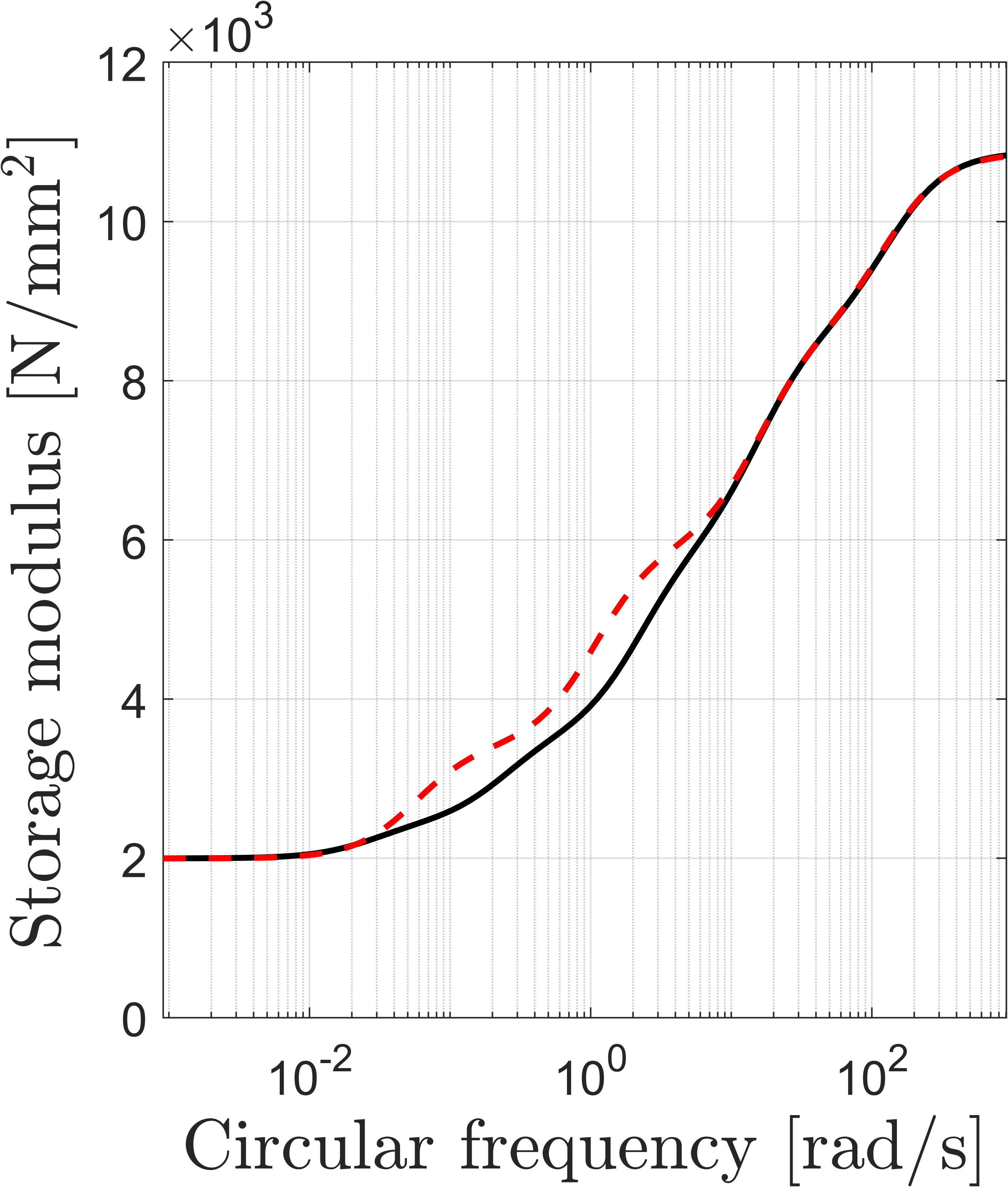} 
\end{subfigure}\\[12pt]
\begin{subfigure}{0.02\textwidth}
\begin{turn}{90} 
5 Clusters
\end{turn}
\end{subfigure}
\hspace{6pt}
\begin{subfigure}{0.2\textwidth}
\centering
\includegraphics[width=1\linewidth]{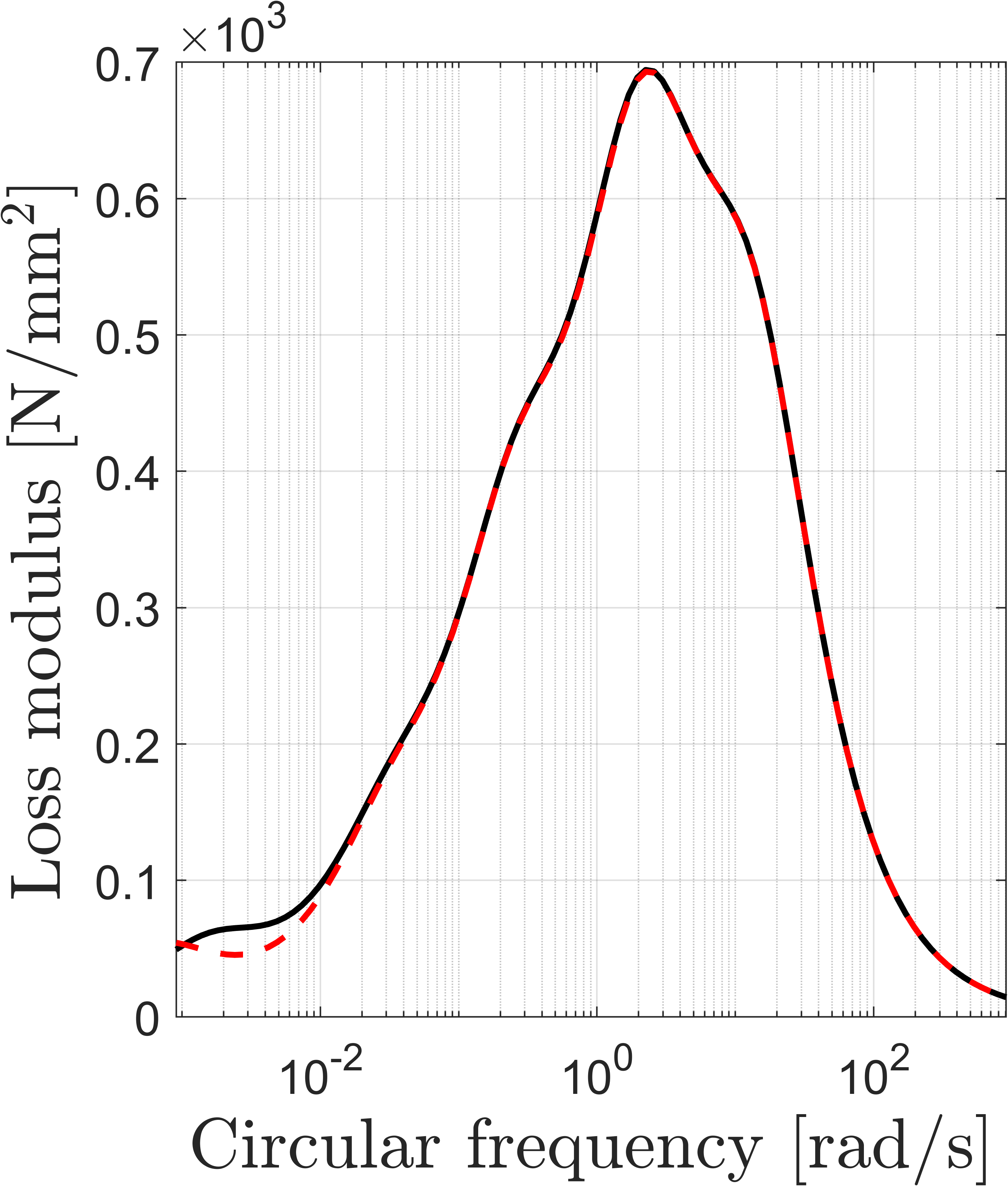} 
\end{subfigure}
\hspace{6pt}
\begin{subfigure}{0.2\textwidth}
\centering
\includegraphics[width=1\linewidth]{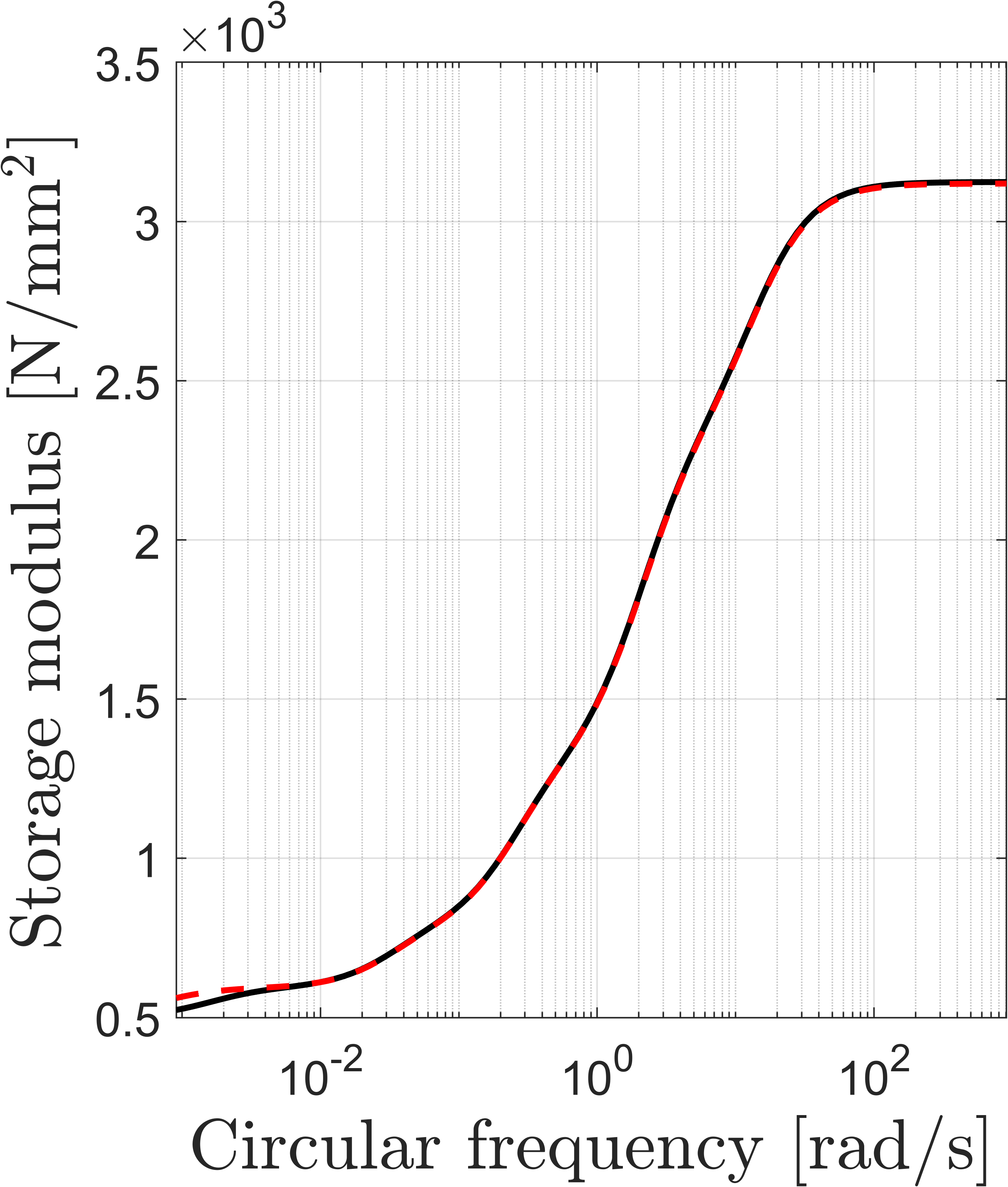} 
\end{subfigure}
\hspace{6pt}
\begin{subfigure}{0.2\textwidth}
\centering
\includegraphics[width=1\linewidth]{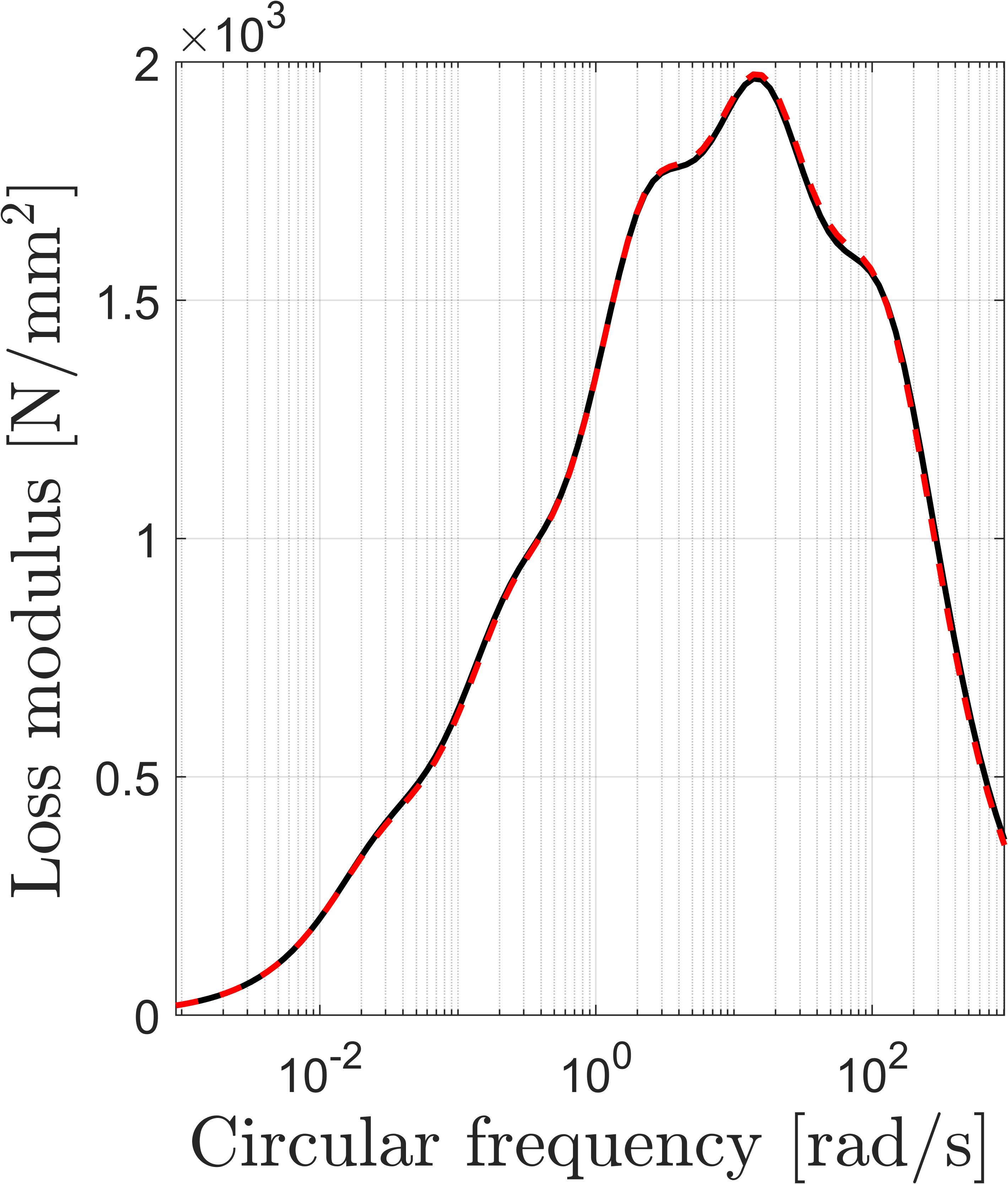} 
\end{subfigure}
\hspace{6pt}
\begin{subfigure}{0.2\textwidth}
\centering
\includegraphics[width=1\linewidth]{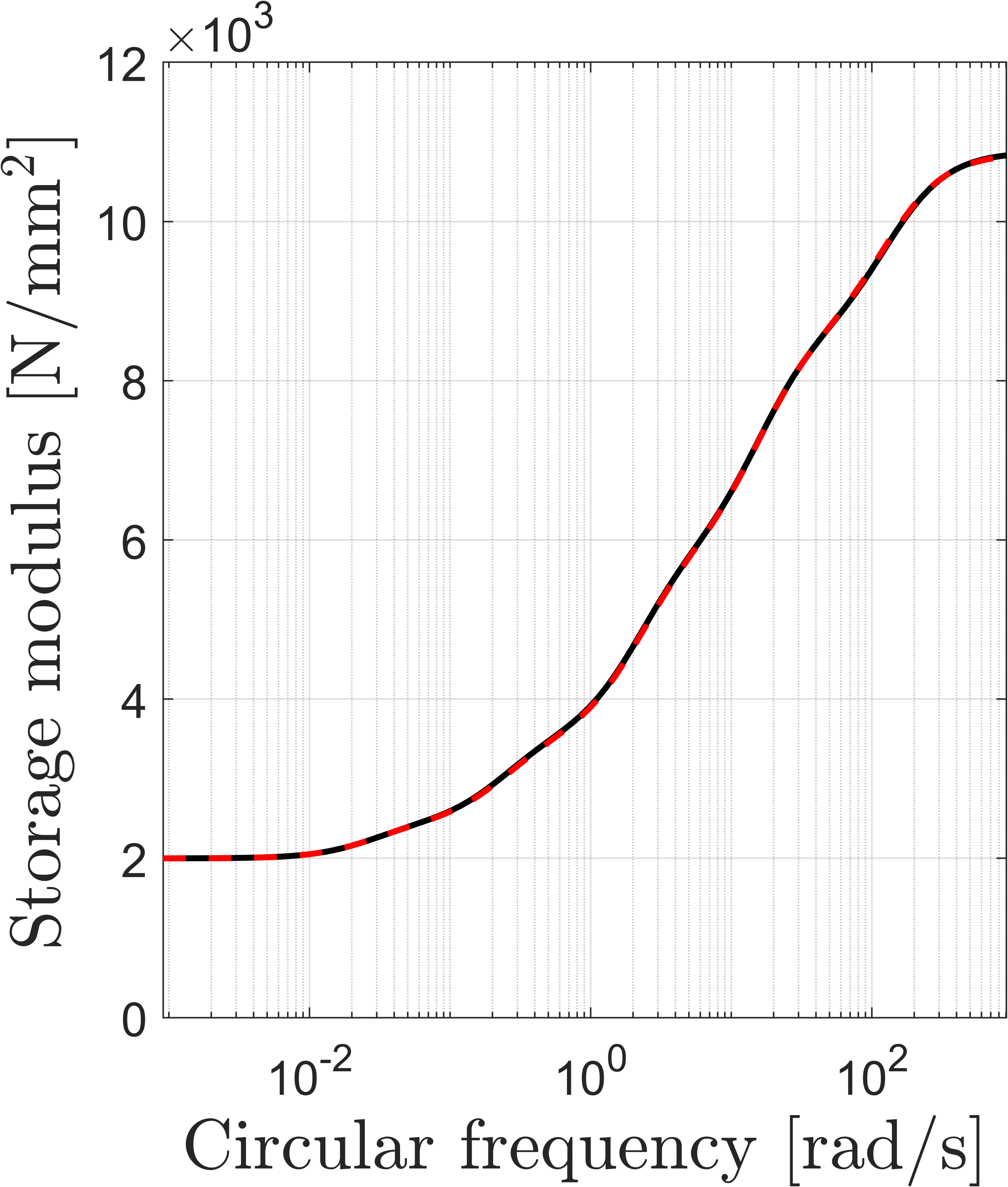} 
\end{subfigure}\\[12pt]
\begin{subfigure}{0.02\textwidth}
\begin{turn}{90} 
6 Clusters
\end{turn}
\end{subfigure}
\hspace{6pt}
\begin{subfigure}{0.2\textwidth}
\centering
\includegraphics[width=1\linewidth]{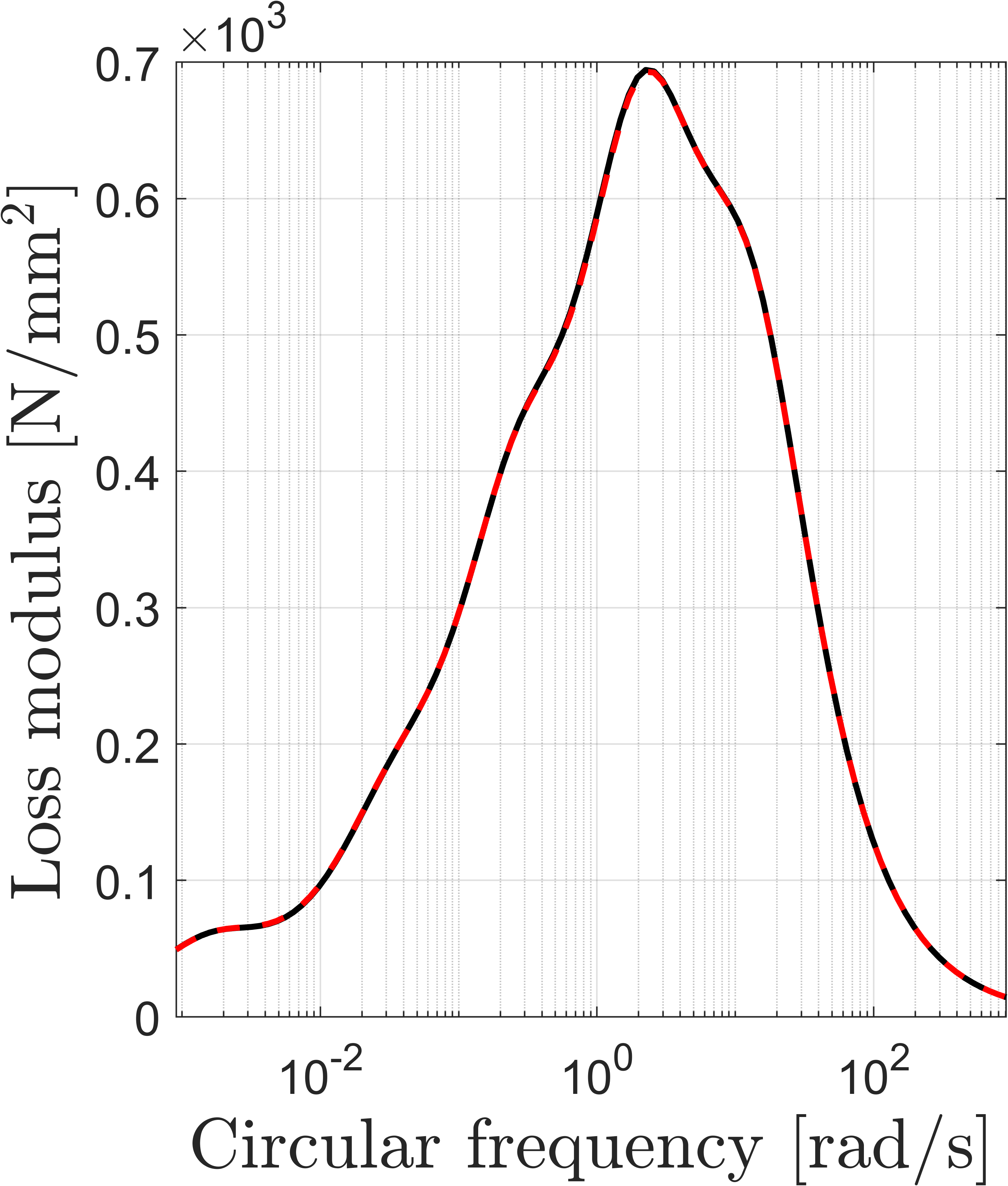} 
\end{subfigure}
\hspace{6pt}
\begin{subfigure}{0.2\textwidth}
\centering
\includegraphics[width=1\linewidth]{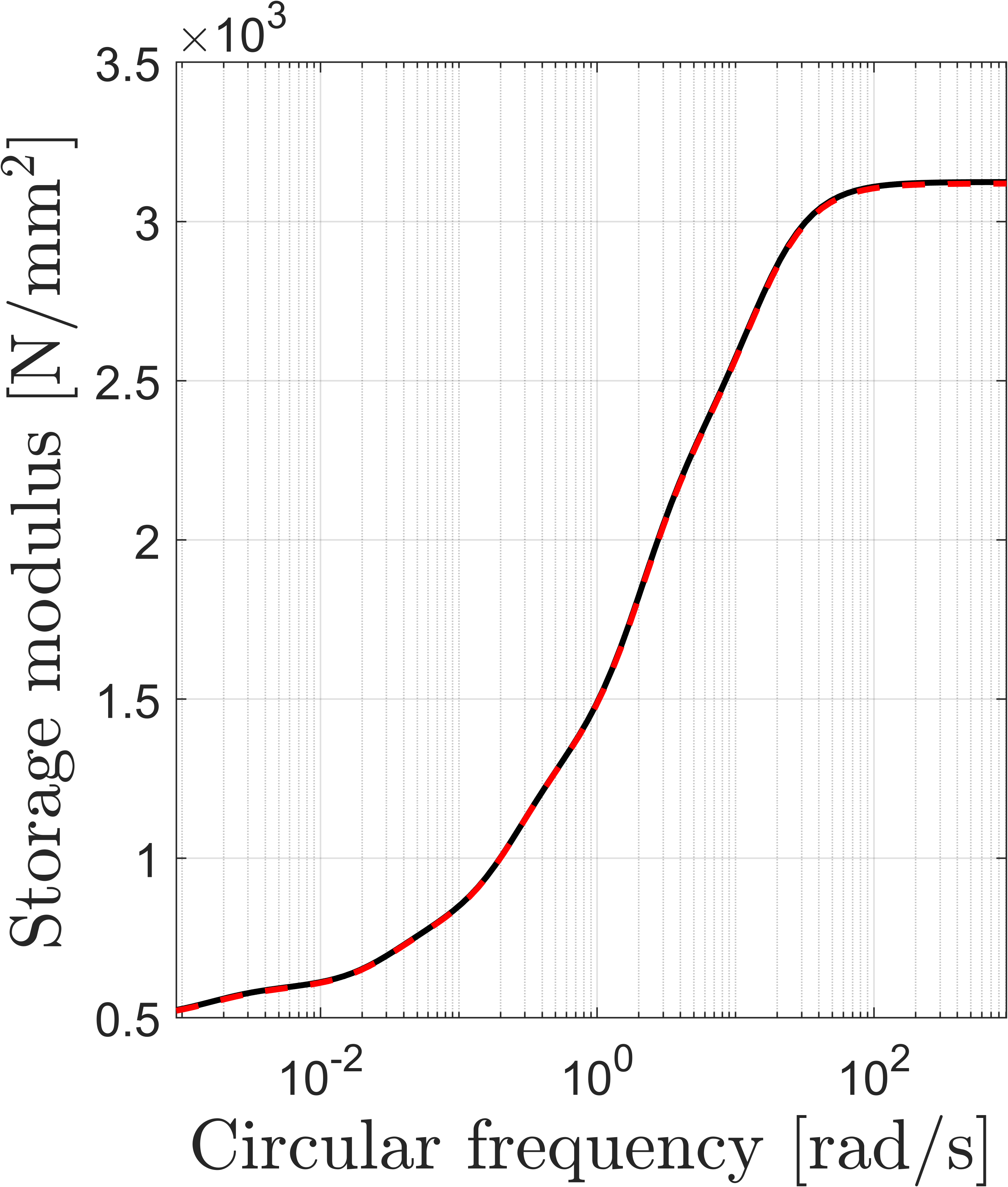} 
\end{subfigure}
\hspace{6pt}
\begin{subfigure}{0.2\textwidth}
\centering
\includegraphics[width=1\linewidth]{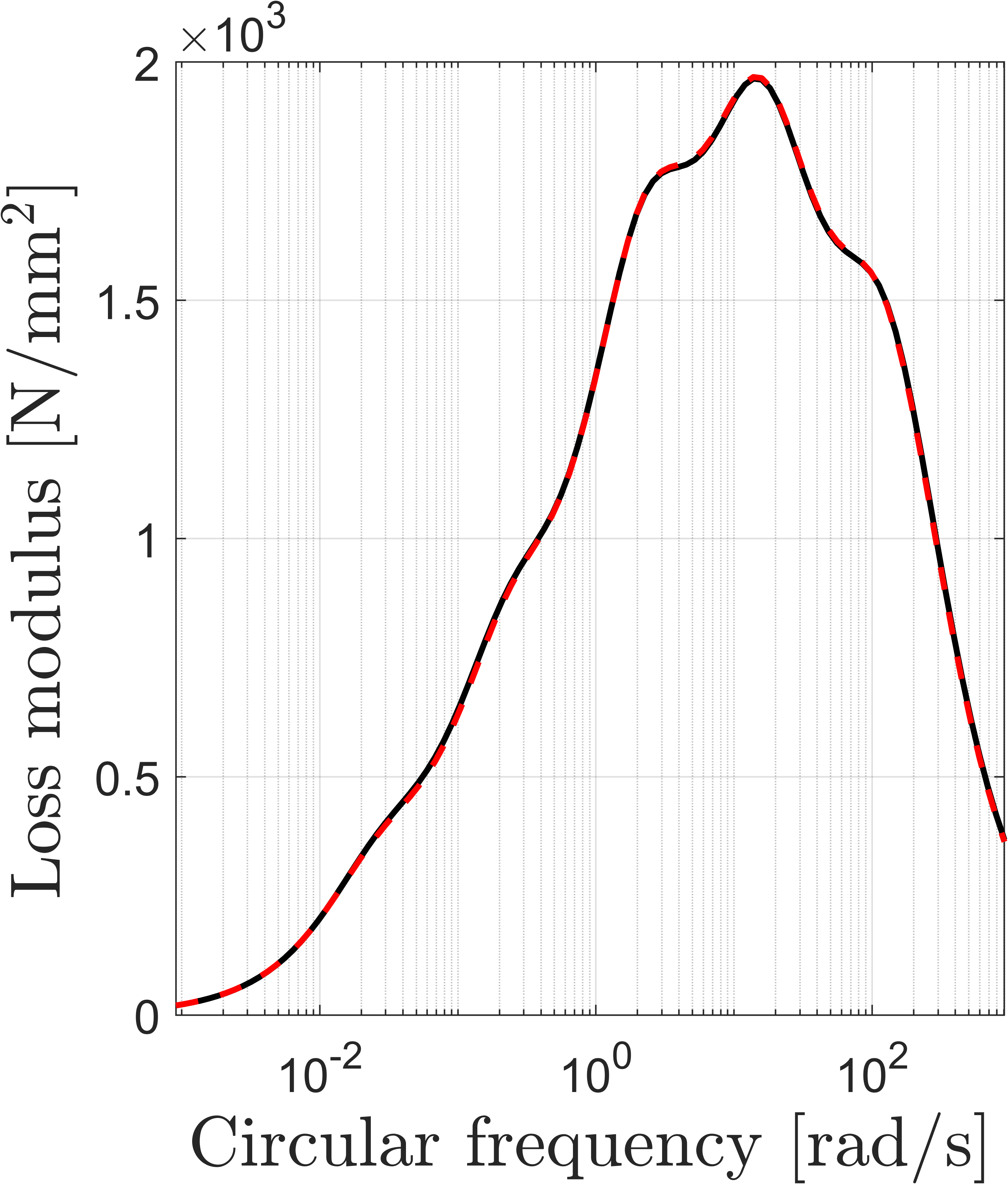} 
\end{subfigure}
\hspace{6pt}
\begin{subfigure}{0.2\textwidth}
\centering
\includegraphics[width=1\linewidth]{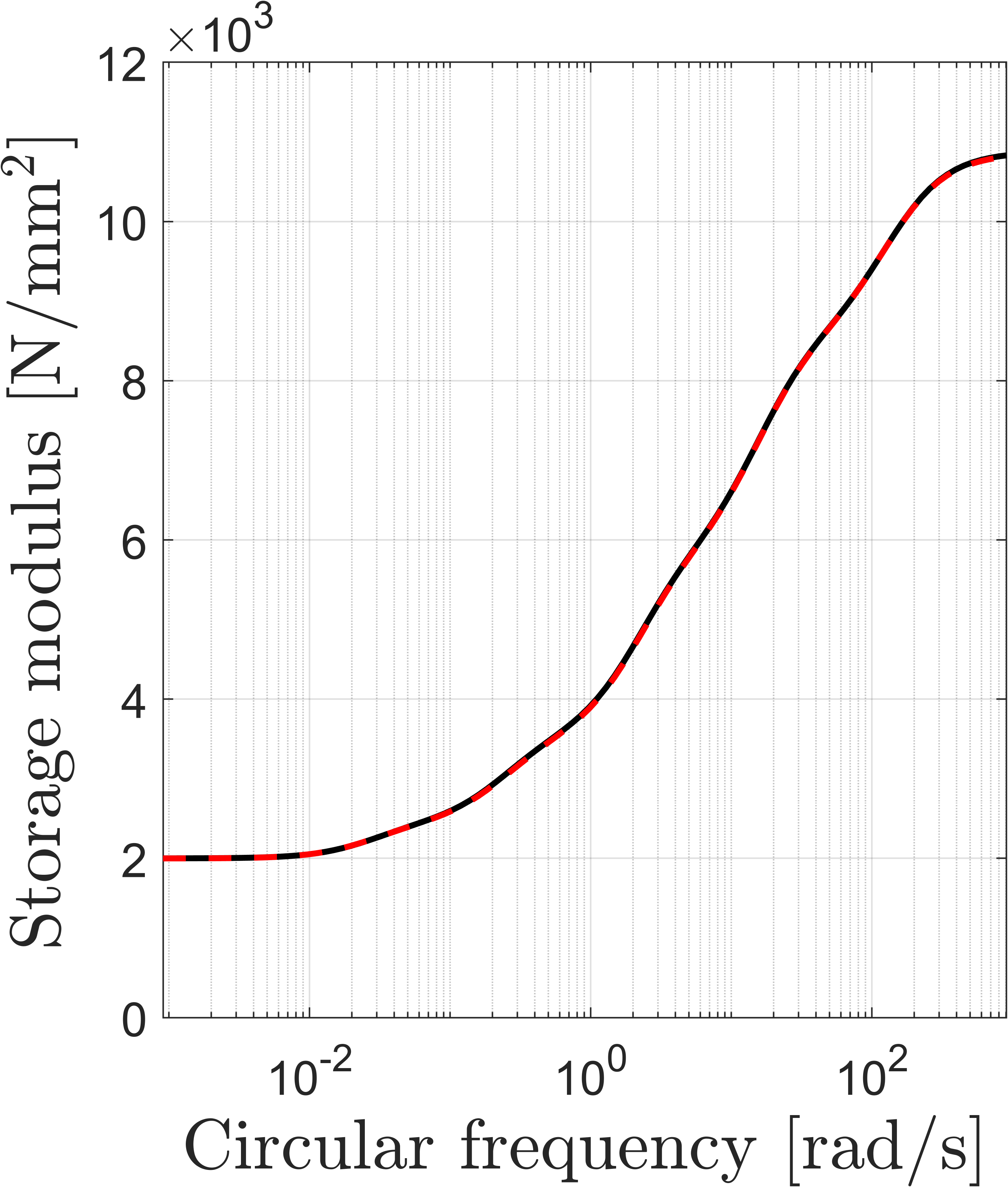} 
\end{subfigure}
\caption{Comparisons of true and identified response functions ordered as: shear loss, shear storage, bulk loss,  bulk storage (rows-wise from left to write) and with increasing number of clusters from 2 to 6 (columns-wise from top to bottom). Noisy data with $\sig = \SI{1e-2}{mm}$.}
\label{fig:comparison_moduli_sig1e-4} 
\end{figure}

Finally, we plot in Figure~\ref{fig:comparison_moduli_sig1e-4} the comparison between identified and true shear/bulk loss/storage functions. As expected, for the 5-cluster solution a small deviation from the true response is observed at very low frequencies (long-term response). Consistently with Figure~\ref{fig:MSElog_vs_clsts_sig5em4+1em5}, as a sixth cluster is added (bottom row), an excellent agreement with the true response is achieved at all frequencies. 

\begin{table}[h!]
	\centering
	\begin{tabular}{r r r r | r r r r}
	 \multicolumn{4}{c}{5 Clusters} &   \multicolumn{4}{c}{6 Clusters}\\
	$\f G$			 & $\f \tau_G$	 	&  $\f K$			& $\f \tau_K$ & $\f G$			 & $\f \tau_G$	 	&  $\f K$			& $\f \tau_K$	\\[-3pt]
	$[\SI{}{N/mm^2}]$	 & $[\SI{}{s}]$ 	&  $[\SI{}{N/mm^2}]$	& $[\SI{}{s}]$ & $[\SI{}{N/mm^2}]$	 & $[\SI{}{s}]$ 	&  $[\SI{}{N/mm^2}]$	& $[\SI{}{s}]$		\\
	\hline
  	  495				 &	  -				&	 1999				&	-&  495				 &	  -				&	 1999				&	-\\		
  	  777				 &  0.0726	  			&	 2237				& 0.008 &777				 &0.0726	  			&	 2110				&  0.0078\\
  	 1017			 &  0.4793  			&	 2709				&  0.0635 & 	 1017			 &0.4793	  			&	  127				&  0.0082\\
  	  531				 &3.9232	  			&	2368				& 0.4541 &    531				 &3.9232	  			&	 2709				&    0.0635\\
  	  201				 &30.4283 				&	 1091				& 4.1405 & 	  201				 &30.4283				&	2368				& 0.4541\\
              99				 &1571.1208	  			&	 457			&       35.7694 &      95				 &622.7509	  			&	 1091			& 4.1405\\
                				 &						&					&	 		&   3				 & 10000	  			&	457			&       35.7694\\
              \hline
	\end{tabular}
	\caption{Identified parameters with 5 clusters after the two-stage procedure with noise $\sig = \SI{1e-2}{mm}$.}\label{tb:discovered_params_sig1em4_5-6cls}
\end{table}

\section{Conclusions}
We extended EUCLID, an automated material model discovery and identification strategy relying on full-field displacement and net force data, to linear viscoelasticity. For this case, we perform \textit{a priori} model selection  and adopt a generalized Maxwell model expressed by a Prony series, which is known to be able to approximate an arbitrarily complex linear viscoelastic behavior if a sufficiently large number of terms is included. For the identification procedure we deploy EUCLID, which consists of four ingredients: i. the data, assumed to be delivered from material testing on a single specimen, using a loading excitation with a sufficiently rich frequency content and monitoring the full-field displacements (e.g. with DIC); ii. a very wide material model library - in our case, a very large number of terms in the Prony series, corresponding to equally spaced relaxation times on a logarithimc scale within a chosen range; iii. the physics constraint of linear momentum balance, enforced weakly on the data both in the interior and on the loaded sides of the specimen; iv. the sparsity constraint, enforced through sparsity-promoting regularization in the optimization problem. 

The devised strategy comprises two stages. Stage 1 relies on sparse regression; starting from a very large number of terms in the Prony series, it enforces linear momentum balance on the data and exploits sparsity-promoting regularization to drastically reduce the number of terms (identifying the few most relevant relaxation times) and simultaneously identify the values of the non-zero material parameters (i.e. the corresponding bulk and shear moduli). Stage 2 relies on k-means clustering; starting from the reduced set of terms in the Prony series from stage 1, it further reduces their number by grouping together Maxwell elements with very close relaxation times and summing the corresponding moduli. Automated procedures are proposed for the choice of the regularization parameter in stage 1 and of the number of clusters in stage 2. The overall strategy is demonstrated on artificial numerical data, both without and with the addition of noise, 
and shown to efficiently and accurately identify a linear viscoelastic model with five relaxation times across four orders of magnitude, out of a library with several hundreds of terms spanning relaxation times across seven orders of magnitude.

Further research should address the application to real experimental data, \r{which is expected to pose challenges related to the quality of the DIC measurements (e.g., loss of grid points during loading, unavailable measurements close to the boundary, non-Gaussian noise). A further interesting and meaningful extension would be that to non-linear viscoelasticity.}

\section*{Acknowledgments}
EM was partially supported by the National Centre for HPC, Big Data and Quantum Computing funded by the European Union within the Next Generation EU recovery plan (CUP B83C22002830001). This support is gratefully acknowledged.

MF and LDL would like to acknowledge funding by SNF through grant N. $200021\_204316$ ``Unsupervised data-driven discovery of material laws''.


\bibliographystyle{elsarticle-num}
\bibliography{viscoel,viscoel_MF}

\end{document}